\documentclass[a4paper,11pt]{article}
\pdfoutput=1 

\usepackage{jheppub} 
\usepackage{bm,amssymb,slashed,graphicx,multirow,soul,mathtools,xspace,array}  
\usepackage{float}                   
\allowdisplaybreaks
\usepackage{ bbold }
\usepackage{subfigure}

\newcommand{\todo}[1]{{\color{red} \ifmmode\else[todo]\fi #1}}
\usepackage[usenames,dvipsnames]{xcolor}
     \definecolor{hgreen}{rgb}{0,.3,0}
      \definecolor{darkgreen}{rgb}{0.3,.8,0.2}
     \definecolor{hred}{rgb}{.3,0,0}
     \definecolor{hblue}{rgb}{0,0,.3}
     \definecolor{LightGray}{gray}{0.95}

\newcommand{\lp}{\left(}
\newcommand{\rp}{\right)}

\newcommand{\beq}{\begin{equation} }
\newcommand{\eeq}{\end{equation}} 
\newcommand{\bi}{\begin{itemize} }
\newcommand{\ei}{\end{itemize} }

\definecolor{Red}{rgb}{1.,0.,0.}
\definecolor{Grn}{rgb}{0.,0.75,0.}
\definecolor{Blu}{rgb}{0.,0.,1.}

\DeclareMathOperator{\diag}{diag}   
\let\Re\relax
\DeclareMathOperator{\Re}{Re}
\let\Im\relax
\DeclareMathOperator{\Im}{Im}
\DeclareMathOperator{\Tr}{Tr}
\newcommand{\lrpartial}{\negthickspace\stackrel{\leftrightarrow}{\partial}\negthickspace{}}

\usepackage[T1]{fontenc} 

\usepackage{amsmath,amssymb,epsfig,ulem,color,slashed}
\allowdisplaybreaks  

\title{\boldmath Anomaly free Froggatt-Nielsen models of flavor }

\author[1]{Aleks Smolkovi{\v c},}
\author[2]{Michele Tammaro,}
\author[2]{Jure Zupan}

\affiliation[1]{Jo{\v z}ef Stefan Institute, Jamova 39, 1000 Ljubljana, Slovenia}
\affiliation[2]{Department of Physics, University of Cincinnati, Cincinnati, Ohio 45221,USA}

\emailAdd{aleks.smolkovic@ijs.si}
\emailAdd{tammarme@mail.uc.edu}
\emailAdd{zupanje@ucmail.uc.edu}

\abstract{We introduce two anomaly free versions of Froggatt-Nielsen (FN) models, based on either $G_{\rm FN}=U(1)^3$ or $G_{\rm FN}=U(1)$ horizontal symmetries, that generate the SM quark and lepton flavor structures. The structure of these ``inverted'' FN models is motivated by the clockwork mechanism: the chiral fields, singlets under $G_{\rm FN}$, are supplemented by chains of vector-like fermions charged under $G_{\rm FN}$. Unlike the traditional FN models the hierarchy of quark and lepton masses is obtained as an expansion in $M/\langle \phi\rangle$, where $M$ is the typical vector-like fermion mass, and $\langle \phi\rangle$ the flavon vacuum expectation value. The models can be searched for through deviations in flavor observables such as $K-\bar K$ mixing, $\mu\to e$ conversion, etc., where the present bounds restrict the masses of vector-like fermions to be above ${\mathcal O}(10^7~{\rm GeV})$. If $G_{\rm FN}$ is gauged, the models can also be probed by searching for the flavorful $Z'$ gauge bosons. In principle, the $Z'$s can  be very light, and can be searched for using precision flavor, astrophysics, and beam dump experiments. }

\begin{document} 

\maketitle

\flushbottom

\section{Introduction}
\label{sec:intro}
The masses of the Standard Model (SM) quarks and leptons exhibit large hierarchies \cite{Tanabashi:2018oca}. The mass of the lightest quark, the up quark, is five orders of magnitude smaller than the mass of the heaviest quark, the top quark. The hierarchies in the leptonic sector are even more striking; the heaviest charged lepton, the tau, has a mass that is at least ten orders of magnitude larger than the mass of the heaviest neutrino. Furthermore, in the quark sector the measurable misalignment between the  gauge and mass-eigenstates is also hierarchical, as encoded in the CKM matrix. For leptons, on the other hand, the three mixing angles in the PMNS matrix are all large. 

This very peculiar flavor structure calls for a dynamical explanation. A very attractive solution to the above SM flavor puzzle is provided by the Froggatt-Nielsen (FN) models of flavor \cite{Froggatt:1978nt} (see also \cite{Leurer:1992wg,Leurer:1993gy}).  In FN models the SM mass hierarchy is due to a horizontal $U(1)_{\rm FN}$, spontaneously broken by the vacuum expectation value (vev) of the flavon field, $\langle \phi \rangle$.  The SM fermions carry horizontal $U(1)_{\rm FN}$ charges. The up and down quark Yukawa matrices are then given by\footnote{To shorten the discussion we focus manly on quarks, but also show in Section \ref{sec:leptons} how the discussion naturally generalizes to leptons.}
\beq
\big(Y_u\big)_{ij}\sim \biggr(\frac{\langle \phi\rangle}{M}\biggr)^{[q_i]-[u_j]},\qquad \big(Y_d\big)_{ij}\sim \biggr(\frac{\langle \phi\rangle}{M}\biggr)^{[q_i]-[d_j]},
\eeq
with $M$ the typical mass of the new FN vector-like fermions,  while $[q_i], [u_i], [d_i]$ are the horizontal charges of the SM  left-handed quark doublets, the up quarks, and the down quarks, respectively, with $i,j=1,2,3$ the generational indices. For $\langle \phi\rangle/M\sim 0.2$ one can reproduce the parametric size of the CKM matrix and the quark masses with the appropriate choice of horizontal charges, $[q_i], [u_i], [d_i]$ (for details, see below). For generic horizontal charges, and without additional field content, such a horizontal $U(1)_{\rm FN}$ is anomalous and thus needs to be a global symmetry. It is also possible to gauge the FN horizontal symmetry for special choices of horizontal charges for which the symmetry is not anomalous. For example, Ref.~\cite{Froggatt:1998he} builds neutrino mass models based on an anomaly free $U(1)_{\rm FN}^2$ FN symmetry, while the full list of anomaly free charge assignments for $U(1)_{\rm FN}$ can be found in  \cite{Allanach:2018vjg,Costa:2019zzy} (for related work, see also \cite{Ellis:2017nrp,DelleRose:2017xil,Correia:2019woz,Correia:2019pnn,Berger:2000sc,Chen:2006hn,Rathsman:2019wyk}). The gauged $U(1)_{\rm FN}$ can also be left naively anomalous, if the cancellation of $U(1)_{\rm FN}$ gauge anomalies occurs through the Green-Schwarz mechanism \cite{Ibanez:1994ig,Binetruy:1994ru,Jain:1994hd,Dudas:1995yu,Dudas:1996fe,Binetruy:1996xk,Irges:1998ax,King:1999mb,King:1999cm,Shafi:2000su,Dreiner:2003yr,Dreiner:2007vp}.

In this paper we explore a twist to the original FN proposal. In our modified set of FN models the horizontal $U(1)_{\rm FN}$ is anomaly free for any horizontal charge assignments, $[q_i], [u_i], [d_i]$, making it much easier to obtain agreement with the measured fermion masses and mixings. In this case up and down quark Yukawa matrices are given by 
\beq
\label{eq:YuYd:newFN}
\big(Y_u\big)_{ij}\sim \biggr(\frac{M}{\langle \phi\rangle}\biggr)^{[q_i]-[u_j]},\qquad \big(Y_d\big)_{ij}\sim \biggr(\frac{M}{\langle \phi\rangle}\biggr)^{[q_i]-[d_j]}.
\eeq
In these ``inverted FN models'' the SM mass hierarchy is obtained for flavon vev that is parametrically larger than the vector-like masses, e.g., for $\langle \phi \rangle / M\sim 5$. Such a realization of the FN flavor model was discussed in Ref.~\cite{Alonso:2018bcg}, where it was given as a possible UV realization of the clockwork mechanism \cite{Giudice:2016yja}. We extend the results of Ref.~\cite{Alonso:2018bcg} in several ways. First of all, two choices for the horizontal symmetry group, $G_{\rm FN}=U(1)_{\rm FN}^3$ and $G_{\rm FN}=U(1)_{\rm FN}$, are explored. 
The choice $G_{\rm FN}=U(1)_{\rm FN}^3$ most closely resembles the analysis in Ref.~\cite{Alonso:2018bcg}, except that we allow ${\mathcal O}(1)$ variations for all the Yukawa couplings. We also assume that $G_{\rm FN}$ is gauged,
and explore the phenomenology of the associated $Z'$ gauge bosons. 

\begin{figure}[t]
	\begin{center}
		\includegraphics[width=0.5\linewidth]{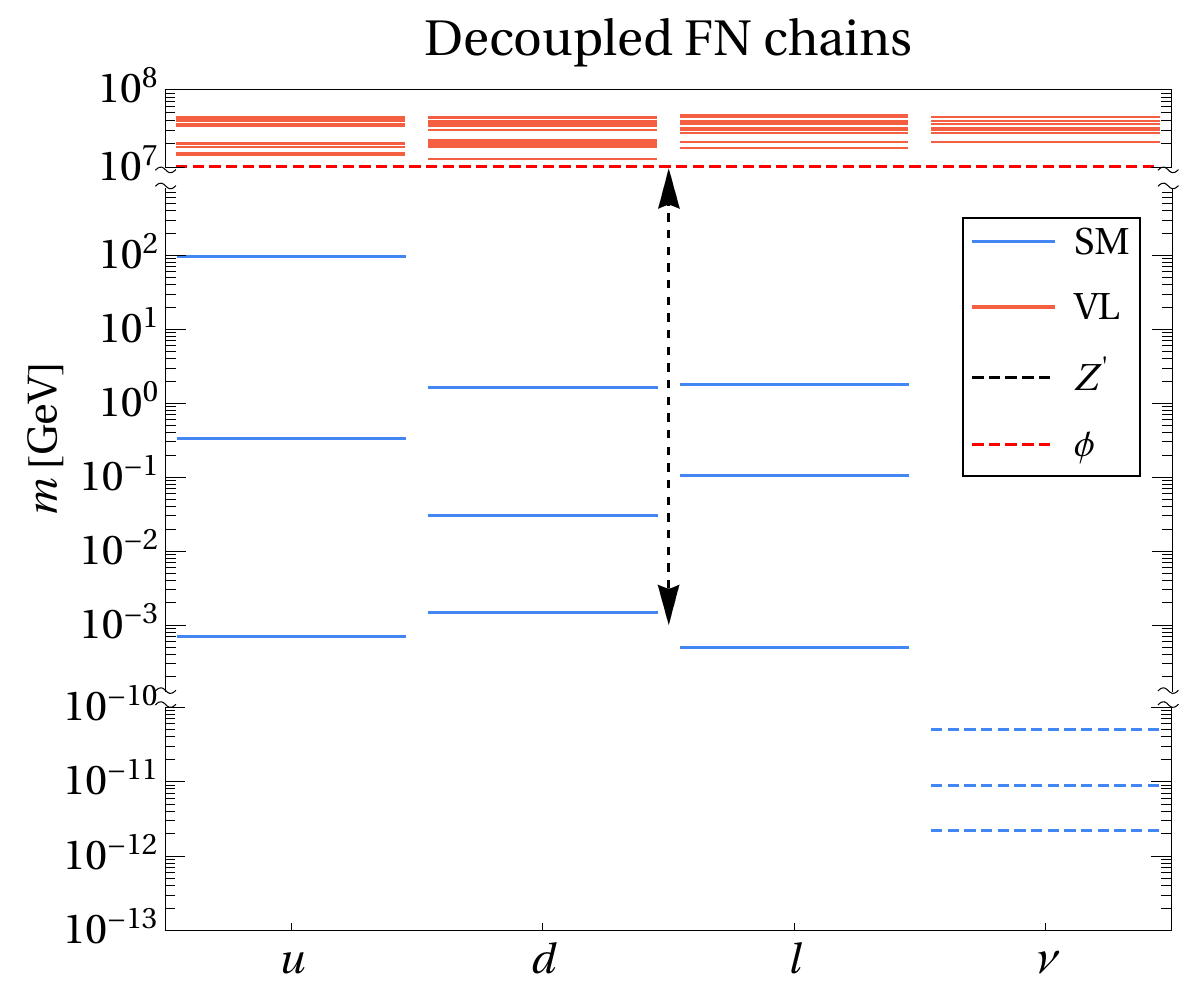}~~~~~~~~
		\includegraphics[width=0.5\linewidth]{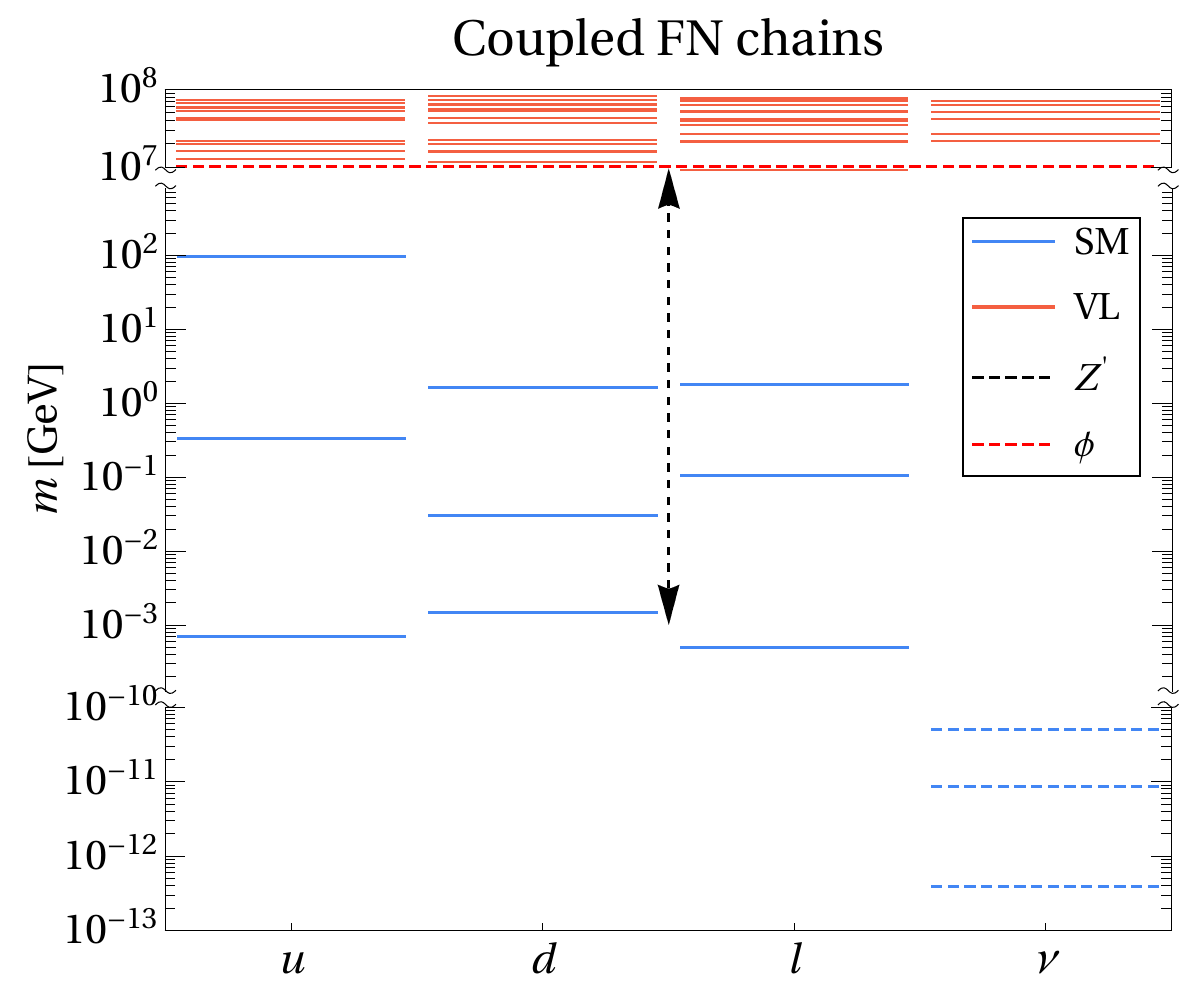}
		\caption{Mass spectra for $G_{\rm FN}=U(1)^3$ (left) and $G_{\rm FN}=U(1)$ models (right), with masses of vector-like masses denoted in red, of SM fermions in blue and the considered range of $Z'$ masses denoted with a dashed line. The spectra correspond to the two benchmarks considered in Section \ref{sec:Z'} and Appendix \ref{app:benchmarks}. The flavon mass (dashed red) is not fixed in the two benchmarks, but is set to $10^7$ GeV in the two panels, indicative of its typical value. 
		\label{fig:spectrum}}
	\end{center}
\end{figure}

The typical spectra of inverted FN models for the two choices of $G_{\rm FN}$ are shown in Fig.~\ref{fig:spectrum}. The flavor constraints require the vector-like fermions and the flavon to be heavy, for instance, the vector-like fermions need to be heavier than about ${\mathcal O}(10^7~{\rm GeV})$. The mass of $Z'$, however, can span many orders of magnitude. For gauge couplings $g'\sim {\mathcal O}(1)$ the $Z'$ is heavy, while for small values of $g'$ it can be light and accessible at experiments. A better part of the present manuscript is devoted to analyzing the experimental possibilities for the light $Z'$ searches.

The paper is organized as follows. In Section \ref{sec:U(1)3} we present the inverted FN model based on $U(1)_{\rm FN}^3$ symmetry, starting with the mass suppression mechanism for a single quark generation, which is then readily extended to the case of three generations. Here we also explain that the scaling \eqref{eq:YuYd:newFN} can be understood using a spurion analysis for an approximate $U(1)_{\rm app}$ that emerges at low energies, and is broken by $M$ rather than $\langle \phi\rangle$.  In subsection 
\ref{sec:numerics:decoupled} we then perform a numerical scan over the parameters of the $U(1)_{\rm FN}^3$ model and compare the results with the measured values of quark masses and mixings. In Section \ref{sec:U(1)}  
we present the $U(1)_{\rm FN}$ model and the respective numerical scan. In Section \ref{sec:leptons} we extend both models to the lepton sector. In Section \ref{sec:Z'} we explore the phenomenological implications of the two models, covering the bounds from a number of flavor conserving and violating processes induced by the exchanges of $Z'_i$ gauge bosons, with implications for precision flavor, astrophysics and beam dump experiments. Section \ref{sec:conclusions} contains our conclusions, while Appendix \ref{sec:app:KKbar} gives further details on $K-\bar K$ bounds for flavorful light $Z'$. The representative benchmarks for the two models are given in Appendix \ref{app:benchmarks}.

\section{Decoupled FN chains - the $U(1)_{\rm FN}^3$ model}
\label{sec:U(1)3}

\subsection{The set-up of the model}
We start by assuming that there is a separate horizontal $U(1)_{\rm FN}$ for each generation of the SM fermions, i.e.,  that the horizontal gauge group is $G_{\rm FN}=U(1)_{1}\times U(1)_2\times U(1)_3$. The field content is that of the SM, supplemented by a number of vector-like fermions whose FN charges differ by one unit and are organized in chains of vector-like fermions.

\begin{figure}[t]
\begin{center}
\includegraphics[width=12.cm]{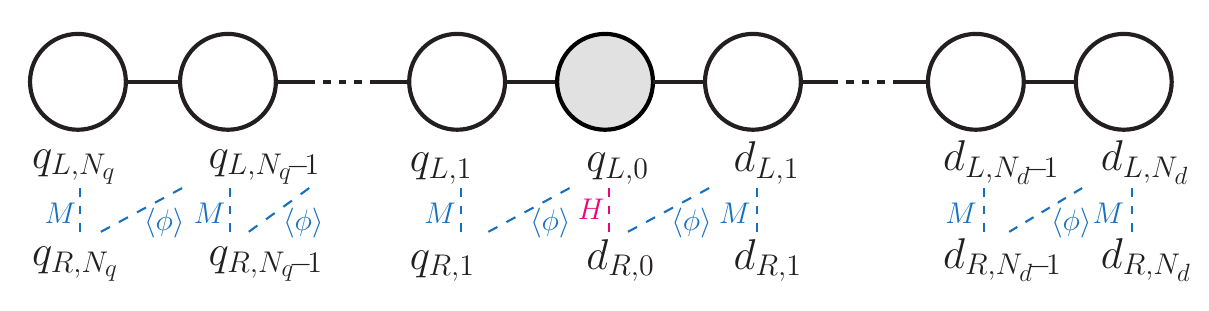}
\caption{\label{fig:single_gen} The  anomaly free inverted Froggatt-Nielsen model for a single generation of down quarks. The quark doublet vector-like chain is to the left, and down quark singlet vector-like chain to the right of the central node, shaded with gray.  The horizontal charges of fields on each node differ by $1$. The central node has chiral quark fields, uncharged under the horizontal $U(1)_{\rm FN}$, coupling to the Higgs. For clarity of presentation the labels on vector-like masses, as well as the Yukawa couplings are not shown. 
}
\end{center}
\end{figure}

We first illustrate the set-up for the case of a single SM generation of down quarks, setting also $G_{\rm FN}=U(1)_{\rm FN}$ for now. The Lagrangian is
\beq
{\cal L}_1={\cal L}_{q}+{\cal L}_{d}+ \big(Y^d_0\bar q_{L,0} d_{R,0} H+{\rm h.c.}\big),
\eeq
where
\begin{align}
\label{eq:L_QL}
{\cal L}_{q}&= i\sum_{n=1}^{N_q} \bar q_{R,n} \slashed D q_{R,n} + i\sum_{n=0}^{N_q} \bar q_{L,n} \slashed D q_{L,n} - \sum_{n=1}^{N_q}\big( M_n^q \bar q_{L,n} q_{R,n}- Y^q_n \phi \bar q_{L,n-1} q_{R,n}+{\rm h.c.}\big),
\\
\label{eq:L_DR}
{\cal L}_{d}&= i\sum_{n=0}^{N_d} \bar d_{R,n} \slashed D d_{R,n} + i\sum_{n=1}^{N_d} \bar d_{L,n} \slashed D d_{L,n} - \sum_{n=1}^{N_d}\big( M_n^d \bar d_{L,n} d_{R,n}- Y^d_n \phi \bar d_{L,n} d_{R,n-1} +{\rm h.c.}\big). 
\end{align} 
The $q_{L/R,j}$ and $d_{L/R,j}$ have the SM quantum numbers of $q_L$ and $d_R$ SM quark fields, respectively. 
The flavon fields, $\phi$, carry the horizontal $U(1)_{\rm FN}$ charge of $[\phi]=+1$, while $d_{R/L,n}$ and $q_{R/L,n}$ have charges $[d_{R/L,n}]=+n$ and $[q_{R/L,n}]=-n$, respectively. The above Lagrangian can be represented as a chain of nodes that contain fields with the same $U(1)_{\rm FN}$ charges, connected by flavon fields, see Fig. \ref{fig:single_gen}. For simplicity we set all the vector-like masses to be equal, $M$. This can be trivially relaxed, without affecting the main properties of the model as long as they remain ${\mathcal O}(M)$ (without loss of generality they can all be made real positive through phase redefinitions). More important is the assumption that the flavon couples on each node only to one field out of each pair of vector-like fermions, cf. Fig. \ref{fig:single_gen}. In general, there could also be terms of the form $\phi^* \bar d_{L,n-1} d_{R,n}$ and $\phi^* \bar q_{R,n-1} q_{L,n}$, which were dropped in \eqref{eq:L_QL} and \eqref{eq:L_DR}. The reason for this choice is that we expect the above model to be supersymmetrized at some higher scale. Such terms are  then absent because the superpotential is holomorphic.

The flavon obtains a vev, $\langle \phi \rangle$, which spontaneously breaks the horizontal $U(1)_{\rm FN}$. Each of the chains of vector-like fermions, $q$ and $d$, has one massless zero mode, and $N_q$ and $N_d$ massive vector-like fermion states with mass ${\mathcal O}(M, \langle \phi\rangle)$. For $\langle \phi\rangle \gg M$, with $Y_n^d, Y_n^q\sim {\mathcal O}(1)$ the two zero modes are mostly localized towards the ends of the respective chains. More precisely, the $q_{L,0}'$ and $d_{R,0}'$ massless mass eigenstates are given by
\beq
\label{eq:field_rotation_chain_diag_decpld}
q_{L,0}'=\sum_{n=0}^{N_q}V_{n0}^{q_L} q_{L,n}, \qquad d_{R,0}'=\sum_{n=0}^{N_d}V_{n0}^{d_R} d_{R,n}.
\eeq
The notation is borrowed from \cite{Alonso:2018bcg}, with $V^{q_L}$ and $V^{q_R}$ the $N_q\times N_q$ and $(N_q+1)\times (N_q+1)$ unitary matrices that diagonalize the $N_q\times (N_q+1)$ mass matrix for the $q$ fields through a bi-unitary transformation (and similarly for $d_{L,R}$ fields with obvious change of notation). For $Y_n^d=Y_n^q=1$, $M_n^d=M_n^q=M,$ the model realizes the clockworking mechanism, for which the zero mode profiles are known analytically \cite{Alonso:2018bcg}, 
\beq
\label{eq:Vj0}
V_{n0}^{q_L}={\cal N}_0^{q_L} \biggr(\frac{M}{\langle \phi\rangle}\biggr)^{N_q-n}, \qquad V_{n0}^{d_R}={\cal N}_0^{d_R} \biggr(\frac{M}{\langle \phi\rangle}\biggr)^{N_d-n},
\eeq
with the normalization constants ${\cal N}_0^{\psi}=\sqrt{\big(q^2-1\big)/(q^2-1/q^{2N_\psi})}\simeq 1$, where $q=\langle \phi \rangle/M\gg 1$.

The zero mode obtains a mass once the electroweak symmetry is broken through the Higgs vev. The mass is suppressed by the zero mode wave functions on the zero node, 
\beq
\label{eq:md}
m_d\simeq f_{q_L} f_{d_R} Y_0^d v/\sqrt 2,
\eeq
where the overlaps with the zero-node are given by
\beq
\label{eq:fpsi}
f_{\psi }\equiv V_{00}^{\psi}\simeq  \biggr(\frac{M}{\langle \phi\rangle}\biggr)^{N_\psi}, 
\eeq
and are thus exponentially suppressed for large $N_\psi$ and $\langle \phi\rangle \gg M$. 

\begin{figure}[t]
\begin{center}
\includegraphics[width=12.cm]{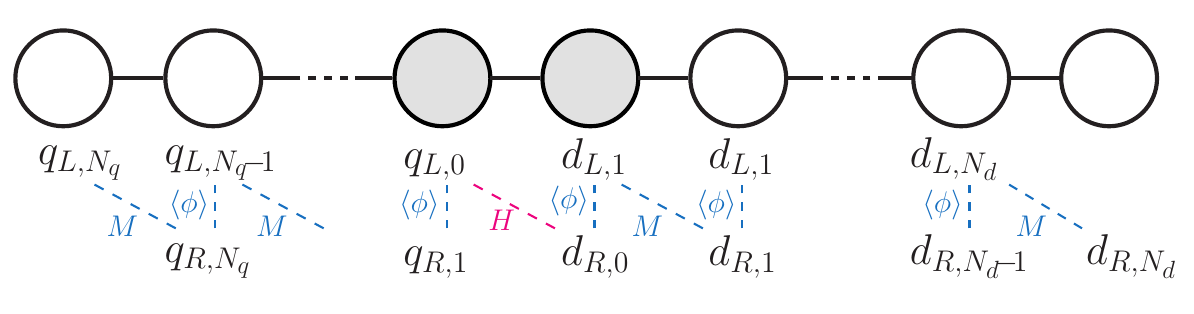}
\caption{\label{fig:single_gen_rearr} The anomaly free inverted Froggatt-Nielsen model for a single generation of down quarks, but with rearranged fields on the nodes, see text for details. 
}
\end{center}
\end{figure}

The suppression can be understood from a spurion analysis. In the limit of a heavy flavon  the fermion mass matrix has an emergent approximate $U(1)_{\rm app}$ global symmetry after the flavon obtains a vev. This is easy to see once the fermion fields are re-arranged in a modified chain of nodes as shown in Fig. \ref{fig:single_gen_rearr}, while the heavy flavon is integrated out and need not be considered here. The fields on a particular node have $U(1)_{\rm FN}$ charges that differ by one unit, but have the same $U(1)_{\rm app}$ charges. The $U(1)_{\rm app}$ charge reduces by one unit when hopping to the right by one node in the chain. That is,  the $q_{L,N_q}$ has the $U(1)_{\rm app}$ charge $N_q$, the $q_{L,N_q-1}$ and $\bar q_{L,N_q}$ the $U(1)_{\rm app}$ charge $N_q-1$, and so on, all the way to  $d_{R,N_d}$  which carries a charge $-N_d$ (the two greyed out nodes in Fig. \ref{fig:single_gen_rearr} should be treated as a single node, with all the corresponding fields uncharged under $U(1)_{\rm app}$). The mass terms proportional to $M$ explicitly break $U(1)_{\rm app}$. Treating $M$ as a spurion with $U(1)_{\rm app}$ charge of $+1$ the fermion mass matrix is formally invariant under $U(1)_{\rm app}$. The zero modes, i.e., the SM fermions are mostly $q_{L,N_q}$ and $d_{R,N_d}$, so that the zero mode mass term is given by
\beq
\label{eq:LSM:effective}
{\cal L}_{\rm SM} \sim Y_0^d \biggr(\frac{M}{\langle \phi\rangle}\biggr)^{N_q+N_d} \bar q_{L,N_q} d_{R,N_d} H,
\eeq
which is $U(1)_{\rm app}$ invariant. This is exactly the suppression found in \eqref{eq:md}. 
Furthermore, note that the effective FN charge of the zero mode is approximately the same as its $U(1)_{\rm app}$ charge. The zero modes are localized toward the ends of the vector-like chains, see Fig. \ref{fig:zero_mode_profile}. To leading order the $q_{L,N_q}$ and $d_{R,N_d}$ zero modes thus carry effective FN charges $N_q$ and $-N_d$, respectively.

\begin{figure}[t]
	\begin{center}
		\includegraphics[width=10.cm]{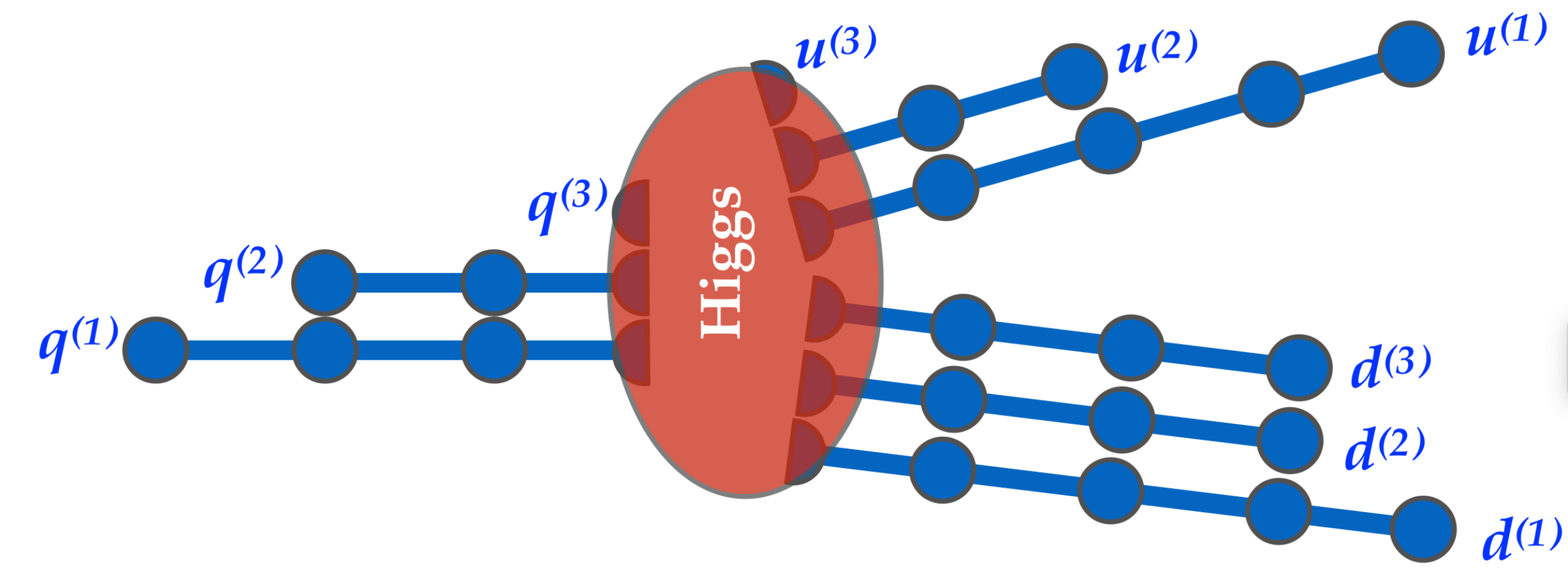}
		\caption{\label{fig:decoupled:setup} The vector-like chains (blue) for the inverted FN model with $G_{\rm FN}=U(1)_{\rm FN}^3$. The chiral fermions are on the zero node, denoted by the red ellipse, which also contains the Higgs. }
	\end{center}
\end{figure}

\begin{figure}[t]
	\begin{center}
		\includegraphics[width=10.cm]{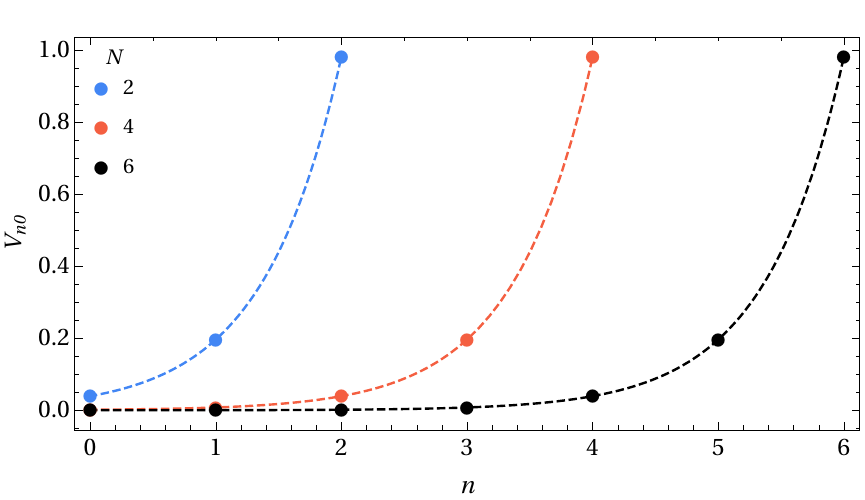}
		\caption{\label{fig:zero_mode_profile} Zero mode profiles for chains of lengths $N=2$ (blue), $N=4$ (red), and $N=6$ (black), setting $q=5$. The overlaps with the zero node, $n=0$, are exponentially suppressed.
		}
	\end{center}
\end{figure}

The above discussion is readily extended to the case of three generations, including both down and up quarks. The horizontal gauge group is taken to be $G_{\rm FN}=U(1)_1\times U(1)_2\times U(1)_3$, with a separate $U(1)_i$ for each of the three generations of fermions. The fermion Lagrangian is then
\beq
\label{eq:Lf:decoupled}
{\cal L}_f=\sum_{i}\Big( {\cal L}_{q{(i)}}+{\cal L}_{d{(i)}}+ {\cal L}_{u{(i)}}\Big)+ \sum_{ij}\Big[\big(Y^d_0\big)_{ij}\bar q_{L,0}^{(i)} d_{R,0}^{(j)} H+\big(Y^u_0\big)_{ij}\bar q_{L,0}^{(i)} u_{R,0}^{(j)} \tilde H+{\rm h.c.}\Big],
\eeq
with $i,j=1,2,3,$ the generation indices, and $\tilde H=i \sigma_2 H^*$. The kinetic Lagrangians are given in \eqref{eq:L_QL} and \eqref{eq:L_DR}, but with extra generational superscript on fermions, $q_a^{(i)}, u_a^{(i)}, d_a^{(i)}$ (replacing $d\to u$ in ${\cal L}_{d{(i)}}$ gives ${\cal L}_{u{(i)}}$), and with $M_n^q\to M_n^{q(i)}$, $Y_n^q\to Y_n^{q(i)}$, $N_q\to N_{q(i)}$, etc., as well as $\phi\to \phi_i$, since each generation is gauged under a separate $U(1)_i$.
 This results in three uncoupled chains, one per each generation, for $q$ fields, for $d$ fields, and for $u$ fields. All of these chains attach to the central node with the Higgs, see Fig. \ref{fig:decoupled:setup}.   The Yukawa matrices $Y_0^d, Y_0^u$ on the zero node are $3\times3$ complex matrices, while the Yukawa couplings on each of the chains, \eqref{eq:L_QL} and \eqref{eq:L_DR}, are just complex numbers.\footnote{The flavon Yukawa couplings $Y_1^{q(i)}$,  connecting the zero node with the first node of the $q-$chain, are in general arbitrary complex three-vectors. To simplify the notation in \eqref{eq:Lf:decoupled} we take them to be orthogonal, and then use the $3\times 3$ unitary field redefinitions of $\bar q_{L,0}^{(i)}$ to ensure the $Y_1^{q(i)}$ have nonzero components only in the $i$-th direction. We assume the same for $Y_1^{d(i)}$ and $Y_1^{u(i)}$.} The Yukawa interactions with the Higgs are thus the only terms mixing different generations, see the last term in \eqref{eq:Lf:decoupled}.

Up to higher corrections in $v/M$ the SM fermions are equal to the zero modes. The zero mode wave functions peak toward the ends of the chains, cf. Eq. \eqref{eq:Vj0} and Fig.~\ref{fig:zero_mode_profile}. This means that the SM fermion, $q_L^{(i)}$, has the largest component that is due to $q_{L,N_q}^{(i)}$, the left-handed part of the vector-like fermion that is the furthest away from the Higgs. For $u_R^{(i)}$ the largest component is  $u_{R,N_u}^{(i)}$, and similarly for $d_R^{(i)}$  it is $d_{R,N_d}^{(i)}$. The SM fermions thus carry large $U(1)_i$ charges, and couplings to the $U(1)_i$ gauge bosons, $Z_i'$s, roughly given by the length of the corresponding vector-like chain, $[\psi]_i\approx N_{\psi{(i)}}$.  However, for the phenomenology of $Z_i'$ also the subleading couplings are important, because they lead to flavor violating transitions. We explore the implications of these in Section \ref{sec:Z'}.

The zero modes, identified with the SM fermions, obtain a nonzero mass after electroweak symmetry breaking from the  overlaps with the Higgs. The Higgs is on the zero node, so that
\beq
\label{eq:md:mu}
(m_d)_{ij}\simeq f_{q_L^{(i)}} f_{d_R^{(j)}} \big(Y_0^d\big)_{ij} \frac{v}{\sqrt 2}, \qquad 
(m_u)_{ij}\simeq f_{q_L^{(i)}} f_{u_R^{(j)}} \big(Y_0^u\big)_{ij} \frac{v}{\sqrt 2}.
\eeq
The structure of the zero node overlaps, $f_a$, explains the hierarchy of SM fermion masses. For $M\ll \langle \phi\rangle$ the zero node overlaps, $f_a$, are more suppressed the longer the corresponding chain of the vector-like fermions. This can be seen analytically, if all the Yukawa couplings are equal to one, see Eq. \eqref{eq:fpsi}, but is also true in general.   

Below we perform a scan over the inputs of the model in order to explore how well the eigenvalues of the quark mass matrices \eqref{eq:md:mu} resemble the observed values. To simplify the discussion, we set
\beq
\label{eq:qdef}
\frac{M_1}{\langle \phi_1\rangle}=\frac{M_2}{\langle \phi_2\rangle}=\frac{M_3}{\langle \phi_3\rangle}=\frac{1}{q}\simeq\lambda= 0.2,
\eeq
where $M_1, M_2,M_3$ are the typical values of vector-like masses for each of the three generations. 
This choice will also make it easier to compare with the single $U(1)_{\rm FN}$ case, to be covered in Section \ref{sec:U(1)}.
The scalings for  the mass matrices in \eqref{eq:md:mu} are then, for $Y_0^d, Y_0^u\sim {\mathcal O}(1)$, 
\beq
\label{eq:md:mu:scaling}
(m_d)_{ij}\simeq \lambda^{N_{q{(i)}}+N_{d{(j)}}} \frac{v}{\sqrt 2}, \qquad 
(m_u)_{ij}\simeq \lambda^{N_{q{(i)}}+N_{u{(j)}}} \frac{v}{\sqrt 2}.
\eeq 

\subsection{Numerical scan}
\label{sec:numerics:decoupled}
\begin{figure}[t]
	\begin{center}
		{\bf Decoupled FN Chains, $\bm{ G_{\mathbf{FN}}=U(1)_{\mathbf{FN}}^3}$}\\
		\includegraphics[width=1\linewidth]{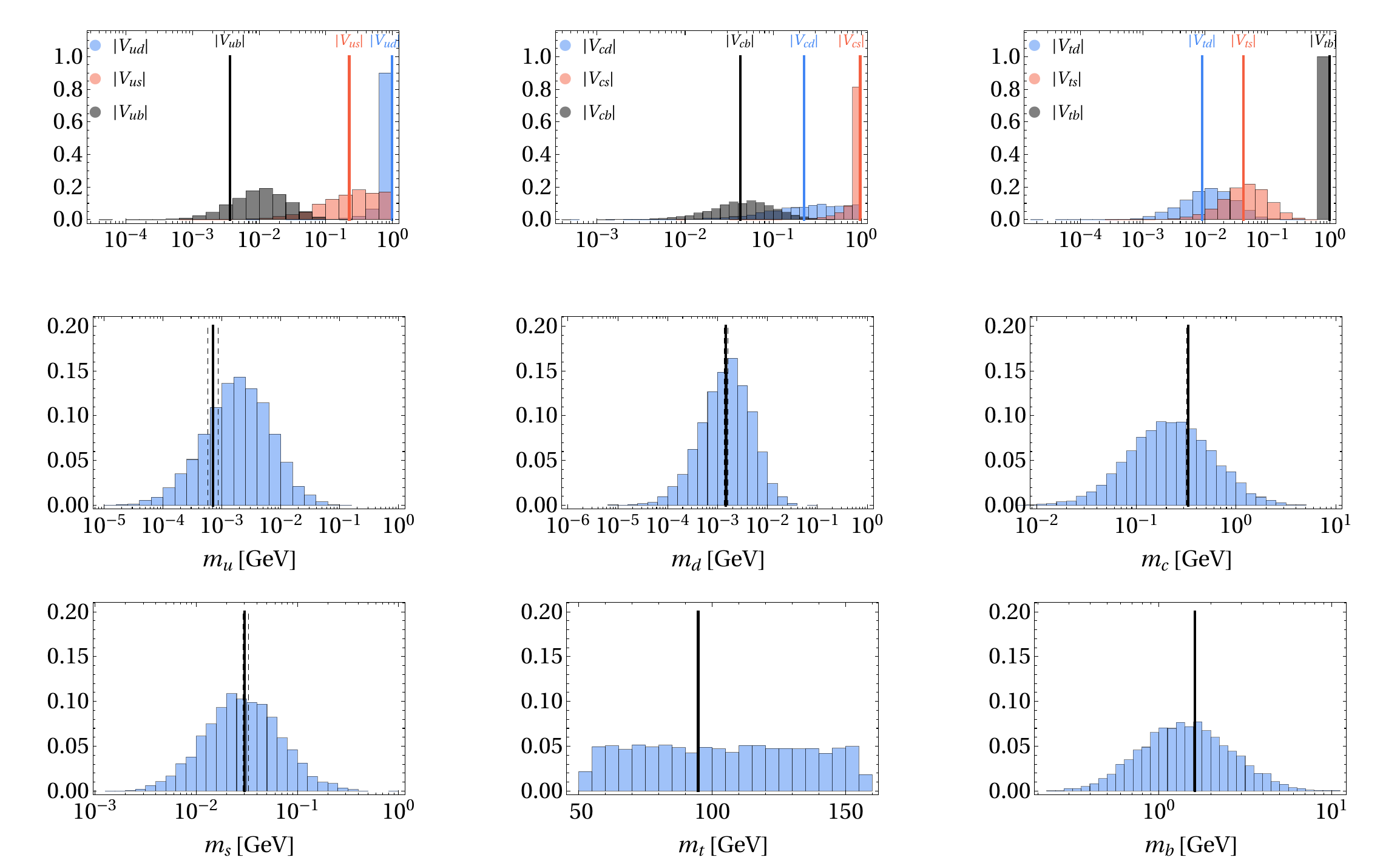}
		\caption{Distributions of the CKM matrix elements and quark masses for inverted FN model with the $G_{\rm FN}=U(1)_{\rm FN}^3$ gauge group, see text for details. The vertical solid (dashed) lines mark the SM values ($1\sigma$ bands) for the observables at $\mu=10^7$ GeV.		\label{fig:numerical_scan_decoupled}
		}
	\end{center}
\end{figure}

For the numerical scan we set $\langle \phi\rangle =10^7$ GeV and $q=5$, cf.~Eq.~\eqref{eq:qdef}. 
The vector-like masses, $M_n^f $, $f=u, d, q$,~Eqs.~\eqref{eq:L_QL}, \eqref{eq:L_DR}, in the units of $10^7$ GeV, and the zero node Yukawa matrix elements, $(Y_0^d)_{ij}, (Y_0^u)_{ij}$,~Eq.~\eqref{eq:Lf:decoupled}, are taken to be random complex numbers of the form $r_a e^{i\varphi_a}$. The Yukawa couplings between the FN fermions and the flavon are taken to be of the form $Y_n^f \langle \phi\rangle= r_a e^{i\varphi_a} q$, so that the ratios $|Y^f\langle \phi\rangle/M^f|$ are on average equal to $q$. The magnitudes and phases are taken to be uniformly distributed over $r_a\in [0.3,0.9]$ and $\varphi_a\in [0,2\pi)$, with $r_a,\varphi_a$ uncorrelated between different couplings. The range  of $r_a$ was chosen such that the top Yukawa, $\bar m_t/(v/\sqrt 2)\simeq 0.55$ is close to the median, while the ratio of the boundaries, $r_{\rm max}/r_{\rm min}$, is smaller than $q$. The predictions are not very sensitive to the precise ranges for $r_a$ since these cancel on average in the ratios $M_n^f/(Y_n^f \langle \phi\rangle)$, which 
control the hierarchies of  the SM quark masses.  The effect of changing the average value of $r_a$ is thus mostly due to a different average values of the $Y_0^{d,u}$ matrix elements. For instance, using $r_a\in [0.6, 1.8]$ changes the distributions for the quark masses in Fig. \ref{sec:numerics:decoupled} by an overall factor of 2.  Relatively larger $r_a$ ranges, on the other hand, lead to wider distributions of quark masses and mixings.

\begin{table}[t]
	\centering
	\begin{tabular}{ccccccccc}
		\hline\hline
		$\overline{m}_u$ & $\overline{m}_d$  & $\overline{m}_c$ & $\overline{m}_s$ & $\overline{m}_t$ & $\overline{m}_b$ & $\overline{m}_e$ & $\overline{m}_\mu$ & $\overline{m}_\tau$  \\
		\hline
		0.0007 & 0.0015  & 0.33 & 0.03 & 95 & 1.6 & 0.0005 & 0.104 & 1.798\\
		\hline\hline
	\end{tabular}
	\caption{ The experimental values of quark and charged lepton masses (in GeV) at $\mu=10^7$ GeV, obtained from NNLO QCD RG evolution, and used for comparison with the numerical scan. Despite the large allowed experimental range for $m_u$ and $m_d$ we show only the central values, which coincide with the choice made  
	in Ref.~\cite{Alonso:2018bcg} at $\mu=2$ TeV.
	\label{tab:numinputs}}
\end{table}

The lengths of the chains with the vector-like quarks are taken to be almost the same as in Ref. \cite{Alonso:2018bcg},\footnote{The difference arises partially from taking significantly higher $\mu$ and different median values of $Y_0^{d,u}$ matrix elements. We take $N_{d(3)}=3$, $N_{u(2)}=2$ while in \cite{Alonso:2018bcg} these were set to $N_{d(3)}=2, N_{u(2)}=1$. We thank A. Kagan for suggesting the new charge assignments that improve the agreement with bottom and charm quark masses.}
\beq
\begin{split}
	\label{eq:chain_lengths_decoupled}
	N_{q(1)}&=3,\quad N_{q(2)}=2, \quad\,\, N_{q(3)}=0,\\
	N_{u(1)}&=4,\quad N_{u(2)}=2, \quad\, N_{u(3)}=0, \\
	N_{d(1)}&=4,\quad N_{d(2)}=3, \quad\,\, N_{d(3)}=3,
\end{split}
\eeq
giving the configuration of the vector-like FN chains shown in Fig. \ref{fig:decoupled:setup}.

In each run of the scan we generate and diagonalize the chains, then calculate the quark masses and the CKM matrix elements. The results of the scan with 5000 runs are shown in Fig.~\ref{fig:numerical_scan_decoupled}, with the CKM matrix elements shown in the first row, and the quark masses in the second and third row. For ease of comparison we also denote the measured central values with vertical solid lines, while the one sigma experimental bands are delineated with dashed lines. The quark masses are the $\overline{\text{MS}}$ masses at $\mu=10^4$ TeV, which we anticipate to be the rough lower bound on the vector-like fermion masses from flavor constraints, see Section \ref{sec:Z'} for details. The values of SM fermion masses at $\mu=10^4$ TeV, obtained through NNLO RG evolution, are listed in Table~\ref{tab:numinputs} (at
$\mu=2$ TeV the values of the quark masses coincide with Ref. \cite{Alonso:2018bcg}). 
For the CKM matrix elements we checked using results in \cite{Babu:1987im} that the effects of RG evolution are negligible, so that we use the unevolved values from the PDG  \cite{Tanabashi:2018oca}.

\section{Coupled FN chains - the $U(1)_{\rm FN}$ model}
\label{sec:U(1)}

\begin{figure}[t]
	\begin{center}
		\includegraphics[width=10.cm]{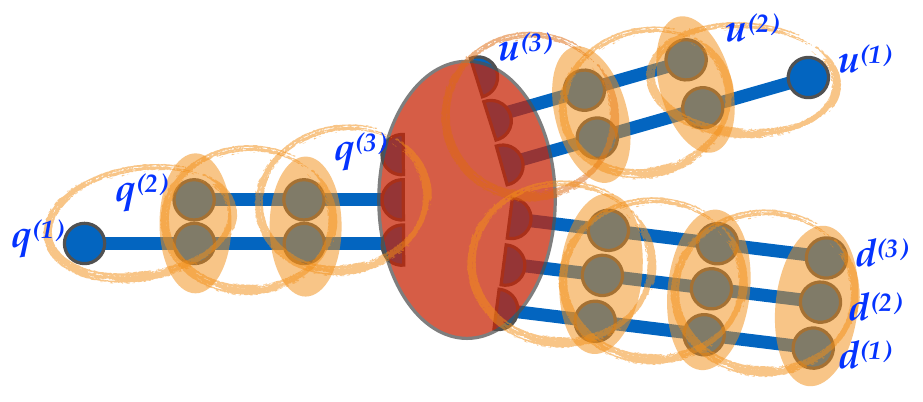}
		\caption{\label{fig:coupled:setup} The vector-like chains (blue) for the inverted FN model with $G_{\rm FN}=U(1)_{\rm FN}$. The chiral fermions are on the zero node, denoted by the red ellipse, which also contains the Higgs. Different generations on each node mix through vector-like masses (filled orange ellipses), as do the fields on neighboring nodes through Yukawa flavon interactions (empty orange ellipses).} 
	\end{center}
\end{figure}

\subsection{The set-up of the model}
We discuss next the coupled FN chains, which are obtained in the case of a single horizontal $U(1)_{\rm FN}$, see Fig. \ref{fig:decoupled:setup}. The Yukawa couplings between vector-like fermions and the flavon field are now complex matrices. 
The Lagrangian is thus
\beq
\label{eq:L_1_coupled}
{\cal L}_1={\cal L}_{q}+{\cal L}_{d}+ {\cal L}_{u}+ \sum_{ij}\Big[\big(Y^d_0\big)_{ij}\bar q_{L,0}^{(i)} d_{R,0}^{(j)} H+\big(Y^u_0\big)_{ij}\bar q_{L,0}^{(i)} u_{R,0}^{(j)} \tilde H+{\rm h.c.}\Big],
\eeq
where
\beq
\begin{split}
\label{eq:L_DR:U(1)}
{\cal L}_{d}=& i\sum_{n=0}^{N_{d(i)}} \bar d_{R,n}^{(i)} \slashed D d_{R,n}^{(i)} + i\sum_{n=1}^{N_{d(i)}} \bar d_{L,n}^{(i)} \slashed D d_{L,n}^{(i)} 
\\
&-\sum_{n=1}^{N_{d(1)}}  \Big( \sum_{i,j=1}^{\hat N_d|_n}\big(M_n^d)_{ij} \bar d_{L,n}^{(i)} d_{R,n}^{(j)}- \sum_{i=1}^{\hat N_d|_n} \sum_{j=1}^{\hat N_d|_{n-1}}(Y^d_n)_{ij} \phi \bar d_{L,n}^{(i)} d_{R,n-1}^{(j)} +{\rm h.c.}\Big), 
\end{split}
\eeq
and similarly for ${\cal L}_{u}$ with the $d\to u$ replacement. For ${\cal L}_{q}$, on the other hand, there is one more left-handed field for each generation than the right-handed fields, so that
\beq
\begin{split}
\label{eq:L_QL:U(1)}
{\cal L}_{q}=& i\sum_{n=1}^{N_{q(i)}} \bar q_{R,n}^{(i)} \slashed D q_{R,n}^{(i)} + i\sum_{n=0}^{N_{q(i)}} \bar q_{L,n}^{(i)} \slashed D q_{L,n}^{(i)}
\\
& - \sum_{n=1}^{N_{q(1)}} \Big(  \sum_{i,j=1}^{\hat N_q|_n}\big(M_n^q\big)_{ji} \bar q_{L,n}^{(j)} q_{R,n}^{(i)}-   \sum_{i=1}^{\hat N_q|_n}\sum_{j=1}^{\hat N_q|_{n-1}} \big(Y^q_n\big)_{ji} \phi \bar q_{L,n-1}^{(j)} q_{R,n}^{(i)}+{\rm h.c.}\big). 
\end{split}
\eeq
The summation is over the nodes and the generations on each node. We label the fermions such that the $i-$th generation fermions have a vector-like chain of length $N_{q(i)}$. The first generation has the longest fermion chain, so that the summation $n=1,\ldots, N_{q(1)}$ sums over all of the nodes. On node $n$ there are $\hat N_q|_n$ generations, which are coupled to the $\hat N_q|_{n-1}$ generations of fermions on the $(n-1)$-th node through an $\hat N_q|_{n-1} \times \hat N_q|_{n}$ complex Yukawa matrix $Y_n^q$.

This produces a set of coupled FN chains, where the mixing between generations is not only through $3\times 3$ Yukawa matrices on the zero node, coupling to the Higgs, but also through the flavon Yukawa couplings, $\big(Y^q_n\big)_{ji}$, and vector-like masses, $\big(M_n^q\big)_{ji}$, cf. Fig.~\ref{fig:coupled:setup}. The zero modes are obtained through unitary transformations
\beq
\label{eq:field_rotation_chain_diag_cpld}
q'{}_{\negmedspace L,0}^{(i)}=\sum_{n=0}^{N_{q(1)}}\sum_{j=1}^{\hat N_{q}|_n}V_{n(j),0(i)}^{q_L} q_{L,n}^{(j)}, \qquad d'{}_{\negmedspace R,0}^{(i)}=\sum_{n=0}^{N_{d(1)}}\sum_{j=1}^{\hat N_{q}|_n}V_{n(j),0(i)}^{d_R}  d_{R,n}^{(j)},
\eeq
and similarly for the up quark zero modes. As in the decoupled case, the Higgs provides the zero modes with nonzero masses after electroweak symmetry breaking from their overlaps with the zero node. However, in the coupled case  the zero mode overlaps with the zero node are described by $3\times 3$ matrices, $V_{0(j),0(i)}^{d_R, u_R}$, due to the inter-generational mixing in the chains.
 Therefore, the quark mass matrices become
\begin{equation}
\label{eq:md:mu:coupled}
(m_d)_{kl} \simeq V^{q_L}_{0(i),0(k)} \big(Y_0^d\big)_{ij} V_{0(j),0(l)}^{d_R}  \frac{v}{\sqrt 2}, \qquad (m_d)_{kl} \simeq V^{q_L}_{0(i),0(k)} \big(Y_0^u\big)_{ij} V_{0(j),0(l)}^{u_R}  \frac{v}{\sqrt 2}.
\end{equation}

For  $\langle \phi\rangle \gg M$
the zero-modes are, also in this case, mostly localized towards the ends of the respective FN chains, with small overlaps with the Higgs on the zero node. Parametrically,
\beq
\label{eq:coupled:overlaps}
f_{q_L^{(i)}}\sim \Big(\frac{M^q}{Y^q \langle \phi\rangle}\Big)^{N_{q(i)}}, \qquad f_{u_R^{(i)}}\sim \Big(\frac{M^u}{Y^u \langle \phi\rangle}\Big)^{N_{u(i)}}, \qquad f_{d_R^{(i)}}\sim \Big(\frac{M^d}{Y^d \langle \phi\rangle}\Big)^{N_{d(i)}},
\eeq
with $M^{q,u,d}$, $Y^{q,u,d}$
denoting the typical values of the corresponding matrix elements.
The hierarchy in the overlaps then translates into the hierarchy of the quark masses, cf. Eq. \eqref{eq:md:mu}. However, due to additional flavour mixing on each node the parametric relations are now even more approximate compared to the case of decoupled of FN chains. We show this by performing a numerical scan. Since the expressions for SM quark masses effectively involve multiplications of a number of random matrices, Eq.~\eqref{eq:coupled:overlaps}, part of the hierarchy comes from the properties of random matrix multiplications \cite{vonGersdorff:2017iym} (see also Section \ref{sec:leptons}).

\subsection{Numerical scan}
\label{sec:scan:coupled}
We choose the chain configurations with the following number of fermion generations on each node 
\beq
\begin{split}
	\label{eq:chain_configuration_coupled}
	\hat N_q|_n=& \quad \{3, 2, 2, 1\}, \\
	\hat N_u|_n=& \quad \{3, 2, 2, 1\}, \\
	\hat N_d|_n=& \quad \{3, 3, 3, 3\}.
\end{split}
\eeq
Here $ n=0,\ldots,3$, respectively, for each of the three cases. This implies that the lengths of the vector-like chains are given by, 
\beq
\begin{split}
	\label{eq:chain_lengths_coupled}
	N_{q(1)}&=3,\quad N_{q(2)}=2, \quad\,\, N_{q(3)}=0,\\
	N_{u(1)}&=3,\quad N_{u(2)}=2, \quad\, N_{u(3)}=0, \\
	N_{d(1)}&=3,\quad N_{d(2)}=3, \quad\,\, N_{d(3)}=3,
\end{split}
\eeq
where $N_a+1$ is the length of the corresponding chain, cf. Eq.~\eqref{eq:L_DR:U(1)}.
The above chain configuration differs from the one for the decoupled FN chains in \eqref{eq:chain_lengths_decoupled}, in that the all three $d_R$ generations now have the same lengths of chains, and  that the chain for the first $u_R$ generation is shorter. 

\begin{figure}[t]
	\begin{center}
		{\bf Coupled FN Chains, $\bm{ G_{\mathbf{FN}}=U(1)_{\rm FN}}$}\\
		\includegraphics[width=1\linewidth]{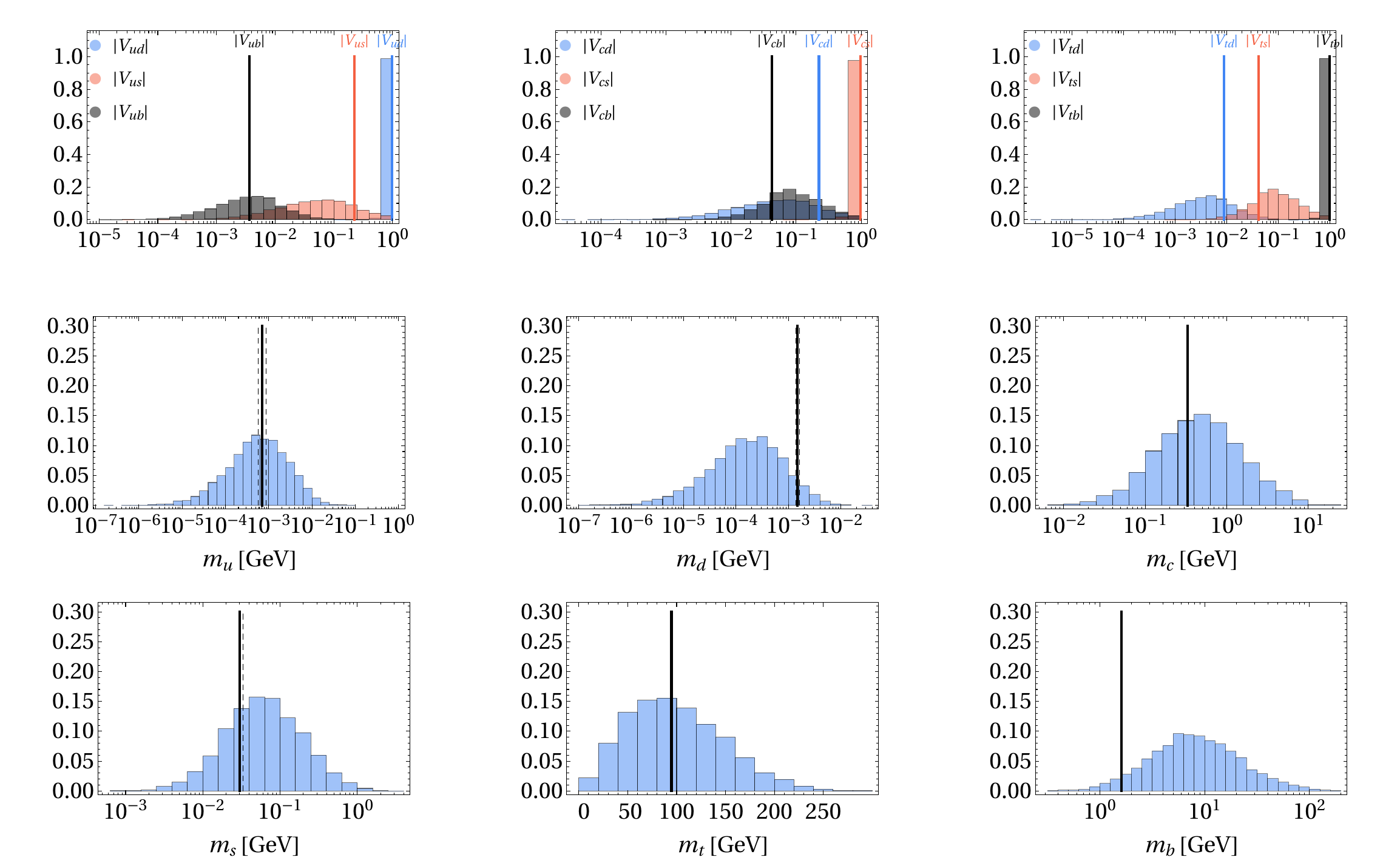}
		\caption{Distributions of the CKM matrix elements and quark masses generated in the numerical scan with 5000 runs in the case of coupled FN chains, see text for details. 
		 \label{fig:numerical_scan_coupled}
		}
	\end{center}
\end{figure}

We use a similar procedure as in Section \ref{sec:numerics:decoupled} to produce the numerical scan for the coupled case. We fix $q=5$, $\langle\phi\rangle=10^7$ GeV, and vary the matrices $M_{ij}^f, (Y_n^f)_{ij}\langle\phi\rangle$, $f=u,d,q$, cf.~Eqs.~\eqref{eq:L_DR:U(1)} \eqref{eq:L_QL:U(1)}, and $(Y_0^d)_{ij},(Y_0^u)_{ij}$, cf.~Eq. \eqref{eq:L_1_coupled}. The vector-like masses are taken to be random complex numbers of the form $r_a e^{i\varphi_a}$ in the units of $10^7$ GeV, while the elements of Yukawa matrices are taken to be of the form $(Y_n^f)_{ij} \langle\phi\rangle= r_a e^{i\varphi_a} q$. The $r_a$ are varied in the range $[0.3,0.9]$ and phases in $[0,2\pi)$. For each scan run we generate and diagonalize the coupled chains and extract the quark masses and the CKM matrix elements. The results of the scan with 5000 runs are shown in Fig.~\ref{fig:numerical_scan_coupled}, using the same layout as in Fig.~\ref{fig:numerical_scan_decoupled}.

\begin{figure}[t]
	\begin{center}
	{\bf Coupled FN Chains, $\bm{ G_{\mathbf{FN}}=U(1)_{\rm FN},~q=10^3}$}\\
		\includegraphics[width=1\linewidth]{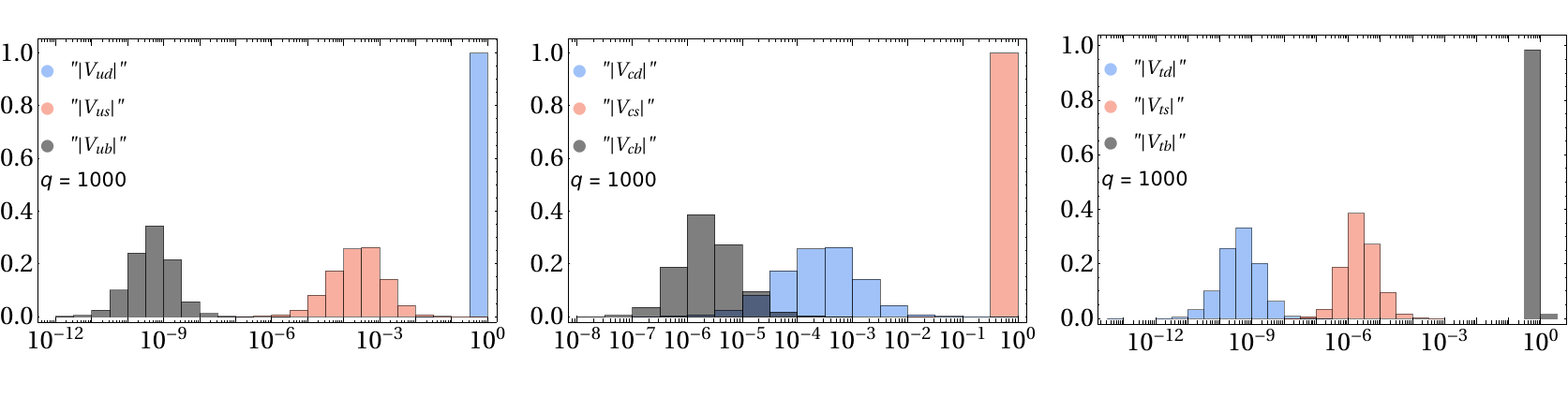}
		\caption{The CKM elements that would be obtained from coupled FN chains for the case of a very large $q = 1000$. This shows the hierarchy between the would-be CKM matrix elements more clearly. 
		\label{fig:numerical_scan_coupled_high_q}
		}
	\end{center}
\end{figure}

Even though the $d_R$ chains are of the same length for all generations, we still have hierarchical masses for down quarks due to different lengths of $q_L$ chains. Part of the required hierarchy also comes from the fact that products of random matrices lead to matrices with hierarchical eigenvalues \cite{vonGersdorff:2017iym}. The resulting mass distributions are much broader than in the $U(1)_{\rm FN}^3$ case. This is particularly apparent in the distributions of CKM matrix elements, where the distributions for the $V_{cd}$ and $V_{cb}$ matrix elements  almost completely overlap. This is a result of relatively small expansion parameter $1/q\simeq0.2$, required to fit the observed values realized in Nature. In Fig.~\ref{fig:numerical_scan_coupled_high_q} we show that a hierarchical structure does appear also for the CKM matrix elements once a much larger value of  $q=10^3$ is taken.

\section{Extension to leptons}
\label{sec:leptons}
It is straightforward to extend the above framework to leptons. To shorten the discussion we assume that the neutrinos have Majorana masses. For the $G_{\rm FN}=U(1)_{\rm FN}^3$ case the Lagrangian for the leptons is then given by replacing $Q\to L$, $d\to e$ in Eq.~\eqref{eq:Lf:decoupled},  and ignore terms that involve $u$. Performing the same replacements in Eq.~\eqref{eq:L_1_coupled} gives the Lagrangian for the $G_{\rm FN}=U(1)_{\rm FN}$ case. 

For the neutrino masses we assume that they come from the dimension 5 Weinberg operator. On the zero node we thus add the mass term 
\beq
{\cal L}_{{\rm dim}\,5}\supset c_{ij} (L_iH)(L_jH)/\Lambda_{\rm LN},
\eeq
 with $c_{ij}\sim {\mathcal O}(1)$. In the $\langle \phi \rangle \gg M_n^L$ limit the neutrino mass matrix takes the form
\beq
\label{eq:Weinberg:op}
m_{ij}^\nu\simeq c_{ij} \frac{v^2}{\Lambda_{\rm LN}} \Big(\frac{M^L}{\langle \phi \rangle}\Big)^{N_{L(i)}+N_{L(j)}},
\eeq
after the vector-like fermions are integrated out.

The FN charge assignments for charged leptons depend crucially on the assumed flavor structure of the neutrino mass matrix, see, e.g., \cite{CRZZ:2019}. We focus on the case where the neutrino masses are completely anarchic, with all the PMNS mixing angles taken to be ${\mathcal O}(1)$. This happens if all the left-handed leptons have the vector-like fermion chains of the same length,
\beq
N_{L(1)}=N_{L(2)}=N_{L(3)}.
\eeq

\begin{figure}[t]
	\begin{center}
		\includegraphics[width=0.45\linewidth]{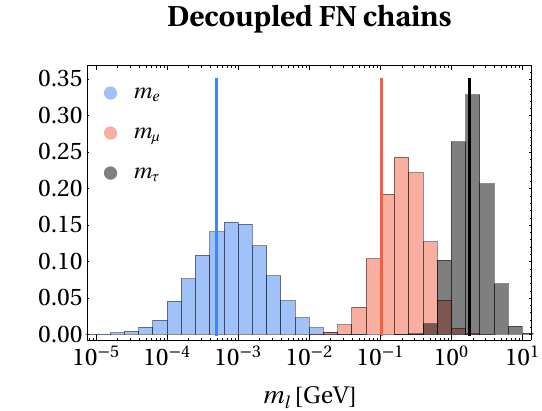}~~~~~~
		\includegraphics[width=0.45\linewidth]{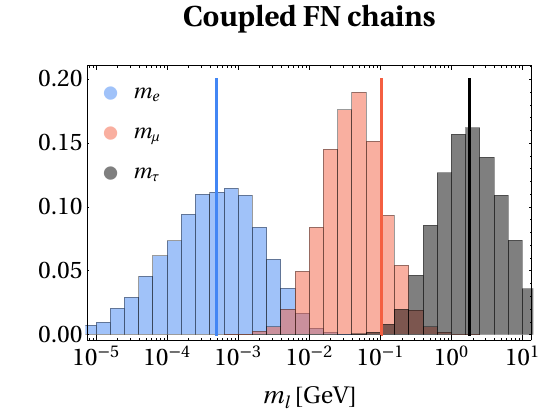}
		\caption{The distributions of the charged lepton masses for the  case of  decoupled FN chains, $G_{\rm FN}=U(1)_{\rm FN}^3$ (left), with charge assignments in Eq.~\eqref{eq:NL:decoupled}, and for the coupled FN chains, $G_{\rm FN}=U(1)_{\rm FN}$ (right),  with charge assignments in Eq.~\eqref{eq:coupled:leptons:N}. 
		\label{fig:leptons:scan} 
		}
	\end{center}
\end{figure}

\begin{figure}[t]
	\begin{center}
		{\bf Decoupled FN Chains, $\bm{ G_{\mathbf{FN}}=U(1)_{\rm FN}^3}$ - neutrinos}\\
		\includegraphics[width=1\linewidth]{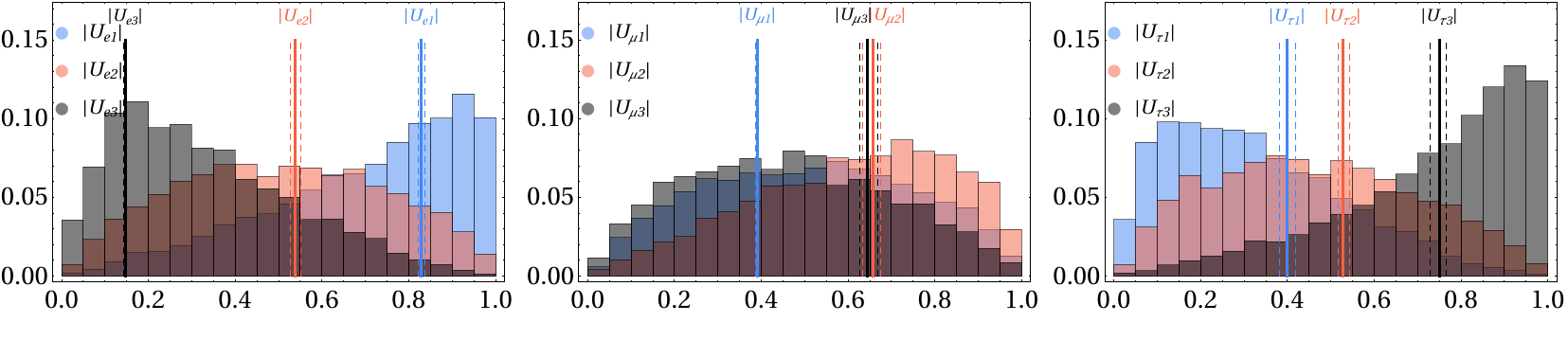}
				\includegraphics[width=0.34\linewidth]{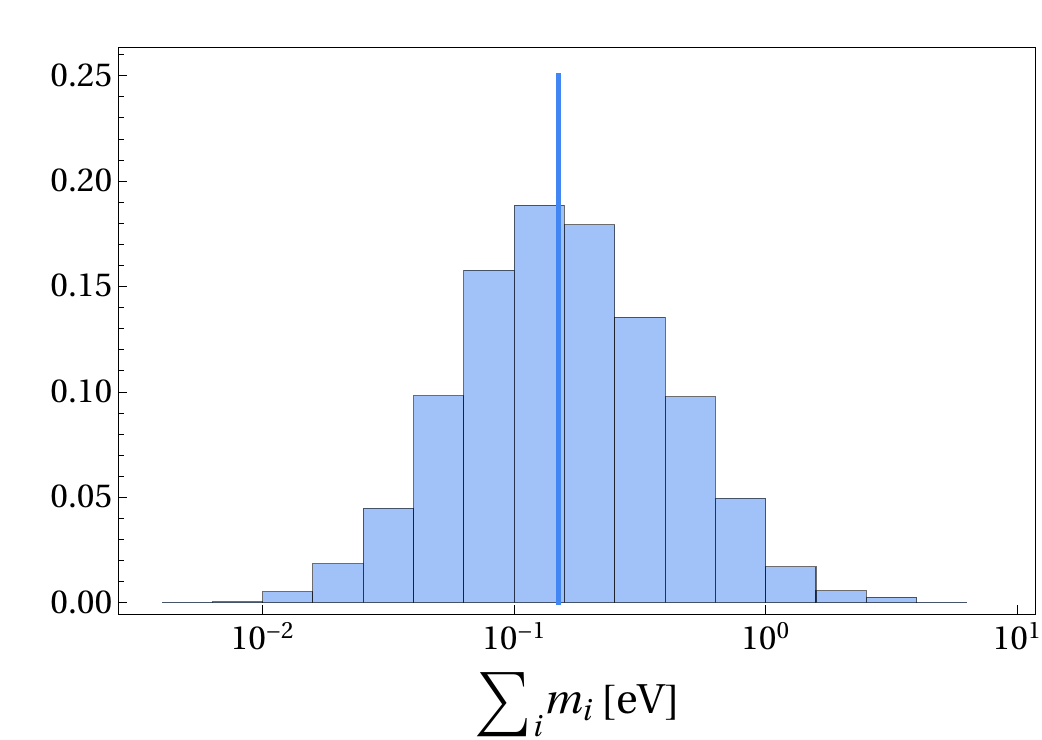}~~~
				\includegraphics[width=0.63\linewidth]{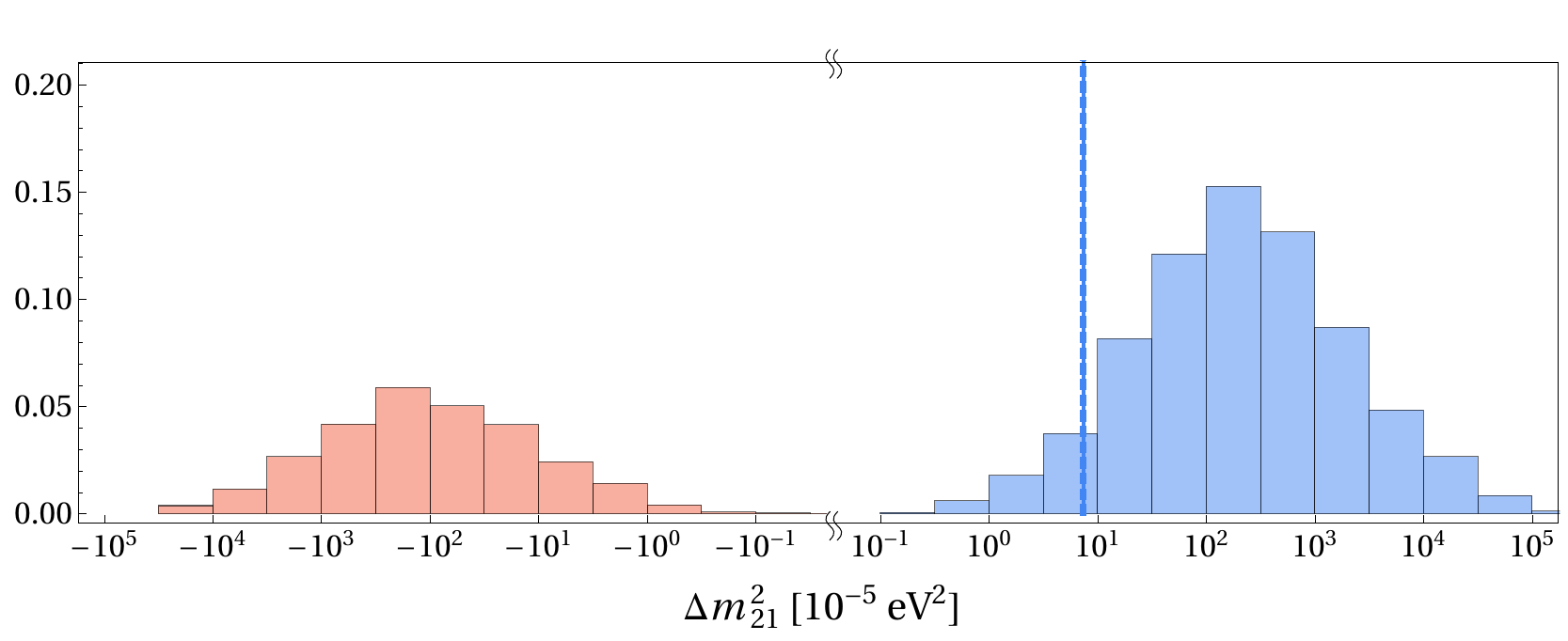}\\
		\includegraphics[width=0.63\linewidth]{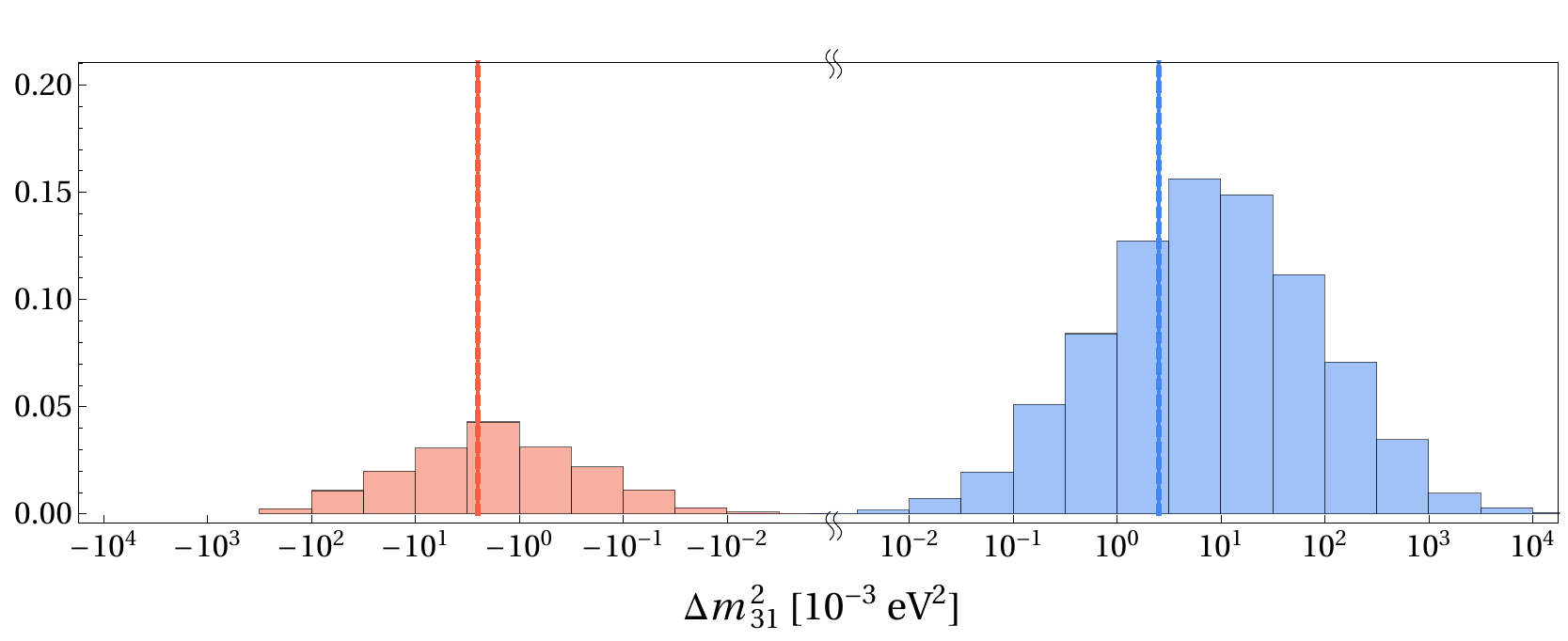}\\
		\caption{
			The distributions of the absolute values of the PMNS matrix elements (top), as well as the sum of neutrino masses $\sum_i m_{\nu_i}$ (middle left)  and neutrino mass squared differences $\Delta m_{12}^2$, $\Delta m_{31}^2$ (middler right and bottom) for the case of decoupled FN chains, $G_{\rm FN}=U(1)_{\rm FN}^3$, setting $\Lambda_{\text{LN}}=10^{11}$ GeV. The measured values are shown as vertical bars \cite{Tanabashi:2018oca,Vagnozzi:2017ovm,Esteban:2018azc}, with dashed lines denoting 1$\sigma$ errors. For $\sum_i m_{\nu_i}$ the upper bound is shown~\cite{Vagnozzi:2017ovm}, while for 
			$\Delta m_{31}^2$ two lines are shown: blue (orange) for normal (inverted)  ordering. 
			\label{fig:leptons:nu:pmns:decoupled} 
		}
	\end{center}
\end{figure}

For $G_{\rm FN}=U(1)_{\rm FN}^3$ the hierarchy among charged lepton masses is then due to different lengths of FN chains for the right-handed leptons, giving
\beq
m_{ij}^e \sim v \Big(\frac{M}{\langle \phi \rangle}\Big)^{N_{L(i)}+N_{e(j)}}.
\eeq
The observed hierarchy between $m_e:m_\mu:m_\tau$ is obtain for 
\beq
N_{e(1)}=N_{e(2)}+3= N_{e(3)}+4.
\eeq 
Even with this identification, there is still significant freedom in phenomenologically viable charge assignment, since one can compensate for a particular choice of $N_{L(i)}$ by adjusting globally the $N_{e(i)}$. In Fig. \ref{fig:leptons:scan} (left) we show the charged lepton masses that are obtained by setting 
\beq
\label{eq:NL:decoupled}
N_{L(1)}=N_{L(2)}=N_{L(3)}=3, \qquad N_{e(1)}=4, \qquad N_{e(2)}=1,\qquad N_{e(3)}=0.
\eeq
Fig.~\ref{fig:leptons:nu:pmns:decoupled} shows the resulting values of PMNS matrix elements and of the neutrino masses, setting $\Lambda_{\rm LN}=10^{11}$~GeV in order to approximately saturate the bound on the sum of neutrino masses from cosmology $\sum_i m_{\nu_i}\lesssim 0.15$ \cite{Vagnozzi:2017ovm}. These are compared with the measured values with solid (dashed) lines denoting the central values (1$\sigma$ bands)  \cite{Tanabashi:2018oca}.
The scan is peformed in the same way as for the quarks in the case of decoupled FN chains, Section \ref{sec:numerics:decoupled}. The vector-like masses are taken to be random complex numbers with magnitudes in the ranges $r_a\in [0.3,0.9]$ in units of $10^7$ GeV, with arbitrary phases, while the Yukawa couplings $Y_n^f \langle\phi\rangle$ are equal to $q$ up to a randomized complex prefactor with magnitude in the range  $[0.3,0.9]$ and a random phase, where $\langle \phi \rangle = 10^7$ GeV. This preferentially leads to  normal hierarchy, see Fig.~\ref{fig:leptons:nu:pmns:decoupled}, and to PMNS phase and Majorana phases that are completely random.

\begin{figure}[t]
	\begin{center}
		{\bf Coupled FN Chains, $\bm{ G_{\mathbf{FN}}=U(1)_{\rm FN}}$ - neutrinos}\\
		\includegraphics[width=1\linewidth]{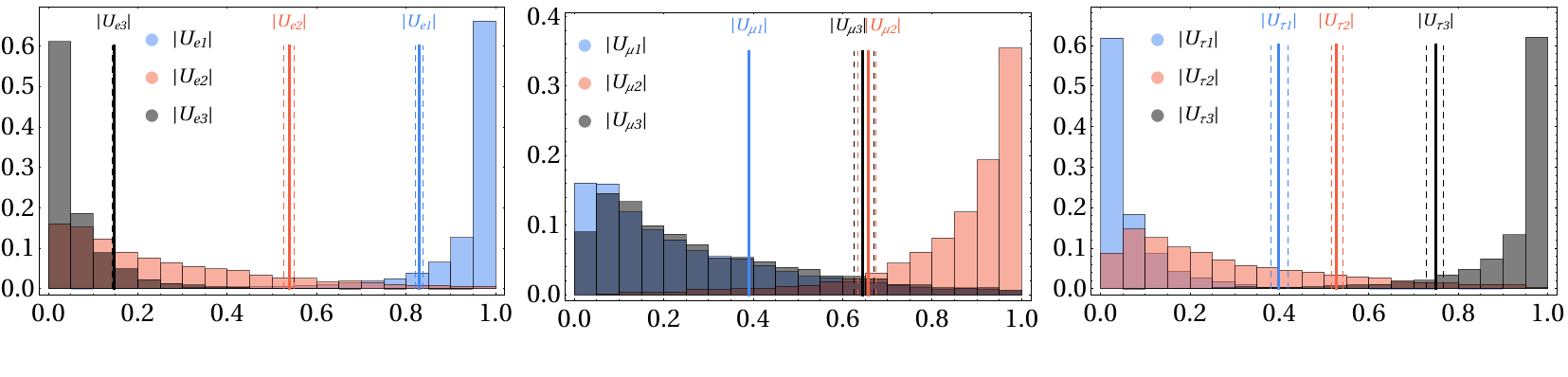}\\
		\includegraphics[width=0.34\linewidth]{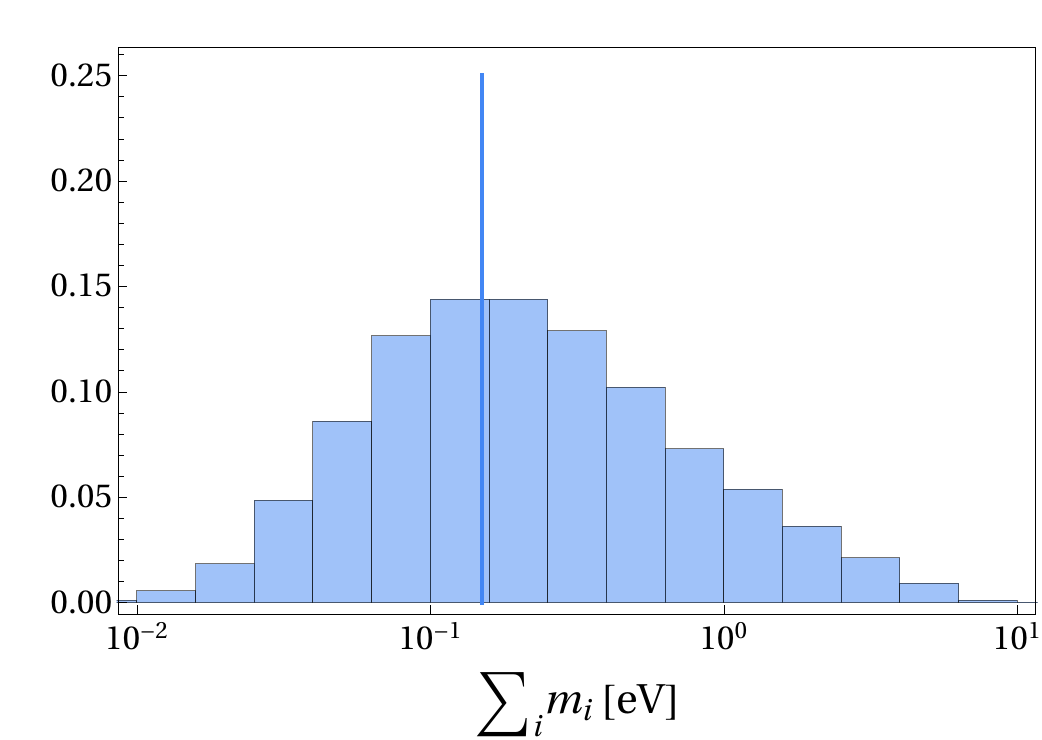}~~~\includegraphics[width=0.63\linewidth]{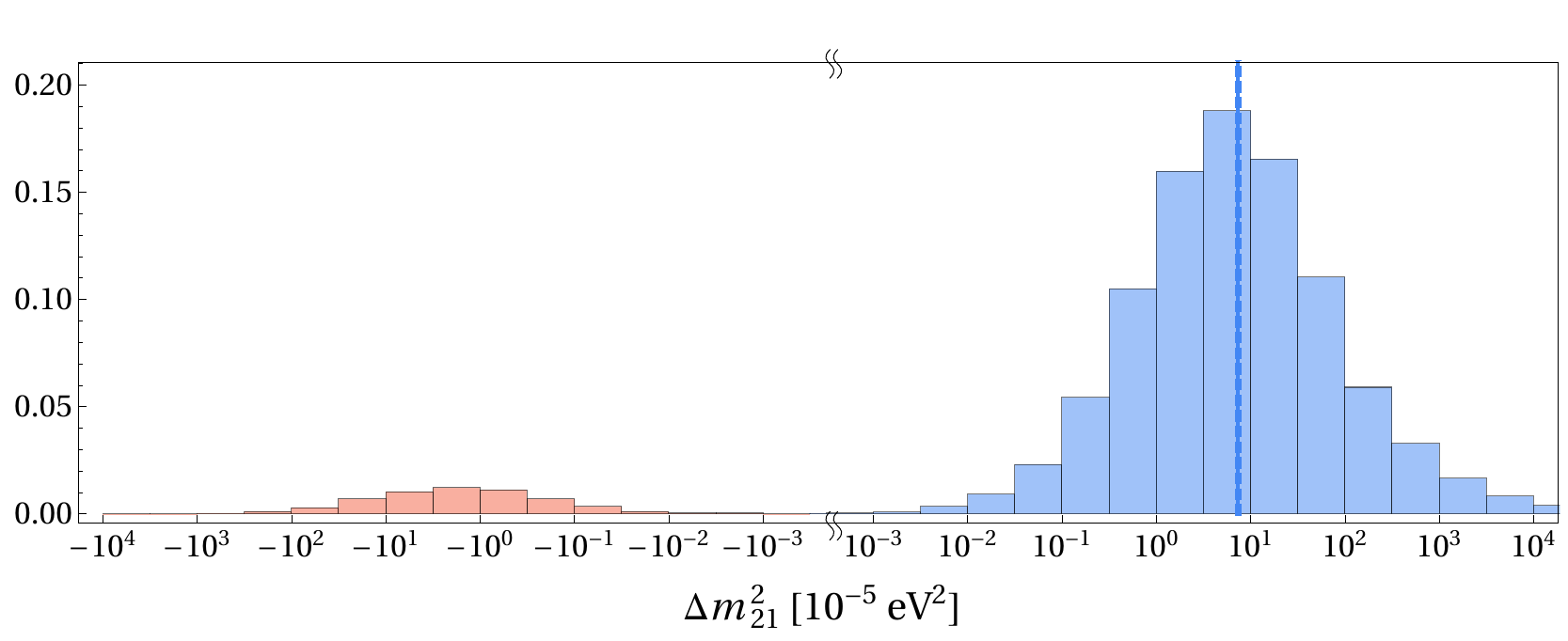}\\
\includegraphics[width=0.63\linewidth]{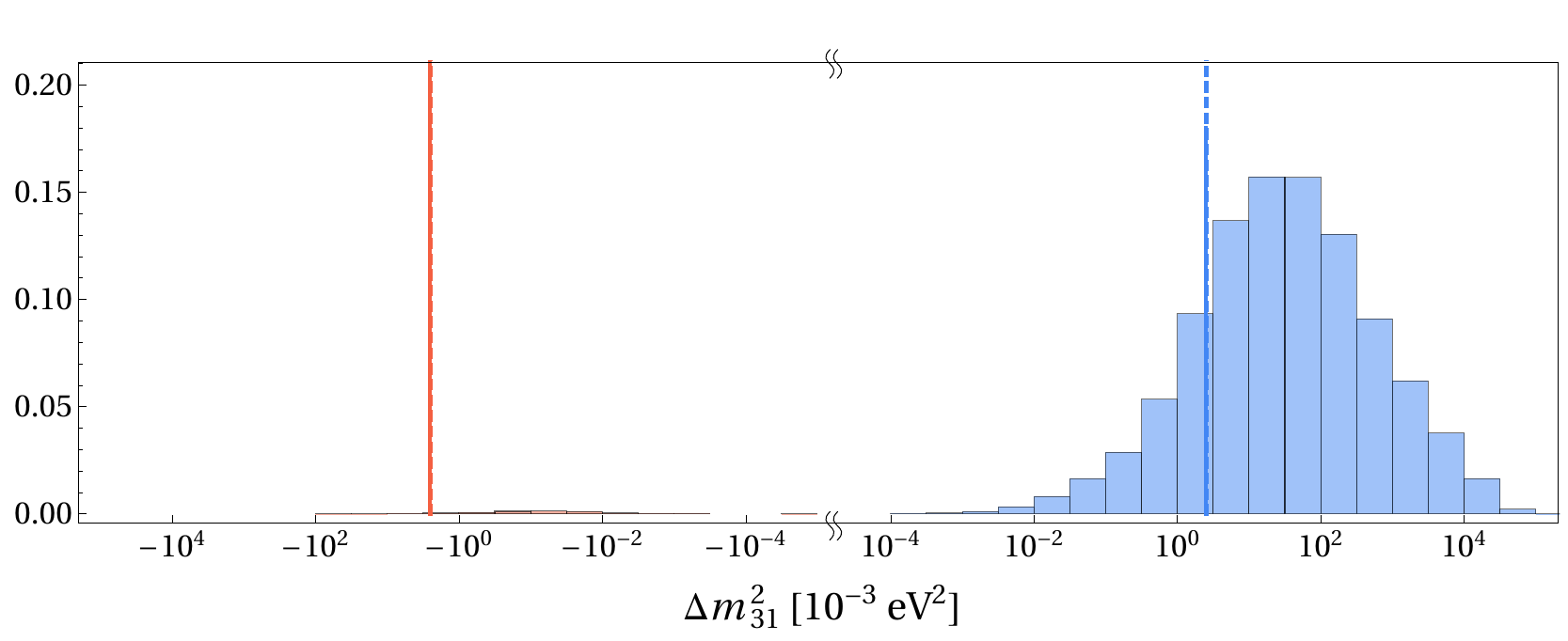}
\caption{The same as in Fig. \ref{fig:leptons:nu:pmns:decoupled},  but for the coupled FN chains, $G_{\rm FN}=U(1)_{\rm FN}$. Here we use $\Lambda_{\text{LN}}=10^{13}$ GeV, see text for details. 
 \label{fig:leptons:nu:pmns:coupled} 
		}
	\end{center}
\end{figure}

For the case of coupled FN chains, $G_{\rm FN}=U(1)_{\rm FN}$, we exploit the fact that products of random matrices have hierarchical eigenvalues. We find that a completely anarchic charge assignment
\beq
\label{eq:coupled:leptons:N}
N_{L(i)}\equiv N_L=2, \qquad N_{e(i)}\equiv N_e=3,
\eeq
 describes well the hierarchy among the charged leptons, see Fig. \ref{fig:leptons:scan} (right). In the $\langle \phi\rangle \gg M$ limit the charged lepton mass matrix is then
\beq
\label{eq:me:coupled}
m_{ij}^e \simeq \frac{v}{\sqrt{2}}M_{N_L}^L \big(Y_{N_L}^L  \langle \phi\rangle\big)^{-1} \cdots \big(Y_{1}^L \langle \phi\rangle\big)^{-1} Y_0^e\big(Y_{1}^e \langle \phi\rangle\big)^{-1} M_{1}^e\cdots \big(Y_{N_e}^e \langle \phi\rangle\big)^{-1} M_{N_e}^e.
\eeq
For $(M^{L,e}_n)_{ij}=(c_{n}^{L,e})_{ij} q\langle \phi\rangle$, with the $Y^{L,e}_n$ and $c^{L,e}_n$ matrix elements randomly distributed, such that their average values vanish, one has   \cite{vonGersdorff:2017iym}
\begin{align}
\label{eq:det:random}
\langle (\det m^e)^2\rangle &=\Big(\frac{v}{\sqrt 2}{q^{-(N_e+N_L)}}\Big)^{2N_f} \big(N_f! \sigma^{2 N_f})^{2(N_e+N_L)+1},
\\
\label{eq:tr:random}
\langle {\rm Tr}\, m^e m^{e\dagger}\rangle&=\frac{v^2}{2}q^{-2(N_e+N_L)} N_f (N_f\sigma^2)^{2(N_e+N_L)+1},
\end{align}
where for simplicity we have assumed that $(c_n^{L,e})^{-1}$ and $Y_n^{L,e}$ all follow the same distribution with variance $\sigma$. Here $N_f=3$ is the number of families, while $2(N_e+N_L)+1$ is the number of random matrices that get multiplied in Eq. \eqref{eq:me:coupled}. From \eqref{eq:det:random} and \eqref{eq:tr:random} we get
\beq\label{eq:memmu:hier}
\frac{\sqrt{m_e m_\mu}}{m_\tau}\lesssim \frac{1}{N_f^{3/4}} \biggr(\frac{N_f!}{N_f^{N_f}}\biggr)^{\frac{2(N_e+N_L)+1}{4}}=\frac{1}{3^{3/4}}\biggr(\frac{2}{9}\biggr)^{\frac{2(N_e+N_L)+1}{4}}.
\eeq
If the lengths of the  FN chains grow the eigenvalues become more hierarchical following \eqref{eq:memmu:hier}, which we also checked numerically. 

The distributions of electron, muon and tau masses that follow from the anarchic charge assignments for the coupled FN chains in Eq.~\eqref{eq:coupled:leptons:N} are shown in Fig.~\ref{fig:leptons:scan} (right). We see that despite the anarchic charges the hierarchy among the  eigenvalues indeed reproduces well  the measured hierarchy of charged lepton masses. In the scan we used the same approach as for quarks in Section \ref{sec:scan:coupled}; the vector-like masses are random complex variables with a magnitude in the range $[0.3,0.9]$ and a random phase, while the elements of Yukawa matrices have entries equal to $q$ up to a similar random complex prefactor with magnitude in the range $[0.3,0.9]$.

The resulting PMNS matrix elements and neutrino masses for $\Lambda_{\text{LN}}=10^{13}$ GeV are shown in Fig.~\ref{fig:leptons:nu:pmns:coupled}. We observe that there is a tendency for $U_{e,2}, U_{\mu,3}, U_{\tau, 2}, U_{\tau,3}$ to be below the observed values, and for $U_{e,1}, U_{\mu,2}, U_{\tau,3}$ to be above. Still, the agreement with the values realized in Nature remains reasonable. For the neutrino masses the normal ordering is heavily favored as can be seen from Fig.~\ref{fig:leptons:nu:pmns:coupled} (middle and bottom). The Majorana and PMNS phases are randomly distributed.

\section{The phenomenology of flavorful $Z'$ bosons}
\label{sec:Z'}
How can one uncover experimentally whether any of the above anomaly free FN models is realized in Nature? The immediate answer is to search for new contributions to Flavor Changing Neutral Currents (FCNCs), e.g., $B-\bar B$, $K-\bar K$, mixing, $\mu\to e\gamma$, etc. The new contributions are due to the exchanges of flavons, heavy vector-like fermions, and, if $G_{\rm FN}$ is gauged, also flavorful $Z$'s. Below we will see that the FCNC constraints on possible tree level $Z'$ exchanges
 bound $\langle \phi \rangle \gtrsim {\mathcal O}(10^7{\rm~GeV})$. This in turn means that the vector-like fermions are heavier than about $10^7$ GeV, since for ${\mathcal O}(1)$ Yukawa couplings their masses are comparable to the flavon vev, $\langle \phi \rangle$.

We will derive the FCNC constraints on inverted FN models assuming the $Z'$ contributions dominate. That is, we will assume that both the vector-like fermions and the flavon have masses  $m_i\sim \langle \phi\rangle$. As a result, their contributions to FCNC transitions are subleading
compared to the $Z'$ contributions. The vector-like fermions contribute to FCNCs only at one-loop, while the FCNC flavon couplings to the SM fermions $f_i, f_j$ are suppressed by $\sqrt{m_{f_i}m_{f_j}}/\langle \phi\rangle$.

In the construction of inverted FN models we made two choices for the anomaly free  horizontal group $G_{\rm FN}$. If  $G_{\rm FN}$ is gauged, 
there is one extra gauge boson, $Z'$, for  the $G_{\rm FN}=U(1)_{\rm FN}$ case, and three new gauge bosons, $Z_i'$,  $i=1,2,3$ for the  $G_{\rm FN}=U(1)_{\rm FN}^3$ case.  The rest of this section is devoted to the phenomenology of these flavorful $Z'$s. If the $G_{\rm FN}$ gauge couplings are small the $Z'$ can be light. This means that beside the indirect searches using FCNCs, the $Z'$ can also be searched for in on-shell production, e.g., in beam dumps or in astrophysical environments.

We first focus on the $G_{\rm FN} = U(1)_{\rm FN}$ case. After the $U(1)_{\rm FN}$ is broken the relevant terms in the Lagrangian are 
\beq\label{eq:coupl:kinmixL}
{\cal L} \supset -\frac{1}{4} B_{\mu\nu} B^{\mu\nu} -\frac{1}{4} Z'_{\mu\nu} Z'{}^{\mu\nu} -\frac{\epsilon}{2} B_{\mu\nu} Z'{}^{\mu\nu} + B_\mu J^\mu_Y + W_\mu^a J_{W^a}^\mu +Z'_{\mu} J^\mu_{\rm FN}\,, 
\eeq
where $B_{\mu\nu}$ and $Z_{\mu\nu}'$ are the field strength tensors of $U_Y(1)$ and $U(1)_{\rm FN}$, respectively, with $B_\mu$, $Z_\mu'$ the corresponding gauge bosons, while $W_\mu^a$ are the $SU(2)_L$ gauge bosons. 
Since the FN fermions are charged both under the SM gauge group and the $U(1)_{\rm FN}$, the fermionic kinetic terms give the couplings of the $B_\mu, W_\mu^a$ and $Z_\mu'$ to the chains of FN fermions. These result in the last three terms in \eqref{eq:coupl:kinmixL}, where
\begin{align}
\label{eq:BCurrent}
J_Y^\mu &= g_Y Y_{q} \sum_{i=1}^{3}  \big(\bar q_{L,0}^{(i)} \gamma^\mu q_{L,0}^{(i)}\big)  + g_Y Y_{q}  \sum_{n=1}^{N_{q(1)}} \sum_{i=1}^{\hat N_q|_n}  \big(\bar q_{L,n}^{(i)} \gamma^\mu q_{L,j}^{(i)}+\bar q_{R,n}^{(i)} \gamma^\mu q_{R,n}^{(i)} \big) +\cdots ,
\\
\label{eq:ZprimeCurrent}
J_{\rm FN}^\mu &= g'   \sum_{n=1}^{N_{q(1)}} \sum_{i=1}^{\hat N_q|_n} n \delta_q \big(\bar q_{L,n}^{(i)} \gamma^\mu q_{L,n}^{(i)}+\bar q_{R,n}^{(i)} \gamma^\mu q_{R,n}^{(i)} \big) +\cdots,
\end{align}
while $J_{W^a}^\mu$ is obtained from $J_Y^\mu$ by replacing $g_Y Y_q$ with $g T^a$, where $g$ is the $SU(2)_L$ coupling constant, and $T^a$ the corresponding generators (now acting inside the quark currents).
The ellipses denote the couplings to leptonic, up- and down-quark FN chains.  These are obtained from the terms explicitly shown in \eqref{eq:BCurrent}, \eqref{eq:ZprimeCurrent} by making replacements  $q_{L}\to L_{L}, e_{R}, u_{R}, d_{R}$ and $q_{R}\to L_{R}, e_{L}, u_{L}, d_{L}$, respectively. Here $g_Y$ and $g^\prime$ are the $U_Y(1)$ and $U(1)_{\rm FN}^\prime$ gauge couplings, while $Y_f$ and $n\delta_f$ are, respectively,  the hypercharge and the horizontal $U(1)_{\rm FN}$ quantum numbers of the fermion $f_{L/R,n}^{(i)}$, where $\delta_q=\delta_L=-1$ and $\delta_u=\delta_d=\delta_e=1$. 

The couplings to the SM fermions are obtained by performing the unitary transformations in \eqref{eq:field_rotation_chain_diag_cpld}, and keeping only the zero modes. This gives
\beq
\begin{split}
\label{eq:BZprimeCurrent:SM}
J_Y^\mu &= g_Y Y_{q} \sum_{i=1}^{3}  \big(\bar q_{L}^{(i)} \gamma^\mu q_{L}^{(i)}\big) +\cdots, \qquad J_{W^a}^\mu = g \sum_{i=1}^{3}  \big(\bar q_{L}^{(i)} \gamma^\mu T^a q_{L}^{(i)}\big) +\cdots, 
\\
J_{\rm FN}^\mu &= g'  \sum_{i,j=1}^{3} c'{}_{\negmedspace ij}^{q_L} \big(\bar q_{L}^{(i)} \gamma^\mu q_{L}^{(j)}\big) +\cdots,
\end{split}
\eeq
where, as before, the ellipses denote the couplings to right-handed quarks and to leptons, obtained through trivial replacements. The two electroweak currents, $J_Y^\mu$ and $J_{W^a}$, are flavor diagonal, since the FN chains carry the same SM charges as the corresponding SM fermions. In contrast, the FN current, $J_{\rm FN}^\mu$, has flavor violating couplings,  
\beq
c'{}_{\negmedspace ij}^{q_L}=\sum_{n=1}^{N_{q(1)}} \sum_{k=1}^{\hat N_q|_n} V_{n(k),0(i)}^{q_L *} V_{n(k),0(j)}^{q_L} n \delta_q,
\eeq
and similarly for the other $c'{}_{\negmedspace ij}^{f}$ matrices. 
 The off-diagonal entries arise because the horizontal charges of fermions on different nodes differ. Note that the zero modes are localized toward the ends of vector-like chains, and thus we expect $c'{}_{\negmedspace ij}^{q_L}$ to have eigenvalues that are ${\mathcal O}(N_{q(i)})$, see also  the discussion below  \eqref{eq:LSM:effective}.

We can get rid of the kinetic mixing between $B_\mu$ and $Z_\mu'$, Eq. \eqref{eq:coupl:kinmixL}, by performing a field redifinition $B_\mu \to B_\mu - \epsilon Z'_\mu$.\footnote{To have canonically normalized fields one also needs to rescale $Z'$ at ${\mathcal O}(\epsilon^2)$, which we can safely ignore since we work to ${\mathcal O}(\epsilon)$.} This induces a coupling of $Z'$ with $J_Y^\mu$ proportional to $\epsilon$,
\beq\label{eq:coupl:Lint}
{\cal L}\supset \lp B_\mu - \epsilon Z'_\mu\rp J_Y^\mu +  Z'_\mu J^\mu_{\rm FN}\,.
\eeq
After the field redefinition the covariant derivative acting on the Higgs also contains $Z'$, $D_\mu H=\big(\partial_\mu+i g_Y (B_\mu-\epsilon Z_\mu')/2+i g T^a W_\mu^a\big)H$.
After FN symmetry and electroweak symmetry are broken, $\phi\to \langle \phi\rangle$, $H\to (0,v/\sqrt 2)$, the scalar kinetic terms, 
\beq
{\cal L}\supset (D_\mu\phi)^\dagger (D_\mu \phi)+(D_\mu H)^\dagger (D_\mu H),
\eeq
mix the $B_\mu, W_\mu^3$ and $Z_\mu'$ (here $D_\mu \phi=\big(\partial_\mu+i g'  Z_\mu' \big)\phi$). This mixing can be thought of as occurring in two steps. In the $\epsilon\to 0$ limit the electroweak breaking mixes $B_\mu, W_\mu^3$ into a massless photon, $A_\mu$, and the massive $Z_\mu$. For nonzero $\epsilon$ the $Z_\mu$ and $Z_\mu'$ further  mix into two mass eigenstates, $\hat Z_\mu'=c_\theta Z_\mu'+s_\theta Z_\mu$, $\hat Z_\mu=-s_\theta Z_\mu'+c_\theta Z_\mu$, where $s_\theta=\sin\theta$, $c_\theta=\cos\theta$. The mixing angle is, up to ${\mathcal O}(\epsilon^2)$ corrections, 
\beq
\label{eq:tan2theta}
\tan 2\theta=2 s_W \epsilon \frac{m_Z^2}{2 g'{}^2\langle \phi\rangle^2-m_Z^2},
\eeq
with $s_W=\sin\theta_W$ the sine of the weak mixing angle. The mass of $\hat Z$ is the same as for the SM $Z$, $m_Z$, up to ${\mathcal O}(\epsilon^2)$ corrections, while the $\hat Z'$ has the mass
\beq
m_{Z'}^2=m_Z^2 (s_\theta +\epsilon s_W)^2+ 2 g'^2 \langle \phi \rangle^2.
\eeq
Note that for $m_Z\gg g'\langle \phi \rangle$ we have $\theta\to -s_W \epsilon$ and $m_{Z'}\to \sqrt 2 g' \langle \phi \rangle$. 

We can finally write down the couplings of $Z'$ to the SM fermions, 
\beq
\label{eq:cfi:Z'}
{\cal L}\supset Z_\mu' \sum_{f,i,j} \Big[ g'c_{f_L}^{ij} \big(\bar f_L^{(i)}\gamma^\mu  f_L^{(j)}\big)+g' c_{f_R}^{ij} \big(\bar f_R^{(i)}\gamma^\mu  f_R^{(j)}\big)\Big]\equiv Z_\mu' J^\mu,
\eeq
where the sum runs over all the SM fermions, $f=u,d,\ell, \nu$, with $c_{\nu_R}^{ij}=0$, and $i,j=1,\ldots,3,$ the generation indices. The couplings receive two contributions, the flavor diagonal one from $J_Y^\mu$ and $J_{W^3}^\mu$, while the contribution from the horizontal current, $J_{\rm FN}^\mu$, also contains the flavor violating couplings, 
\begin{align}
\label{eq:cu}
c_{u_L}^{ij}&=\hat \epsilon_{u_L} \delta_{ij}+\big(V_{u_L}^\dagger c'{}^{q_L} V_{u_L}\big)_{ij},  &c_{u_R}^{ij}&=\hat \epsilon_{u_R}\delta_{ij}+\big(V_{u_R}^\dagger c'{}^{u_R} V_{u_R}\big)_{ij}, 
\end{align}
and similarly for $d_{L,R}$ with $u\to d$ replacements, for $\ell_{L,R}$ with $q\to L$, $u\to \ell$ replacements, and for $\nu_L$ with $q\to L$, $u_L\to \nu_L$ replacements in the above expressions. For later convenience we also introduce the vector and axial couplings 
\beq
\label{eq:VA:couplings}
c_{fV}^{ij} = \frac{1}{2}\big(c_{f_R}^{ij} + c_{f_L}^{ij}\big)\,, \qquad\qquad c_{fA}^{ij} =\frac{1}{2} \big(c_{f_R}^{ij} - c_{f_L}^{ij}\big)\,, \qquad f=u,d,\ell.
\eeq
The flavor diagonal $\hat \epsilon_f$ term is of ${\mathcal O}(\epsilon)$, 
\beq
g' \hat \epsilon_f=-e Q_f \epsilon c_W+(s_\theta+s_W \epsilon)\big(-g_Y Y_f s_W+g T_3^f c_W\big),
\eeq
where $T_3^f$ is the weak isospin for fermion $f$.
Note that the second term vanishes in the $m_Z\to \infty$ limit, cf.~Eq.~\eqref{eq:tan2theta}, while the first term is the contribution from the kinetic mixing between $Z'$ and the photon. 
The unitary matrices $V_f$ in \eqref{eq:cu} diagonalize the corresponding SM fermion Yukawa matrices (or in the case of neutrinos the Weinberg operator mass term).

There are two distinct regimes for the couplings of $Z'$ to the SM fermions. If the $c_f$ are dominated by $\hat \epsilon_f$, then the phenomenology of $Z'$ is the same as for the dark photon. In the opposite limit, when $\hat \epsilon_f$ is negligible, the couplings of the $Z'$ are governed by the $U(1)_{\rm FN}$ charges, giving both flavor diagonal and off-diagonal couplings of comparable strength. In the numerical examples below we set $\epsilon\to 0$, see Table \ref{table:U1diagcoupl} and Appendix \ref{app:benchmarks}.  
In this limit the $Z'$ mostly couples through axial vector couplings, since left-handed and right-handed zero modes carry opposite effective $U(1)_{\rm FN}$ charges, cf. Appendix \ref{app:benchmarks}.

The above derivations generalize straightforwardly to the case of $G_{\rm FN} = U(1)_{\rm FN}^3$, i.e., the decoupled FN chains. 
The $Z'$ terms in the  $G_{\rm FN} = U(1)_{\rm FN}$ Lagrangian, Eq. \eqref{eq:coupl:kinmixL}, are replaced by 
\beq\label{eq:decoupl:kinmixL}
{\cal L} \supset \sum_{i=1}^3\lp - \frac{1}{4} Z'_{i,\mu\nu} Z'_i{}^{\mu\nu} -\frac{\epsilon_i}{2} B_{\mu\nu} Z'_i{}^{\mu\nu}+ Z'_{i,\mu} J^\mu_{i,FN}\rp + \sum_{i > j}^3 \frac{\epsilon_{ij}}{2}Z'_{i,\mu\nu}Z'_j{}^{\mu\nu}\,,
\eeq
where $J^\mu_{i,FN}$ are the currents corresponding to the horizontal $U(1)_{i}$ symmetries for the $i$-th generation,
\beq
J_{i,{\rm FN}}^\mu = g_i'   \sum_{n=1}^{N_{q(1)}}n \delta_q \big(\bar q_{L,n}^{(i)} \gamma^\mu q_{L,n}^{(i)}+\bar q_{R,n}^{(i)} \gamma^\mu q_{R,n}^{(i)} \big) +\cdots,
\eeq
with $g_i'$ the gauge coupling of $U(1)_i$.
The last term in Eq. \eqref{eq:decoupl:kinmixL} contains kinetic mixings between different $Z_i'$. These enter the interactions with the SM fermions only at order ${\mathcal O}(\epsilon g')$ or ${\mathcal O}(\epsilon^2)$ and can be safely ignored. The field redefinition $B_\mu \to B_\mu - \sum_{i=1}^3 \epsilon_i  Z'_{i,\mu}$ gets rid of the kinetic mixing between $B_\mu$ and $Z_i'$ and trades it for mixing between $Z$ and $Z_i'$ through the gauge boson mass terms, mirroring the discussion for the $G_{\rm FN}=U(1)_{\rm FN}$ case above. This results in ${\mathcal O}(\epsilon)$ flavor diagonal couplings of $Z_i'$ to the SM fermions, while the $J^\mu_{i,{\rm FN}}$ result in both flavor diagonal and off-diagonal couplings. In general we can write
\beq
\label{eq:Z':decoupled}
{\cal L}\supset  \sum_{f,i,j} Z_{k,\mu}' \Big[ g_k'c_{f_L,k}^{ij} \big(\bar f_L^{(i)}\gamma^\mu  f_L^{(j)}\big)+g_k' c_{f_R,k}^{ij} \big(\bar f_R^{(i)}\gamma^\mu  f_R^{(j)}\big)\Big],
\eeq
where the coefficients have the same general form as in Eq. \eqref{eq:cu}, but now for each $Z_i'$ separately, i.e., 
\begin{align}
\label{eq:cu:decoupled}
c_{u_L,k}^{ij}&=\hat \epsilon_{u_L,k} \delta_{ij}+\big(V_{u_L}^\dagger c'{}_{\negmedspace k}^{q_L} V_{u_L}\big)_{ij},  &c_{u_R,k}^{ij}&=\hat \epsilon_{u_R,k}\delta_{ij}+\big(V_{u_R}^\dagger c'{}_{\negmedspace k}^{u_R} V_{u_R}\big)_{ij}, 
\end{align}
The $V_{u_L}$ and $V_{u_R}$ diagonalize the up quark mass matrix, while the $c'{}_{\negmedspace k}^{q_L}$ matrix has only one nonzero entry,
\beq
\big(c'{}_{\negmedspace k}^{q_L}\big)_{kk}=\sum_{n=1}^{N_{q(k)}} V_{n,0}^{q_L(k) *} V_{n,0}^{q_L(k)} n \delta_q,
\eeq
and similarly for the other $c'{}_{\negmedspace k}^{f}$ matrices. Since the zero modes are localized toward the ends of vector-like chains we expect $\big(c'{}_{\negmedspace k}^{q_L}\big)_{kk}\sim N_{q(k)}$. The $\hat \epsilon_{f,k}$ coefficients are of ${\mathcal O}(\epsilon_i)$ and vanish in the limit $\epsilon_i\to 0$. Since this is the limit we will work we do not display them explicitly. For later convenience we also define the vector and axial couplings as
\beq
\label{eq:VA:couplings:coupled}
c_{fV,k}^{ij} = \frac{1}{2}\big(c_{f_R,k}^{ij} + c_{f_L,k}^{ij}\big)\,, \qquad\qquad c_{fA,k}^{ij} =\frac{1}{2} \big(c_{f_R,k}^{ij} - c_{f_L,k}^{ij}\big)\,, \qquad f=u,d,\ell.
\eeq

The bounds on flavorful $Z'$s come from a variety of experimental observables. They can be grouped into four broad categories: the bounds that come from $Z'$ couplings to quarks or from $Z'$ couplings to leptons, in each case either due to flavor diagonal or from flavor violating couplings. In the rest of this section we work out the relevant bounds for two representative benchmarks, one for the coupled and one for the uncoupled FN chains. The numerical inputs as well as the resulting $Z'$ couplings for the two benchmarks are given in Appendix \ref{app:benchmarks}. In both benchmarks we take the $\epsilon\to0$ limit, therefore the $Z'$ couplings are completely dictated by $J_{\rm FN}^\mu$, Eq. \eqref{eq:BZprimeCurrent:SM}.  The various experimental bounds in the $(m_{Z'},g')$ plane are compiled in Figs. \ref{fig:strongestbounds}-\ref{fig:strongestboundsZ1}. Note that the couplings of $Z'$ are mostly axial, cf. Table \ref{table:U1diagcoupl}. For other phenomenological analysis of light axial vectors, see \cite{Kahn:2016vjr,Kozaczuk:2016nma}.

\subsection{The bounds on flavorful $Z'$ for the $G_{\rm FN}=U(1)_{\rm FN}$ benchmark}
\label{sec:bounds:flavored:Z'}
We first derive the experimental bounds on the $G_{\rm FN}=U(1)_{\rm FN}$ benchmark, see Appendix \ref{sec:app:coupledFN}. 
As we will show below, the $K-\bar K$ mixing bounds the flavon vev to be very large, $\langle \phi \rangle \gtrsim 10^7$ GeV.
This means that the FN fermions are very heavy, with masses $m_F\sim {\mathcal O}(\langle \phi\rangle)$. They only give suppressed contributions to flavor observables at one loop level and can be safely ignored in our analysis. To simplify the discussion, we also assume that the flavon is very massive $m_{\phi}\sim {\mathcal O}( \langle \phi\rangle)$, giving $\sim m_{d_i} m_{d_j}/m_{\phi}^2$ suppressed contributions to the flavor observables compared to the $Z'$ and can thus be ignored. This assumption can be relaxed in the future, since $m_\phi$ is a free parameter. To make the analysis tractable, we also limit the mass of the $Z'$ to be above, $m_{Z'}>10$ MeV, an assumption that could also be relaxed in future studies.
The results obtained for $G_{\rm FN}=U(1)_{\rm FN}$  are straightforward to extend to the $U(1)_{\rm FN}^3$ model, which we do in Section \ref{sec:Zpr:U(1)3}.

\subsubsection{Flavor diagonal couplings to quarks and/or leptons}
We start by deriving constraints on $Z'$ couplings from flavor conserving processes. Representative Feynman diagrams are shown in Fig.~\ref{fig:diag:flavorconserving}, while the numerical values of $Z'$ couplings with fermions in our benchmark are listed in Appendix \ref{sec:app:coupledFN}. The $Z'$  couplings are almost flavor diagonal, see Eqs.~\eqref{eq:RecuL}-\eqref{eq:Recnu}. Furthermore, the couplings to top and bottom are highly suppressed, except the coupling to $b_R$. This is easy to understand from the $U(1)_{\rm FN}$ charge assignments, since the $q_L^{(3)}$ and $u_R^{(3)}$ are not charged under $U(1)_{\rm FN}$, see Eq. \eqref{eq:chain_lengths_coupled} and  Fig. \ref{fig:coupled:setup}. 
 
 \begin{table}
	\begin{center} 
		\begin{tabular}{cccccccc}
			\hline\hline
			$i$ & $c_{u_L}^{ii}$&$c_{u_R}^{ii}$& $c_{d_L}^{ii}$& $c_{d_R}^{ii}$ & $c_{\ell_L}^{ii}$ & $c_{\ell_R}^{ii}$ & $c_{\nu_L}^{ii}$\\
			\hline
			$1$   &-2.3 & 3 & -2 & 3 & -2 & 3 & -2 \\
			$2$   &-2.7 & 1.9 & -2.9 & 2.9 & -1.9 & 3 & -1.9 \\
			
			$3$   &-0.001 & 0.026 & -0.002 & 2.5 & -1.9 & 2.1 & -1.9\\
			\hline\hline
		\end{tabular}
		\caption{The flavor diagonal couplings of $Z^\prime$ to the SM fermions, Eq. \eqref{eq:cfi:Z'}, for the $G_{\rm FN}=U(1)_{\rm FN}$ benchmark. For a complete list see Appendix  \ref{app:benchmarks}.
		}
		\label{table:U1diagcoupl}
	\end{center}
\end{table}

\begin{figure}[t] 
	\begin{center}
		\includegraphics[width=0.75\linewidth]{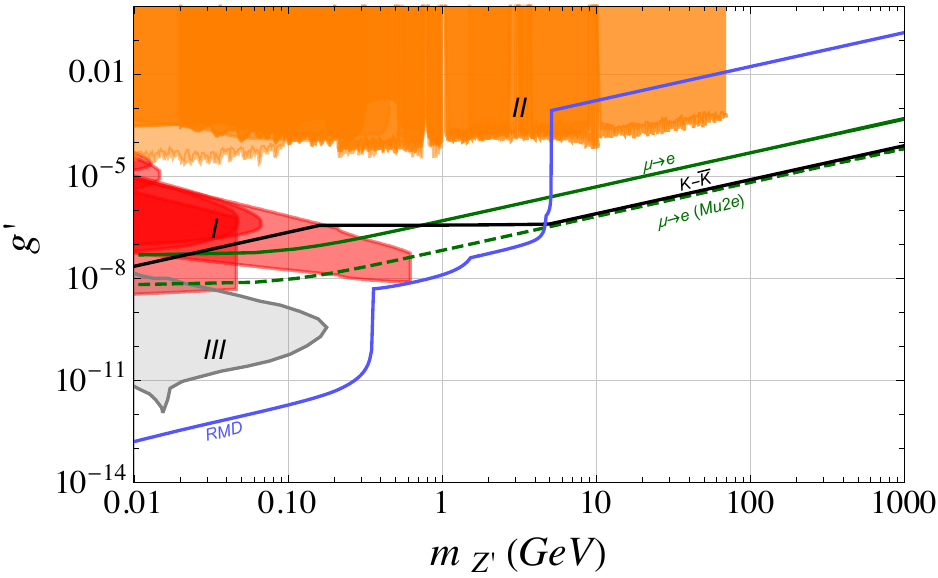}
		\caption{The strongest bounds on the $Z'$ mass as a function of gauge coupling $g'$ of the horizontal $G_{\rm FN}=U(1)_{\rm FN}$ group, setting kinetic mixing to zero. {\it Region I (II)} collects bounds due to the direct production of $Z'$ in proton and electron beam dumps (at $e^+e^-$ colliders), while {\it Region III} shows the exclusion from SN1987A. The solid (dashed) dark green line depicts the present (future) $\mu\to e$ bound, the solid black line the bound from $K - \bar K$ mixing, the solid blue line the limits from rare meson decays (RMD), see text for details. 
		}. \label{fig:strongestbounds}
	\end{center}
\end{figure}

\paragraph{Direct $Z'$ production:} 
\label{subsec:directZ'}
A number of experiments have searched for a dark photon, a heavy gauge boson that kinematically mixes with the hypercharge. These searches were recast in Ref.~\cite{Ilten:2018crw} for a generic $Z'$ with flavor diagonal vector couplings to quarks and leptons, or to new invisible states, with the results available in the form of a public code,  {\tt Darkcast}. 

\begin{figure}[t] 
	\begin{center}
		\includegraphics[width=0.49\linewidth]{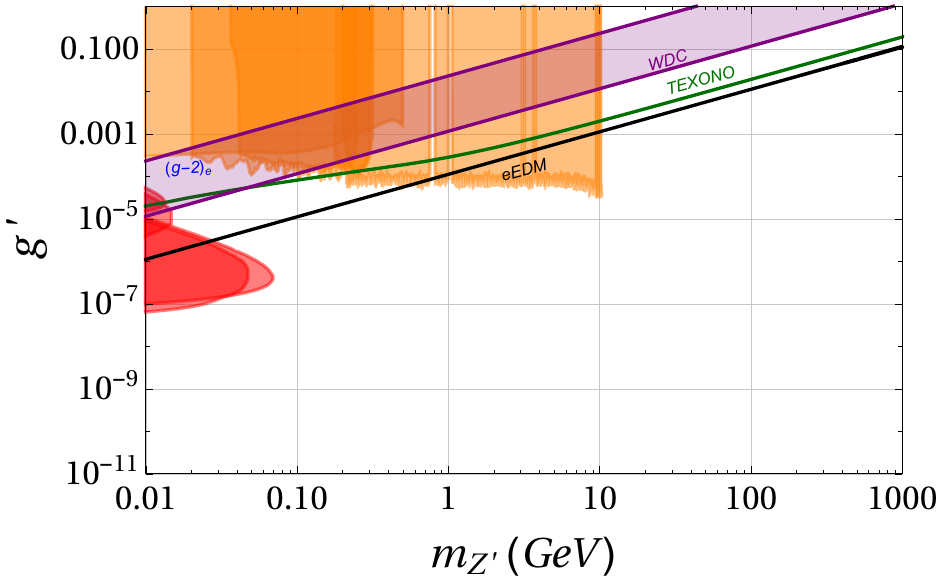}
		\includegraphics[width=0.49\linewidth]{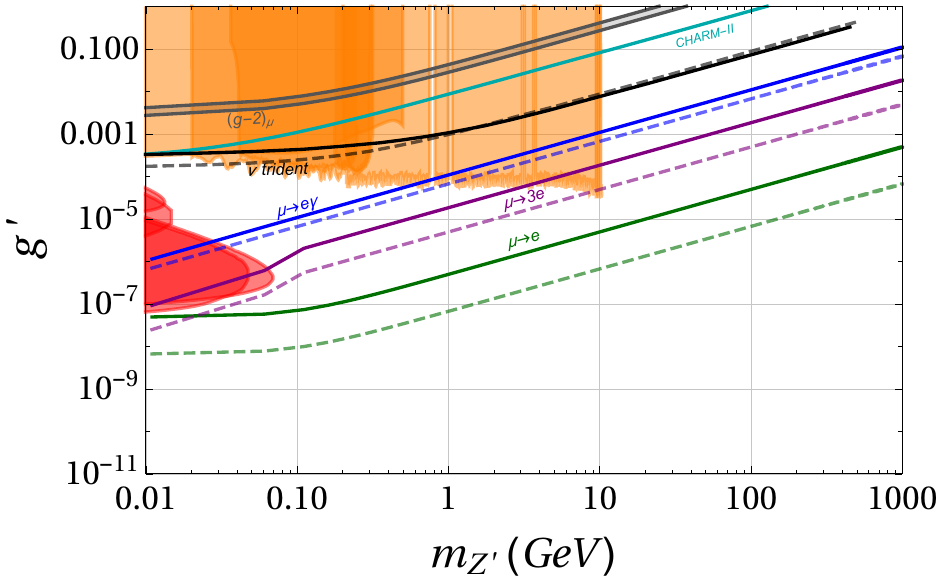}
		\includegraphics[width=0.49\linewidth]{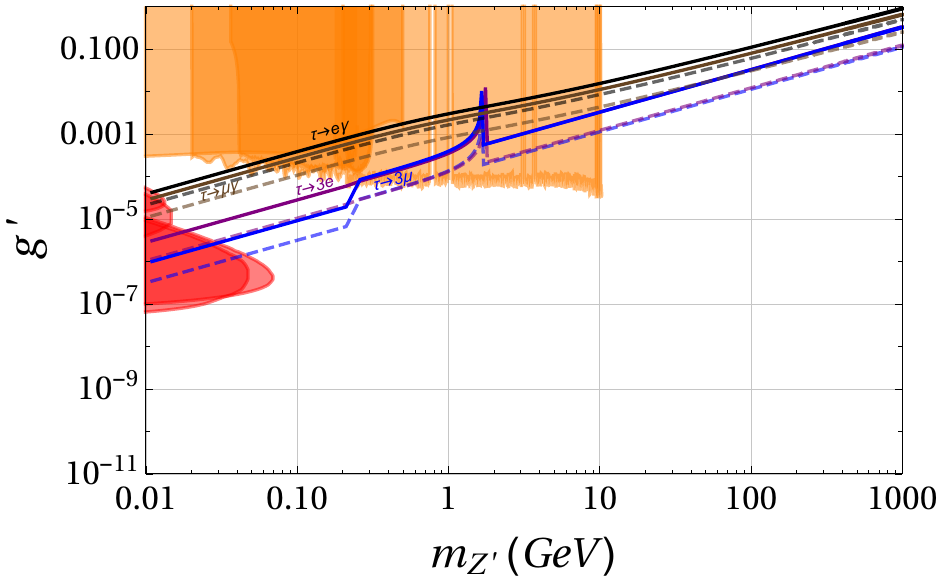}
		\includegraphics[width=0.49\linewidth]{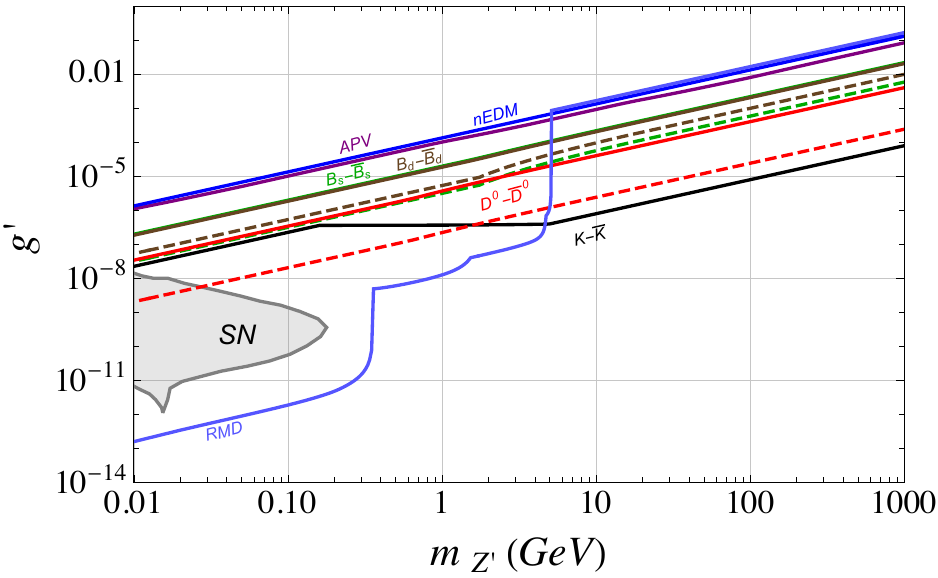}
		\caption{The top panels and bottom left (right) panel show bounds on the $Z'$ mass and the $g'$ gauge coupling constant for the $G_{\rm FN}=U(1)_{\rm FN}'$ model benchmark in the limit of no couplings to quarks (leptons).
		Direct $Z'$ production searches exclude orange and red regions, cf.~Fig.~\ref{fig:strongestbounds}. \textit{Top left}: bounds involving electrons, from the electron EDM ($\bar \nu_e-e$ scattering, $(g-2)_e$, white dwarf cooling), shown as black line (green line, blue line, purple excluded region). \textit{Top right}: bounds involving muons, from  $\mu \to e $ conversion, ($\mu\to3e$, $\mu\to e\gamma$,  neutrino trident production, $\nu_\mu-e $ scattering), shown as green (purple, blue, black, cyan) lines, while the gray region is consistent with $(g-2)_\mu$. 
\textit{Bottom left}: bounds involving taus, from $\tau \to 3\mu$ ($\tau\to 3e$, $\tau\to \mu\gamma$, $\tau\to e\gamma$) shown with blue (purple, brown, black) lines.
 \textit{Bottom right}: bounds involving quarks only, from $K-\bar K$ mixing ($D - \bar D$, $B_s - \bar B_s$, $B_d - \bar B_d$ mixing, neutron EDM, APV, RMD), shown as black (red, green, brown, blue, purple, light blue) lines, while the gray region is excluded by SN1987A. Solid (dashed) lines give present (future) bounds, see text for details. Note that the RMD bounds  involve both lepton and quark couplings.
 } \label{fig:plotonlyleptel} 
	\end{center}
\end{figure}

\begin{figure}[t] 
\begin{center}
\begin{minipage}{2.7cm}
	\includegraphics[width=1\linewidth]{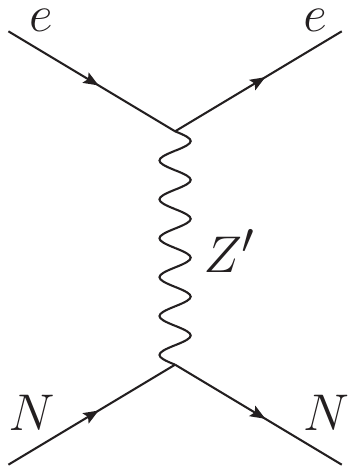}
\end{minipage}
\hspace{0.9cm}		
\begin{minipage}{4.3cm}
	\includegraphics[width=1\linewidth]{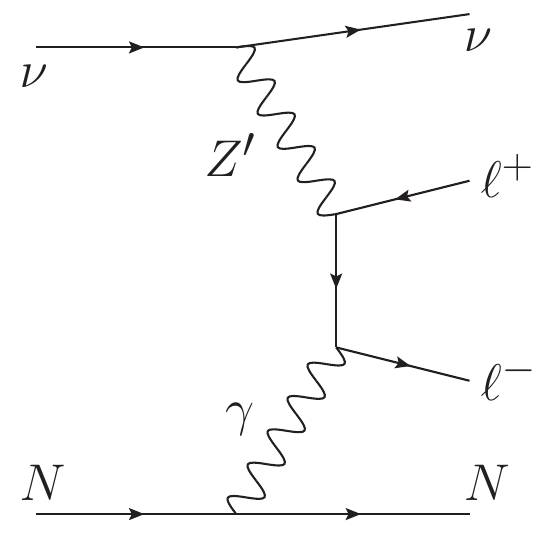}
\end{minipage}		
\hspace{0.9cm}		
\begin{minipage}{4.8cm}
	\includegraphics[width=1\linewidth]{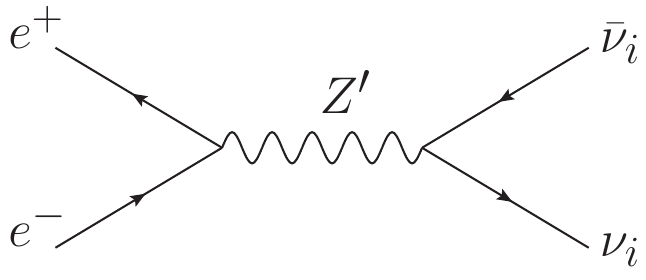}
\end{minipage}
\caption{Representative diagrams relevant for (from left to right) atomic parity violation, neutrino trident production, and white dwarf cooling bounds. 
}\label{fig:diag:flavorconserving}
	\end{center}
\end{figure}

We use {\tt Darkcast} to obtain the limits on $Z'$ from direct production, adapting it to the case in hand. The $U(1)_{\rm FN}$ $Z'$ has predominantly axial vector couplings, cf. Table \ref{table:U1diagcoupl}, unlike dark photon, which only has flavor diagonal vector couplings. For $e^+e^-\to Z'\gamma$ searches at BaBar \cite{Lees:2014xha,Lees:2017lec}, KLOE \cite{Anastasi:2015qla} and LEP \cite{Abdallah:2003np,Abdallah:2008aa} in $Z'\to e^+e^-, \mu^+\mu^-$,{\sl inv} channels, as well as for the  $Z'$ bremsstrahlung \cite{Bjorken:2009mm,Andreas:2012mt,Blumlein:2013cua} in electron beam dump searches at A1 \cite{Merkel:2014avp}, APEX \cite{Abrahamyan:2011gv}, E137 \cite{Bjorken:1988as}, E141 \cite{Riordan:1987aw}, E774 \cite{Bross:1989mp}, Orsay \cite{Davier:1989wz}, KEK \cite{Konaka:1986cb},  NA64 \cite{Banerjee:2018vgk},  and in proton beam dump search at $\nu$-CAL I \cite{Blumlein:1991xh}, with $Z'\to e^+e^-$, the {\tt Darkcast} recast of bounds applies to our $Z'$ benchmark with the replacement $\epsilon e Q_f \to ((c_{f V}^{11})^2 + (c_{f A}^{11})^2)^{1/2}$ for $f=e, u,d$, taking into account the change in the branching ratios due to possible decays to neutrinos and heavier charged fermions.  The induced uncertainties due to this identification are of roughly the same size as the uncertainties due to the approximations done in the original recast of the experiments by Ref.~\cite{Ilten:2018crw}. In the same way, the LHCb searches for dark photon \cite{Aaij:2017rft} can be recast for $m_{Z'}>1$ GeV region, where the production is dominated by Drell-Yan production, $\bar q_i q_i\to Z'$. We can also safely neglect off-diagonal couplings in direct $Z'$ production, which only lead to highly suppressed corrections. The resulting bounds are shown as red (from beam dumps) and orange (from $e^+e^-$ colliders) excluded regions in Fig.~\ref{fig:strongestbounds} and, assuming only couplings to leptons, in the first three panels in Fig. \ref{fig:plotonlyleptel}. 

 In addition, there are a number of searches for light new particles that are harder to recast for axial $Z'$; by LHCb \cite{Aaij:2017rft} for $m_{Z'}<1$ GeV, by NA60 \cite{Arnaldi:2016pzu}, CHARM \cite{Bergsma:1985qz}, $\nu$-CAL I  \cite{Blumlein:1990ay} and KLOE \cite{Archilli:2011zc}. For instance, for dark photon with mass below $1$ GeV the production in $pp$ collision is dominated by production from $\pi^0\to \gamma_d \gamma, \eta\to \gamma_d \gamma, \omega\to \gamma_d \pi^0$ decays, which can be well estimated using vector meson dominance. For axial vector it is not clear what is the dominant production channel, and would require a dedicated phenomenological analysis (for $\pi^0\to Z'\gamma$ see \cite{Kahn:2016vjr}). Using NDA we expect that the exclusions from these remaining experiments are likely to fall within or close to the red and orange exclusion regions in Fig.~\ref{fig:strongestbounds}, the same as they do for the dark photon.

\paragraph{Atomic Parity Violation (APV):}  The APV measurements bound the parity violating combination of couplings $g'{}^2 |c_{\ell A}^{11} c_N^V|$ (see Fig. \ref{fig:diag:flavorconserving} left), as a function of $m_{Z'}$, where $c_{\ell A}^{11}$ is the axial coupling to electrons, while $c_N^V= c_{uV}^{11} (2Z+N)/{A}+c_{dV}^{11}(2N+Z)/A\simeq 1.41 c_{uV}^{11} +1.59 c_{dV}^{11}$ 
is the average vector coupling of $Z'$ to the nucleon. In the last equality we evaluated the average for the Cs nucleus, which has $Z=55$, $A=133$, $N=A-Z=78$, since the most stringent bounds on NP contributions to APV come from measurements of the $6s-7s$ transition in Cs \cite{Wood:1997zq}. 
Translating the results of Ref. \cite{Dzuba:2017puc} to our notation
gives
 $g'{}^2 c_{\ell A}^{11} c_N^V> 3.9 \cdot 10^{-8} \big(m_{Z'}^2/{\rm GeV}^2\big)$, which is not very stringent and is comparable to the nEDM bound in Fig.~\ref{fig:plotonlyleptel} (bottom right).

\paragraph{Constraints from SN1987a:} 
The $Z'$ bosons can be copiously produced in a core of a Supernova (SN) if they are light enough. If $g'$ is small enough the $Z'$ bosons can escape the SN core and contribute to the cooling of the proto--neutron star. Demanding that this cooling mechanism does not lead to an instantaneous energy flux that is bigger than the one from neutrinos when the SN core reaches peak density, gives the bound shown as the gray region in Fig.~\ref{fig:strongestbounds}. For smaller $g'$ the $Z'$ bosons are not produced efficiently, while for larger $g'$ the $Z'$ are efficiently trapped inside the SN. We restrict the analysis to  $m_{Z'} \geq 10$ MeV, in which case two simplifications occur that make it easy to rescale reliably the results for dark photon from Ref. \cite{Chang:2016ntp} to our case of a $Z'$ with predominantly axial vector couplings. First of all, for $m_{Z'} \geq 10$ MeV we can neglect the corrections due to the coupling of $Z'$ with the electron plasma, an important effect for lighter gauge bosons. Furthermore, for $m_{Z'} \geq 10$ MeV the main mechanism of production and absorption of $Z'$ in the SN core is bremsstrahlung in neutron--proton scattering. The bounds can then be obtained by simply rescaling the results from \cite{Chang:2016ntp}. We use that the $pn\to pnZ'$ cross sections, $\sigma_{A,V}$, induced by the axial  or vector couplings of $Z'$ to proton, respectively, satisfy the numerical relation  $\sigma_A/\sigma_V\sim 3 c_{pA}^2/c_{pV}^2$, where $c_{pA,V}= 2 c_{uA,V}^{11}+c_{dA,V}^{11}$. 
We can thus translate 
the bounds in \cite{Chang:2016ntp} by replacing $\epsilon'$ with $\sqrt{3}g' c_{pA}$, giving the gray regions in Figs.~\ref{fig:strongestbounds} and~\ref{fig:plotonlyleptel}.

\paragraph{Neutrino trident production:} Neutrino scattering on nucleus, $A$, can produce lepton pairs through electroweak interactions. Such a trident process, $\nu_i A \to\nu_j\ell_k^+\ell_l^- A$, can receive a contribution from a tree level exchange of an extra $Z'$, see Fig. \ref{fig:diag:flavorconserving} (middle).  This would result in a deviation of the total cross section from the SM prediction, signaling new physics \cite{Altmannshofer:2014pba}. For small $Z'$ couplings the main correction to the trident cross section comes from the interference with the leading SM contribution, which is due to the $Z$ exchange. This means that the results of Ref. \cite{Altmannshofer:2019zhy} for the $L_\mu-L_\tau$ gauge boson directly translate to our case (see also \cite{Ballett:2019xoj}). The bound on $g'$ from Fig.~8 of \cite{Altmannshofer:2019zhy} needs only to be re-interpreted as the bound on $g' \big[c_{\nu_L}^{22} \big((-\frac{1}{4}+s_W^2) c_{\ell_L}^{22}+s_W^2 c_{\ell_R}^{22}\big)/\big(-\frac{1}{4}+2s_W^2\big)\big]^{1/2}$, with $s_W\equiv\sin\theta_W$ the sine of the weak mixing angle.  The resulting bounds from the CCFR experiment \cite{Mishra:1991bv} and the future projections for DUNE \cite{Altmannshofer:2019zhy}  are shown in Fig.~\ref{fig:plotonlyleptel} (top right) as solid black and dashed black lines, respectively.

\paragraph{Electron-neutrino scattering experiments:} Measurements of $\nu_i-e^-$ and $\bar\nu_i-e^-$ scattering cross sections can bound the couplings $c_{\ell A}^{11} c_{\nu L}^{ii}$ and $c_{\ell V}^{11} c_{\nu L}^{ii}$, where $i$ is the flavor of the neutrino in the beam. The most stringent constraints are due to CHARM-II \cite{Vilain:1994qy}, which used ${\cal O}(10~{\rm GeV})$ $\nu_\mu$ and $\bar \nu_\mu$ beams, and from TEXONO \cite{Deniz:2009mu}, which used ${\cal O}(1~{\rm MeV})$ reactor $\bar\nu_e$ beam. 
We use the CHARM-II measured ratio
${\sigma(\nu_\mu e)}/{\sigma(\bar\nu_\mu e)} = 0.910\pm0.113$ \cite{Vilain:1994qy}, 
where $\sigma(\nu_\mu e)$ and $\sigma(\bar \nu_\mu e)$ are the total  $\nu_\mu e$ and  $\bar\nu_\mu e$ scattering cross sections, respectively. Using this result we set the bound shown as a cyan line in Fig. \ref{fig:plotonlyleptel} (top right). The TEXONO measured total $\bar \nu_e$ scattering rate, normalized to the SM prediction, ${R_{\rm exp}}/{R_{\rm SM}}=1.08\pm0.36$ \cite{Deniz:2009mu} results in the bound on $g'$ shown as a green solid line in Fig. \ref{fig:plotonlyleptel} (top left). To derive these constraints we used the analytical results from Ref.~\cite{Lindner:2018kjo}.

\paragraph{M\o ller scattering:} Measurements of parity violating contributions to the M\o ller scattering, $e^- e^-\to e^- e^-,$ bound the product $c_{\ell V}^{11} c_{\ell A}^{11}$. The most precise measurements were performed by E158 at SLAC at  $q^2\simeq (0.16~{\rm GeV})^2$ \cite{Anthony:2005pm}. Comparison with the SM gives 
for $m_{Z'}\lesssim 100$ MeV the bound $g'{}^2|c_{\ell V}^{11} c_{\ell A}^{11}|\lesssim 10^{-8}$, and for $m_{Z'}\gtrsim 100$ MeV the bound
$g'{}^2 |c_{\ell V}^{11} c_{\ell A}^{11}| \lesssim 3\cdot 10^{-8} \cdot (m_{Z'}/200~{\rm MeV})^2$ (see also \cite{Kahn:2016vjr}).
In our benchmark $c_{\ell V}^{11} c_{\ell A}^{11}\simeq 0.30$, giving relatively weak bounds $g'\lesssim 2 \cdot 10^{-4}$ and $g'\lesssim 2 \cdot 10^{-4} (m_{Z'}/100~{\rm MeV})$, respectively, which we thus do not plot in Fig. \ref{fig:plotonlyleptel}.

\paragraph{Isotope shift spectroscopy:} The isotope shift spectroscopy constrains vector couplings of $Z'$, giving $ g'{}^2 |c_{\ell V}^{11} c_{fV}^{11}| \lesssim 10^{-7} \cdot {m_{Z'}^2}/{(10~{\rm MeV})^2}$ for $f=u,d$, when charge radius determination from Lamb shift in muonic atoms is used \cite{Delaunay:2017dku} (see also \cite{Delaunay:2016brc,Frugiuele:2016rii,Berengut:2017zuo}). Since the $Z'$ has suppressed vector couplings, these bounds are not very constraining, and we do not consider them further.

\paragraph{White dwarf cooling:} The tree level $Z'$ exchange contributes to the $e^+ e^-\to \nu\bar \nu$ process which can increase the cooling rate of the white dwarf (WD) core \cite{Dreiner:2013tja}. The cooling rate due to this additional cooling mechanism should not exceed the SM cooling rate due to the plasmon decaying into neutrinos. For our $Z'$ benchmarks the contribution to the star cooling is described by the effective Lagrangian
\beq
{\cal L} = \frac{g'{}^2 c_{\nu_L}^{ij}}{m_{Z'}^2} \lp c_{\ell V}^{11}  \big(\bar e\gamma_\mu e\big)\big( \bar\nu_L^i\gamma^\mu\nu_L^j\big) + c_{\ell A}^{11}  \big(\bar e\gamma_\mu \gamma_5 e\big)\big( \bar\nu_L^i\gamma^\mu\nu_L^j\big) \rp\,,
\eeq
since the $Z'$ is much heavier than the WD internal temperature of a few keV, and can be integrated out. Translating the limits from \cite{Dreiner:2013tja} to our notation gives (see also \cite{Bauer:2018onh}),
\beq
\frac{1.12\cdot10^{-5}}{{\rm GeV}^{-2}} < \frac{g'{}^2 c_{\nu_L}^{\rm eff} c_{e}^{\rm eff}}{m_{Z'}^2} < \frac{4.50\cdot10^{-3}}{{\rm GeV}^{-2}}\,,
\eeq
with $(c_{\nu_L}^{\rm eff})^2=\sum_{ij} |c_{\nu_L}^{ij}|^2$, and $(c_{e}^{\rm eff})^2=|c_{\ell A}^{11}|^2+|c_{\ell V}^{11}|^2$. In our benchmark, $c_{\nu L}^{\rm eff}\simeq 3.4$, and $c_{e}^{\rm eff}\simeq 2.5$, giving the exclusion shown in 
 Fig. \ref{fig:plotonlyleptel} (top left) as a purple region.

\paragraph{Anomalous magnetic moments:} At one loop the $Z'$ exchange contributes to the lepton anomalous magnetic moment, $(g-2)_\ell$, cf. Fig. \ref{fig:diag:LFV} (with $\ell_i=\ell_j$). For our $Z'$ benchmark we only need to keep the contribution to $(g-2)_\mu$ from the diagonal couplings, with $\mu$ running in the loop. A similar diagram with a $\tau$ running in the loop does get a chirality flip enhancement of $m_\tau/m_\mu \sim 10$, but is also suppressed by two off-diagonal couplings, $|(c_{\ell_L}^{23} c_{\ell_R}^{23})/(c_{\ell_L}^{22})^2| \sim 10^{-3}$. For the $(g-2)_e$ the two corresponding factors are $m_\tau/m_e \sim 3\cdot 10^3$ and $|(c_{\ell_L}^{13} c_{\ell_R}^{13})/(c_{\ell_L}^{11})^2| \sim 6\cdot 10^{-5}$, so that again the diagram with diagonal couplings dominates. Using the results of Ref. \cite{Fayet:2007ua}  with a trivial change of notation gives (see also \cite{Kozaczuk:2016nma}),
\beq
\delta a_\ell = \frac{(g'c_{\ell V}^{ii})^2}{12 \pi^2} \frac{m_\ell^2}{m_{Z'}^2} F(m_{Z'}/m_\ell) - \frac{(g'c_{\ell A}^{ii})^2 }{4\pi^2} \frac{m_\ell^2}{m_{Z'}^2}H(m_{Z'}/m_\ell)\,,
\eeq
with $i=1,2$ for $\ell=e, \mu$, respectively.
The normalization of the loop functions,
\beq
F(u) = 3 u^2 \int_0^1 dx\frac{x^2(1-x)}{x^2 + u^2(1-x)} ,\qquad  H(u)= \int_0^1 dx\frac{2 x^3  + (x-x^2)(4-x) u}{x^2 + u (1-x)}\,,
\eeq
is such that for heavy $Z'$, $\lim_{u\to \infty}F(u)=1$, $\lim_{u\to \infty}H(u)=5/3$, whereas for $m_{Z'}=m_\ell$, $F(1)\simeq 0.31, H(1)\simeq 1.31$, and for light $Z'$, i.e., for $u\ll 1$,  $F(u)\simeq 3 u^2/2 $, $H(u)\simeq 1$.

In Fig. \ref{fig:plotonlyleptel} (top right) we show in gray the $1\sigma$ band in the $g', m_{Z'}$ parameter space that gives $(g-2)_\mu$ in agreement with experiment. The required value of $g'$ is excluded by a number of other measurements, among others also by the limit on the allowed NP contribution to $(g-2)_e$, denoted with a blue line in Fig. \ref{fig:plotonlyleptel} (top left).

Note that here we consider only the contributions to $(g-2)_f$ from $Z'$ running in the loop. The contributions from heavy vector-like fermions running in the loop are expected to give contributions that are parametrically of the same order. We do not attempt to include these contributions given that the $(g-2)_f$ measurements put only weak bounds on our inverted FN benchmark. However, because of this approximation one should view the $(g-2)_f$ bounds shown in Fig. \ref{fig:plotonlyleptel} only as indicative. Similar comments apply to electron EDM, which we discuss next.

\paragraph{Electric dipole moments:} The complex off-diagonal couplings of $Z'$ generate at one loop the electric dipole moments (EDMs). The contribution to the EDM of fermion $f_i$ from the fermion $f_j$ running in the loop is given by \cite{Hollik:1998wk} 
\beq\label{eq:edm}
\frac{d_{f_i}}{e} = \frac{g'{}^2}{4\pi^2}\sum_j  m_{f_j}{\rm Im}\lp c_{f V}^{ij} c_{f A}^{ij\,*} -  c_{f A}^{ij} c_{f V}^{ij\,*} \rp \lp 2 C_1^+ - C_0 \rp_{jjZ'},
\eeq 
where $C_1^+, C_0,$ are the three-point one loop integrals arising from the evaluation of the Feynman diagram in Fig.~\ref{fig:diag:LFV} (right), the analytic expression for which can be found in \cite{Beenakker:1991ca}.
The chromo-EDMs, $\tilde d_{f_i}$, are obtained from \eqref{eq:edm} by replacing $g \to g_s$.
The experimental bound on electron EDM, $|d_e| < 1.1\cdot10^{-29} e\,{\rm cm}$~\cite{Andreev:2018ayy}, results in a bound shown as a black line in Fig.~\ref{fig:plotonlyleptel} (top left). 
The bound on neutron EDM, $d_n$,  on the other hand, translates to a bound denoted with a blue line in Fig.~\ref{fig:plotonlyleptel} (bottom right).
The neutron EDM receives contributions both from quark EDMs, $d_q$, and quark chromo-EDMs, $\tilde d_q$, 
so that $d_n = \sum_{q=u,d,s}\big(\beta_q d_q +\tilde \beta_q \tilde d_q\big)$.
For matrix elements of quark EDM operators, $\beta_q$, a recent lattice QCD calculation obtained   $\beta_u=0.784(30)$, $\beta_d= -0.204(15)$, $\beta_s=-0.0027(16)$  at $\mu=2$ GeV in $\overline{\rm MS}$ scheme \cite{Gupta:2018lvp}. In contrast, the matrix elements of chromo-EDM operators are poorly known with the estimates  ranging over an order of magnitude, $\tilde \beta_u=-(0.09-0.9)$, $\tilde \beta_d=-(0.2-1.8)$~\cite{Engel:2013lsa}. In Fig.~\ref{fig:plotonlyleptel} (bottom right) we used the estimated best values, $\tilde \beta_u=-0.35$, $\tilde \beta_d=-0.7$~\cite{Engel:2013lsa}. 

\subsubsection{Flavor violating couplings to quarks}
\begin{figure}[t] 
	\begin{center}
		\includegraphics[width=0.37\linewidth]{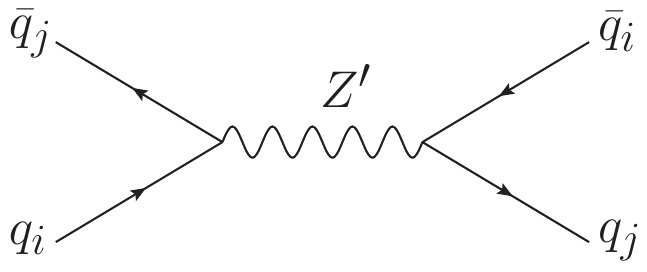} \\\vspace{10pt}
		\caption{Meson mixing process induced by a tree level $Z'$ exchange.} \label{fig:diag:mesonmixing}
	\end{center}
\end{figure}

The exchange of $Z'$ induces Flavor Changing Neutral Currents (FCNCs) already at tree level, as shown in Fig. \ref{fig:diag:mesonmixing}. Strong constraints are obtained from the meson mixings, which we derive in detail in the next paragraphs. 

\paragraph{$K^0 - \bar K^0$ mixing:} To estimate the contributions of tree level $Z'$ exchanges to $K-\bar K$ mixing we distinguish two regimes. For light $Z'$, $m_{Z'}\lesssim m_K$, we can use ChPT with the $Z'$ acting as an external field. Using ChPT expansion we have $(\bar s\gamma^\mu \gamma_5 d)\to -f_K\partial^\mu K^0+\cdots $, where the ellipses denote higher orders in momentum expansion. Using this in Eq. \eqref{eq:cfi:Z'} gives for the tree level $Z'$ exchange contribution to $K-\bar K$ mixing (cf. Eq. \eqref{eq:VA:couplings}) 
\beq
\label{eq:M12:KKbar:light}
M_{12}^{Z'}= \langle K^0| {\cal H}^{Z'}_{\rm eff} | \bar K^0 \rangle=  g'{}^2(c_{dA}^{12}\big)^2 \frac{f_K^2}{2m_K} \frac{m_K^2}{m_{Z'}^2}, \qquad\qquad\text{[light~}Z'],
\eeq
where $f_K\simeq 156$ MeV is the kaon decay constant, and we use the same phase conventions for $M_{12}$ as in \cite{Buras:1998raa,Buras:2010pza}. Note that this contribution is proportional to $1/m_{Z'}^2$ even though the momentum flowing in the $Z'$ propagator is ${\mathcal O}(m_K)$, so that $M_{12}^{Z'} \propto 1/\langle \phi\rangle^2$. This is most easily understood in the Feynman gauge, where the dominant contribution to $M_{12}^{Z'}$ due to the exchange of the $\arg(\phi)$ Goldstone boson, whose couplings to SM fermions are $\propto 1/\langle \phi \rangle$ and do not depend on $g'$. In the unitary gauge the dominant contribution arises from the longitudinal component of the $Z'$ propagator.  
The vector couplings of $Z'$ contribute only at higher order in ChPT and are relatively suppressed by ${\mathcal O}(m_K^2/(4\pi f_K)^2)\sim {\mathcal O}(0.2)$, on top of the suppression of vector couplings themselves. For completeness we calculate these contributions in Appendix \ref{sec:app:KKbar}.

For $m_{Z'} \gg m_K$ we can integrate out the $Z'$ at the scale $\mu\simeq m_{Z'}$ and match onto the effective weak Hamiltonian, $H_{\rm eff}= \sum_a C_a Q_a$, where the sum runs over the operators in Ref. \cite{Bona:2007vi}. The nonzero NP contributions to the Wilson coefficients are, 
\beq\label{eq:Ca:KKbar}
C_1^{sd}=\frac{g'{}^2 }{m_{Z'}^2} (c_{d_L}^{12})^2, \quad \tilde C_1^{sd}=\frac{g'{}^2 }{m_{Z'}^2} (c_{d_R}^{12})^2,
\quad C_5^{sd}=-4\frac{g'{}^2  }{m_{Z'}^2} c_{d_L}^{12} c_{d_R}^{12} \qquad \text{[heavy~}Z'],
\eeq
where $Q_1^{sd}=(\bar d\gamma^\mu s_L)^2$, $\tilde Q_1^{sd}=(\bar d\gamma^\mu s_R)^2$, $Q_5^{sd}=(\bar d^\alpha s_R^\beta) (\bar d^\beta s_L^\alpha)$. 
To compare with the experimental results we run from $\mu\simeq m_{Z'}$ down to $2$ GeV using the results of \cite{Bona:2007vi,Ciuchini:1998ix}. 

The comparison of the SM predictions and the experimental measurements for indirect CP violation parameter $\epsilon_K$ is made through a ratio
\beq
C_{\varepsilon_K} = \frac{\epsilon_K^{\rm exp}}{\epsilon_K^{\rm SM}}=\frac{{\rm Im}\,\langle K^0| H_{\rm eff}^{\text{SM}+Z'} | \bar K^0 \rangle}{{\rm Im}\, \langle K^0| H_{\rm eff}^{\rm SM} | \bar K^0 \rangle}.
\eeq
 Estimating the errors due to charm loop contributions in the SM prediction of $\epsilon_K$ is nontrivial and leads to differences between CKMfitter and UTFit collaborations \cite{Charles:2015gya,UTFit} (see also the introduction in \cite{Ligeti:2016qpi}). For consistency we use the UTFit collaboration extraction $0.87< C_{\epsilon_K}<1.39$  at 95\% C.L. \cite{UTFit} as well as their prediction for $\epsilon_K^{\rm SM}$ using NLO charm contribution \cite{Ciuchini:2000de}. 
 We denote the resulting bounds in Fig. \ref{fig:strongestbounds} and \ref{fig:plotonlyleptel} (bottom right) with black lines, where we use the heavy $Z'$ solution \eqref{eq:Ca:KKbar} for  $m_{Z'}> 3 m_K$, the light $Z'$ solution \eqref{eq:M12:KKbar:light} for $m_{Z'}< m_K/3$, and connect the two with naive linear interpolation to guide the eye.
 The $Z'$ contribution to the $K-\bar K$ mixing is $\propto 1/\langle \phi\rangle^2$. For heavy $Z'$, $m_{Z'}\gg m_K$, this then leads to a bound $\langle\phi\rangle=m_{Z'}/g'>1.2\cdot10^7$ GeV for our benchmark.

\paragraph{$B_q^0 - \bar B_q^0$ mixing:} We distinguish two limits, the heavy and light $Z'$. For heavy $Z'$, $m_{Z'}\gg m_{B_q}$, the $Z'$ is integrated out  at $\mu\simeq m_{Z'}$. The matching onto $H_{\rm eff}= \sum_a C_a Q_a$, gives, analogously to \eqref{eq:Ca:KKbar},
\beq
\label{eq:Bmixing:heavyZ'}
C_1^{qb}=\frac{g'{}^2 }{m_{Z'}^2} (c_{d_L}^{i3})^2, \quad \tilde C_1^{qb}=\frac{g'{}^2 }{m_{Z'}^2} (c_{d_R}^{i3})^2,
\quad C_5^{qb}=-4\frac{g'{}^2  }{m_{Z'}^2} c_{d_L}^{i3} c_{d_R}^{i3} \qquad \text{[heavy~}Z'],
\eeq
where $q=d,s$, with $i=1,2$, respectively. We RG evolve the above Wilson coefficients from $\mu\simeq m_{Z'}$ down to $\mu=4.2{\rm ~GeV}\simeq m_b$ using the results of \cite{Bona:2007vi,Ciuchini:1998ix}, and compare with the allowed deviations in the mixing matrix elements,
\beq
C_{B_q} e^{2 i \phi_{B_q}}= \frac{\langle B_q^0| H_{\rm eff}^{\text{SM}+Z'} | \bar B_q^0 \rangle}{ \langle B_q^0| H_{\rm eff}^{\rm SM} | \bar B_q^0 \rangle}.
\eeq
The latest fit results from UTFit collaboration give $0.942< C_{B_s}<1.288$, $-1.35^\circ<\phi_{B_s}<2.21^\circ$, and $0.83< C_{B_d}<1.29$, $-6.0^\circ<\phi_{B_d}<1.5^\circ$ \cite{UTFit}.

For light $Z'$ we perform an operator product expansion, where, to leading order in $\Lambda_{\rm QCD}/m_b$, the dominant Wilson coefficients of the $B_q^0-\bar B_q^0$ operators are at $\mu\simeq m_b$ given by 
\beq\label{eq:Bmixing:lightZ'}
C_2^{qb}=-\frac{g'{}^2 }{m_{Z'}^2} (c_{d_L}^{i3})^2, \quad \tilde C_2^{qb}=-\frac{g'{}^2 }{m_{Z'}^2} (c_{d_R}^{i3})^2,
\quad C_4^{qb}=-2\frac{g'{}^2  }{m_{Z'}^2} c_{d_L}^{i3} c_{d_R}^{i3} \qquad \text{[light~}Z'],
\eeq
with $m_b\simeq 4.2$ GeV the $b$ quark mass, while the remaining nonzero Wilson coefficients, $C_2^{qb}=-C_1^{qb}m_{Z'}^2 /m_b^2 $, $\tilde C_2^{qb}=-\tilde C_1^{qb} m_{Z'}^2 /m_b^2 $, $C_5^{qb}=2 C_4^{qb} m_{Z'}^2 /m_b^2$, are parametrically suppressed.

We denote the resulting bound in Fig.~\ref{fig:plotonlyleptel} (bottom right) with a green (brown) solid line for $B_s-\bar B_s$ ($B_d-\bar B_d$) mixing, while the corresponding dashed lines shows the projected bounds 
after Belle II and LHCb Upgrade II \cite{Cerri:2018ypt}.

\paragraph{Charm mixing:} The $Z'$ contributions to $D-\bar D$ mixing take the same form as for $B_q^0-\bar B_q^0$, after making the replacements $i\to 1$, $3\to2$ and $d,q\to u, b\to c$ in Eqs.~\eqref{eq:Bmixing:heavyZ'}, \eqref{eq:Bmixing:lightZ'}. The expressions for the case of light $Z'$ are only approximate, valid to the extent that one can use the operator product expansion despite the relatively light charm quark mass. The resulting present bound  \cite{UTFit} and projected sensitivity \cite{Cerri:2018ypt} are shown in Fig.~\ref{fig:plotonlyleptel} (bottom right) as solid and dashed red lines, respectively.

\subsubsection{Flavor violating  couplings to leptons}
\begin{figure}[t] 
	\begin{center}
		\includegraphics[width=0.23\linewidth]{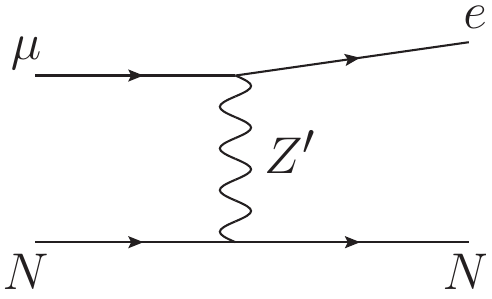}~~~~~~~~~~~
		\includegraphics[width=0.23\linewidth]{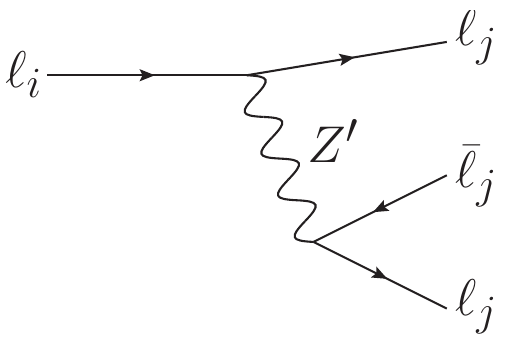}~~~~~~~~~~~~~
		\includegraphics[width=0.23\linewidth]{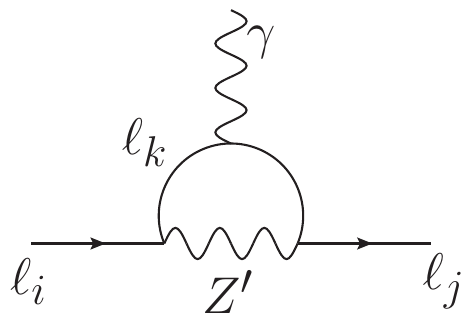}
		\caption{Lepton flavor violating diagrams induced by a $Z'$ exchange. \textit{Left}: $\mu\to e$ conversion. \textit{Center}: decay to three leptons, $\ell_i\to \ell_j\ell_j\bar \ell_j$. \textit{Right}: radiative decay, $\ell_i\to \ell_j\gamma$.} \label{fig:diag:LFV}
	\end{center}
\end{figure}

In this subsection we derive the experimental bounds from lepton flavor violating transitions. The relevant diagrams are  summarized in Fig. \ref{fig:diag:LFV}. 

\paragraph{$\mu\to e$ conversion:} Stopped muons can undergo coherent $\mu\to e$ conversion in the field of the nucleus. The most stringent limits to date were obtained by the SINDRUM II experiment  using $^{197}$Au \cite{Bertl:2006up}
\beq
\label{eq:mutoe}
{\rm Br}(\mu\to e) = \frac{\Gamma(\mu^- \text{Au}\to e^- \text{Au})}{\Gamma_{\rm capt}(\mu^-\text{Au})} < 7\cdot10^{-13}\,,\quad \text{at $90\%$ C.L.}
\eeq
 Here $\Gamma(\mu^- \text{Au}\to e^- \text{Au})$ is the conversion rate, and $\Gamma_{\rm capt}(\mu^-\text{Au})$ the muon capture rate. State of the art estimates for both of these can be found in Ref. \cite{Kitano:2002mt} for a whole range of nuclei, including gold. The tree level $Z'$ exchange induces four-fermion effective interaction for $\mu\to e$ conversion, see Fig.~\ref{fig:diag:LFV}, which in the notation of Refs.~\cite{Kitano:2002mt,Kuno:1999jp} takes the form $ {\cal L}_{\rm int}=-\big[(G_F/\sqrt{2})\sum_{f=u,d} g_{LV(f)} (\bar e\gamma_\mu P_L\mu) (\bar f \gamma^\mu f) +g_{RV(f)} (\bar e\gamma_\mu P_R\mu) (\bar f \gamma^\mu f)\big] + \big[(\bar f \gamma^\mu f)\to (\bar f \gamma^\mu\gamma_5 f), V\to A\big]$, with
\beq
 g_{LV(f)} = \frac{\sqrt{2}}{G_F} \frac{g'{}^2}{m_\mu^2+m_{Z'}^2} c_{\ell_{L}}^{12} c_{fV}^{11}, \qquad f=u,d,
\eeq
and similarly for $g_{RV(f)}$, but with $L\to R$ replacement, while $g_{LA(f)}$ and $g_{RA(f)}$ trivially follow from $V\to A$ replacement. 

The $g_{RV(f)}$ and $g_{LV(f)}$ operators induce coherent $\mu\to e$ conversion for which we use the results of Ref. \cite{Kitano:2002mt}. The $g_{RA(f)}$ and $g_{LA(f)}$ operators induce spin dependent $\mu\to e$ conversion, for which we use the results of \cite{Cirigliano:2017azj}, where the Wilson coefficients in the notation of \cite{Cirigliano:2017azj} are $C_{A,L}^{ff}=g_{LA(f)}/4$,  $C_{A,L}^{ff}=g_{LA(f)}/4$, while the axial structure factors $S_A$ we take from \cite{Engel:1995gw}.
Note that the momentum exchange in $\mu\to e$ conversion is $q^2\simeq -m_\mu^2$. This is small enough that it is easily absorbed by the nucleus without changing its nuclear structure. On the other hand, $q^2\simeq -m_\mu^2$ is also large enough that even for very light $Z'$, $m_{Z'}\ll m_\mu$, the interaction can still be viewed as point-like in the calculation of conversion rate, at least for the heavy nuclei. The neutron and proton density distributions of heavy nuclei have radial extent which is parametrically bigger than the muon Compton wavelengths, for gold roughly  $r_{\max}\sim 4/m_\mu$. This means that we can still use the calculations of overlap integrals from Ref. \cite{Kitano:2002mt} even for light $Z'$, up to corrections of about ${\mathcal O}(30\%)$. 

Spin-dependent conversion starts to be dominant at very low masses. While the spin-independent conversion rate is coherently enhanced and at low masses goes as $\sim A^2m_\mu$, where $A$ is the atomic number, the spin-dependent rate increases as $m_\mu^3/m_{Z'}^2$. For heavy nuclei, such as $^{197}$Au used by the SINDRUM II experiment, the spin-dependent contribution starts to dominate around $m_{Z'}\sim 0.5$ MeV, while for lighter nuclei, such as $^{26}$Al used by Mu2e, it begins at $m_{Z'}\sim 10$ MeV. In both cases, the spin-dependent rate is important for masses outside our range of interest, so we only consider the spin-independent rate in obtaining the bounds on the $Z'$ parameter space.

The constraint from Eq. \eqref{eq:mutoe} is denoted in Fig.~\ref{fig:strongestbounds} with a solid green line, while in Fig.~\ref{fig:plotonlyleptel} (top right) we also show the comparison with the other flavor violating processes involving muons.
The Mu2e collaboration plans to achieve the sensitivity ${\rm Br}(\mu\to e) < 8\cdot10^{-17}$ for $\mu\to e$ conversion on Al nuclei~\cite{Bernstein:2019fyh}, which gives the projected sensitivity on $g'$ denoted with a dashed green line in Figs. \ref{fig:strongestbounds} and \ref{fig:plotonlyleptel} (top right).

\paragraph{Decays to three leptons:} The tree level $Z'$ exchanges mediate the lepton flavor violating decays such at $\tau \to 3\mu, e2\mu, \ldots$, see Fig.~\ref{fig:diag:LFV}. To shorten the discussion we focus on the decays of the form $\ell_i\to3\ell_j$: $\tau \to 3\mu$, $\tau \to 3e$ and $\mu\to 3e$. Similar constraints are obtained also from the remaining modes, $\tau^- \to \mu^- e^+e^-, e^-\mu^+ e^-, \mu^-\mu^+ e^-, \mu^- e^+\mu^-$, etc. 

Depending on the value of $m_{Z'}$ 
there are three different regimes for $\Gamma(\ell_i\to3\ell_j)$.
For $m_{Z'} \gg m_{\ell_i}$, the $Z'$ can be integrated out, leading to  effective four fermion interaction. In this limit, the $\ell_i\to3\ell_j$  decay is a genuine three body decay,  with the decay width \cite{Heeck:2016xkh} 
\beq
\Gamma(\ell_i\to3\ell_j)\simeq \frac{g'{}^4 m_{\ell_i}^5}{768 \pi^3 m_{Z'}^4} \Big[4{\rm Re}
\lp c_{\ell V}^{ji}  c_{\ell A}^{ji} c_{\ell V}^{jj*} c_{\ell A}^{jj*}\rp
 +3\lp |c_{\ell V}^{ji}|^2 + |c_{\ell A}^{ji}|^2 \rp \lp |c_{\ell V}^{jj}|^2 + |c_{\ell A}^{jj}|^2\rp\Big]\,.
\eeq
where we neglected the terms proportional to $m_{\ell_j}$.

In the intermediate mass regime, $2 m_{\ell_j} < m_{Z'} < m_{\ell_i} - m_{\ell_j}$, the $Z'$ can be produced on shell. The $\ell_i\to 3\ell_j$ decay is thus a cascade decay, $\ell_i\to Z' \ell_j$ followed by $Z'\to \ell_j^+ \ell_j^-$, so that 
\beq
\Gamma(\ell_i\to3\ell_j)=\Gamma(\ell_i\to\ell_jZ') {\rm Br}(Z'\to\ell_j\ell_j)\,,
\eeq
where 
\beq \label{eq:ellitoelljZ}
\Gamma(\ell_i\to\ell_jZ')= \Big[(c_{\ell V}^{ij})^2+(c_{\ell A}^{ij})^2\Big] \frac{g'{}^2}{16\pi } \frac{m_i^3}{m_{Z'}^2}\lp 1- \frac{m_{Z'}^2}{m_i^2} \rp^2 \lp 1 + \frac{2 m_{Z'}^2}{m_i^2} \rp \,,
\eeq
where we neglected the mass of $\ell_j$, while  ${\rm Br}(Z'\to\ell_j\ell_j)$ is the branching ratio for the $Z'\to \ell_j^+ \ell_j^-$ decay. 
In our benchmark
${\rm Br}(Z'\to\mu\mu)\sim 0.16$, 
while 
${\rm Br}(Z'\to e e)\sim 0.23$ for $m_\mu<m_{Z'}<m_\tau$ and  ${\rm Br}(Z'\to e e)\sim 0.28$ for $m_{Z'}<m_\mu$.
Note that $\Gamma(\ell_i\to\ell_jZ')\propto g'{}^2/m_{Z'}^2=1/\langle\phi\rangle^2$ and thus does not diverge for $m_{Z'}\to 0$.

Finally, if $m_{Z'} < m_{\ell_j}$ the $Z'$ is off-shell and we again have a genuine three body decay. For $m_{Z'} \ll m_{\ell_j}$ the axial current contributions are $1/m_{Z'}^4$ enhanced, giving \cite{Heeck:2016xkh} 
\beq
\Gamma(\ell_i\to3\ell_j)\simeq \frac{g'{}^4m_{\ell_i}^3m_{\ell_j}^2}{128\pi^3m_{Z'}^4} |c_{\ell A}^{jj}|^2 \lp |c_{\ell V}^{ij}|^2 + |c_{\ell A}^{ij}|^2\rp {\rm log}^2\lp\frac{m_{\ell_j}}{m_{\ell_i}}\rp\,.
\eeq 
The contributions from flavor diagonal vector couplings of $Z'$ are subleading and we can safely neglect them. 

The above expressions constrain $g'$ as a function of  $m_{Z'}$ given present bounds on the $\tau \to 3\mu, 3e$ and $\mu\to 3 e$ decays. For our benchmark the most stringent is the $90\%$ C.L. bound ${\rm Br}(\mu\to 3 e) < 1.0\cdot 10^{-12}$ \cite{Bellgardt:1987du}, shown as the purple solid line in Fig.~\ref{fig:plotonlyleptel} (top right), while the the purple dashed  line corresponds to the projected reach by Mu3e of $5.2\cdot10^{-15}$  \cite{Perrevoort:2018ttp}. 
The present (projected Belle II) bounds ${\rm Br}(\tau\to 3 \mu) < 2.1(0.03)\cdot 10^{-8}$  and ${\rm Br}(\tau\to 3 e) < 2.7(0.04)\cdot 10^{-8}$ \cite{Hayasaka:2010np,Kou:2018nap} are shown as blue and purple solid (dashed) lines in Fig.~\ref{fig:plotonlyleptel} (bottom left). The different thresholds appear as the breaks in the slopes of the lines.

 \paragraph{Rare meson decays (RMD):}  The tree level $Z'$ exchanges induce rare FCNC meson decays such as $M_1\to M_2 \ell_i^+ \ell_j^-$, where $M_1(M_2)$ is the initial (final) state meson. An example of such an exotic decay is $K^+\to\pi^+\mu^+ e^-$ which arises at one loop  in the SM but with a tremendous suppression, with a matrix element proportional to the neutrino masses, and are thus vanishingly small in practice. 
	The tree level transition matrix element due to a $Z'$ exchange instead has no suppression due to neutrino masses.  For heavy $Z'$ the off-shell contribution is proportional to a product of  two off-diagonal couplings, one for the quark flavor changing current and one for the lepton current. The resulting bounds are therefore expected to be weaker than the bounds that follow from the lepton FCNC transitions or from meson mixings. In contrast, when the $Z'$ is light enough to be produced on-shell, the  $M_1 \to M_2 Z'$ decays depend only on quark current off-diagonal couplings. In this case the bounds from rare meson decays in general become the most stringent ones.

	For the $M_1 \to M_2$ transitions 
	the strongest experimental bounds follow from  the $M_1 \to M_2 Z'$ decays with $M_{1,2}$ both pseudoscalars, for which the matrix element is
	\beq
	\langle M_2 | J_\mu | M_1 \rangle = g' c_{qV}^{kl} \big[ f_+(q^2) (p_1 + p_2)_\mu + f_-(q^2) q_\mu \big]\,,
	\eeq
with $J^\mu$ defined in \eqref{eq:cfi:Z'}, and $p_{1,2}$ the momenta of mesons $M_{1,2}$, such that $p_{1,2}^2=m_{1,2}^2$, with $m_{1,2}$ the corresponding meson masses. The quark flavor-violating coupling $c_{qV}^{kl}$ encodes the strength of the $q_l\to q_k$ transition,
e.g., the $c_{dV}^{23}$ coupling is obtained for the $B\to K$  and $c_{dV}^{13}$ for the $B\to \pi$ matrix elements, respectively. The form factors $f_{+,-}(q^2)$ depend on the momentum exchange squared, $q^2$, with $q_\mu = (p_1 - p_2)_\mu$. We follow the common practice and trade the form factor  $f_-(q^2)$ for 
	\beq
	f_0(q^2) = f_+(q^2) + f_-(q^2) \frac{q^2}{m_1^2 - m_2^2}\,.
	\eeq
We use the $z$-expansion based parametrization of the form factors \cite{Bourrely:2008za},
	\beq
	f_i(q^2) = \frac{1}{1 - q^2/m_{\rm pole}^2} \sum_{n}^{N-1} a_{i,n} \left[ z^n(q^2) - (-1)^{n-N} \frac{n}{N} z^N(q^2)  \right]\,, \qquad i=+,0,
	\eeq
	where 
	\beq
	z(q^2) = \frac{\sqrt{t_+ - q^2} - \sqrt{t_+ - t_0}}{\sqrt{t_+ - q^2} + \sqrt{t_+ - t_0}}\,,
	\eeq
with	$t_+ = (m_1 + m_2)^2$\,,  $t_0 = (m_1 + m_2) (\sqrt m_1 - \sqrt m_2)^2$.

	For light $Z'$, with a mass below the lepton pair production threshold,  $m_{Z'} \leq m_{\ell_i} + m_{\ell_j}$, the $M_1\to M_2 \ell_i^- \ell_j^+$ transition is a genuine 3-body  decay, mediated by an off-shell $Z'$. The longitudinal $Z'$ contribution dominates, leading to a decay width proportional to $\propto g'{}^4/m_{Z'}^4\simeq 1/(4 \langle \phi \rangle^4)$.
	The expression for the differential width is well approximated by taking the $m_{Z'}\to0$ limit, giving 
	\beq
	\begin{split}
	\frac{{\rm d}\Gamma(M_1\to M_2\ell_i^-\ell_j^+)}{{\rm d}m_{M_1\ell_i}^2{\rm d}m_{\ell_i\ell_j}^2} &\simeq \frac{g'^4 |c_{qV}^{kl}|^2 f_0^2(m_{\ell_i\ell_j}^2)}{128\pi^3 m_{\ell_i\ell_j}^4 m_1^3 m_{Z'}^4} \Big[|c_{\ell V}^{ij}|^2 \lp m_{\ell_i} - m_{\ell_j} \rp^2 \lp m_{\ell_i\ell_j}^2 - (m_{\ell_i} + m_{\ell_j})^2 \rp \\
	&+ |c_{\ell A}^{ij}|^2 \lp m_{\ell_i} + m_{\ell_j} \rp^2 \lp m_{\ell_i\ell_j}^2 - (m_{\ell_i} - m_{\ell_j})^2 \rp \Big] \lp m_1^2 - m_2^2 \rp^2 \,,
	\end{split}
	\eeq
	where $m_{\ell_i\ell_j}^2 = (p_{\ell_i} + p_{\ell_j})^2$ is the  invariant mass squared of the leptonic pair, and similarly $m_{M_1 \ell_i}^2= (p_{1} + p_{\ell_i})^2$.

For heavy $Z'$, with a mass $m_{Z'} \gg m_1$, the $Z'$ can be integrated out. This also gives a decay width proportional to $\propto g'{}^4/m_{Z'}^4\simeq 1/(4 \langle \phi \rangle^4)$. 
In the limit of massless final states, $m_2,m_{\ell_i},m_{\ell_j}\to0$, the differential rate is given by a rather compact expression,
	\beq
	\frac{{\rm d}\Gamma(M_1\to M_2\ell_i^-\ell_j^+)}{{\rm d}m_{M_1\ell_i}^2{\rm d}m_{\ell_i\ell_j}^2} \simeq \frac{g'^4 |c_{qV}^{kl}|^2 \lp |c_{\ell V}^{ij}|^2 + |c_{\ell A}^{ij}|^2 \rp f_+^2(m_{\ell_i\ell_j}^2)}{32\pi^3  m_1^3 m_{Z'}^4} m_{M_2\ell_i}^2 \lp m_1^2 - m_{M_2\ell_i}^2 - m_{\ell_i \ell_j}^2 \rp\,.
	\eeq
In the numerical analysis we work with the full dependence on $m_2, m_{\ell_i}, m_{\ell_j}$.

	Phenomenologically, the most relevant is the intermediate mass regime, $m_{\ell_i} + m_{\ell_j} <m_Z' < m_1 - m_2$. In this case the $Z'$ is produced on-shell in the 2-body decay, $M_1 \to M_2 Z'$, followed by a decay into two leptons, $Z'\to\ell_i\bar\ell_j$. The total decay width is given by
	\beq
	\Gamma(M_1\to M_2\ell_i^+\ell_j^- = \Gamma(M_1 \to M_2 Z')\times{\rm Br}(Z'\to\ell_i^+\ell_j^-)\,.
	\eeq
The two-body decay width for $M_1 \to M_2 Z'$ transition depends only on the quark current matrix element, and is given by
	\beq
	\Gamma(M_1 \to M_2 Z') = g'^2 |c_{qV}^{kl}|^2 f_+(m_{Z'}^2)^2 \frac{m_1}{16\pi }\frac{\lp y_2^4 + (y_{Z'}^2- 1)^2 - 2 y_2^2 (1 + y_{Z'}^2) \rp^{3/2}}{y_{Z'}^2}\,,
	\eeq
	where $y_i = m_i/m_1$.
	The branching fractions for $Z'\to\ell_i\bar \ell_j$ are easily calculated. Ignoring the kinematical factors for final state particles they are given by ${\rm Br}( Z'\to\ell_i\bar \ell_j)=(|c_{\ell V}^{ij}|^2+ |c_{\ell A}^{ij}|^2)/(\sum_{ff'}|c_{\ell V}^{ff'}|^2+ |c_{\ell A}^{ff'}|^2)$. For our benchmark we have ${\rm Br}(Z'\to ee,\mu\mu)\simeq 0.1$ and ${\rm Br}(Z'\to \sum_\alpha \nu_\alpha\nu_\alpha)\simeq 0.4$ for the flavor conserving channels, and ${\rm Br}(Z'\to e\mu)\sim 10^{-5}$ for the flavor violating decay.
	
	In the numerical analysis we use the PDG values for the measurements of the branching ratios and the upper bounds~\cite{10.1093/ptep/ptaa104}, while the numerical inputs for the form factors are taken from Ref.~\cite{Bondarenko_2018}. The envelope of constraints on $g'$ as a function of $Z'$ mass due to rare meson decays (RMD) is shown as a blue solid line in Figs.~\ref{fig:strongestbounds} and \ref{fig:plotonlyleptel}. We see that the RMD constraints are the most stringent ones up to several GeV $Z'$ masses, i.e., as long as the $B\to \pi Z'$ decays are still kinematically allowed.

The kinematical thresholds from $K, D, B$ decays are clearly visible in Figs.~\ref{fig:strongestbounds} and \ref{fig:plotonlyleptel}. At the lowest $Z'$ masses 
the  $K^+\to\pi^+ Z',~Z'\to\nu\nu$ decay leads to the strongest constraints, a result of $K\to \pi+{\rm inv}$ being well constrained experimentally, as well as the large $Z'\to \nu\bar \nu$ branching ratio.  Conservatively, we use the integrated branching ratio 
	$ {\rm Br}(K^+\to\pi^+\nu\nu) < 2.3\times10^{-10}\,$
	to set the limit on $g'$ in Figs.~\ref{fig:strongestbounds} and \ref{fig:plotonlyleptel}. A more careful analysis of missing mass differential rates, taking care of spill-overs between different bins, would further improve these bounds. 
	
The bounds on $K^+\to\pi^+ \mu^+e^-$ decays, ${\rm Br}(K^+\to\pi^+\mu^+e^-) < 1.3\times10^{-11}\,, {\rm Br}(K^+\to\pi^+\mu^-e^+) < 6.6\times10^{-11}$ lead to the most stringent constraints on $g'$ in the intermediate $Z'$ mass region, despite the small ${\rm Br}(Z'\to e\mu)$. 
 The related decays,  $K_L\to\mu e$, with the decay width given by
 \beq
\Gamma(K_L\to e^-\mu^+) = |{\rm Re}(g_{sd}^A)|^2(|g_{e\mu}^V|^2 + |g_{e\mu}^A|^2) \frac{m_\mu^2 (m_K^2 - m_\mu^2)^2}{8\pi m_{Z'}^4 m_K^3}\,,
\eeq
are always less constraining. When $K\to \pi Z'$ decays are kinematically forbidden the decays of heavier mesons become relevant. In Figs.~\ref{fig:strongestbounds} and \ref{fig:plotonlyleptel}.
  we show the limits that follow from the bounds on the decays,  $
    {\rm Br}(D^+\to\pi^+\mu\mu) < 7.3\times10^{-8}\,,  {\rm Br}(B^+\to\pi^+\mu\mu) < 6.9\times10^{-8}\,,$ and  $ {\rm Br}(B^+\to K^+\mu\mu)  < 4.8\times10^{-7}\,$.

\paragraph{Flavor violating radiative decays:} At one loop the off diagonal couplings of $Z'$ to leptons generate the 
$\ell_i\to\ell_j\gamma$ transition with the decay width~\cite{Lavoura:2003xp}, see also Fig.~\ref{fig:diag:LFV},
\beq
\Gamma(\ell_i\to\ell_j\gamma) = \frac{\alpha g'{}^4 }{4\pi }\lp1- x_{j/i}^2\rp^3 x_{i/Z'}^4 \lp| c_L^\gamma|^2 + |c_R^\gamma|^2\rp m_i,
\eeq
where  $x_{a/b}=m_a/m_b$, and
\beq
c_L^\gamma = Q_k  \lp c_{\ell_R}^{jk*} c_{\ell_R}^{ik} y_{RR} + c_{\ell_L}^{jk *} c_{\ell_L}^{ik} y_{LL} + c_{\ell_R}^{jk*} c_{\ell_L}^{ik} y_{RL} + c_{\ell_L}^{jk*} c_{\ell_R}^{ik} y_{LR}  \rp\,, 
\eeq
with $Q_k$ the charge of the fermion $\ell_k$ running in the loop, while $c_R^\gamma$ can be obtained by making the replacement $L\leftrightarrow R$ in the above expression. The loop functions are
\begin{align}
y_{RR} &=y_{LL}/x_{j/i}=  2 f_1 + 6 f_2 + 3 (1+ x_{k/Z'}^2) f_3, \\
y_{RL} /x_{k/i} &= -4 f_1 - 2(3-x_{k/Z'}^2) f_2 + 3 f_3  - x_{i/Z'}^2 \lp f_2 + \tfrac{3}{2}f_3 \rp - x_{j/Z'}^2\lp f_2 + \tfrac{3 }{2}f_3 \rp\,, \\
y_{LR} &= -3x_{j/Z'}x_{k/Z'} f_3\,,
\end{align}
where (below we shorten $x_{k/Z'}\to x$)
\beq
\begin{split}
f_1 &= \frac{1}{16\pi^2} \left[1/(1-x^2) +2 \log(x)/(1-x^2)^2 \right]\,,\\
f_2 &= - \frac{1}{4} \frac{1}{16\pi^2} \left[(3-x^2)/(1-x^2)^2 + 4\log(x)/(1-x^2)^3 \right]\,,\\
f_3 &= \frac{1}{18}\frac{1}{16\pi^2} \left[(2x^4 - 7x^2 +11)/(1-x^2)^3 + 12 \log(x)/(1-x^2)^4 \right].
\end{split}
\eeq
The present bounds on 
${\rm Br}(\tau\to\mu(e)\gamma) <4.4(3.3) \cdot10^{-8}$  \cite{Aubert:2009ag} and ${\rm Br}(\mu\to e\gamma) <4.2\cdot10^{-13}$ \cite{TheMEG:2016wtm}
are denoted in Fig.~\ref{fig:plotonlyleptel} (bottom left) by a brown(black) solid line and in Fig.~\ref{fig:plotonlyleptel} (top right) by a blue solid line. The projected sensitivities at Belle II \cite{Kou:2018nap} and MEG-II \cite{Baldini:2018nnn} 
${\rm Br}(\tau\to\mu(e)\gamma) < 1(3)\cdot 10^{-9}\,, {\rm Br}(\mu\to e\gamma) <6\cdot10^{-14}$,
are shown with the corresponding dashed lines. Note that in our benchmark for all three transitions the largest contribution comes from a diagram with a $\tau$ running in the loop. Note also, that the contributions from vector-like fermions, which we do not take into account, are parametrically of the same order. The resulting bounds in Fig.~\ref{fig:plotonlyleptel} should thus be taken only as indicative.

\subsection{Bounds on the $Z'_i$ in the $G_{\rm FN}=U(1)_{\rm FN}^3$ model}
\label{sec:Zpr:U(1)3}

\begin{figure}[t] 
	\begin{center}
		\includegraphics[width=0.49\linewidth]{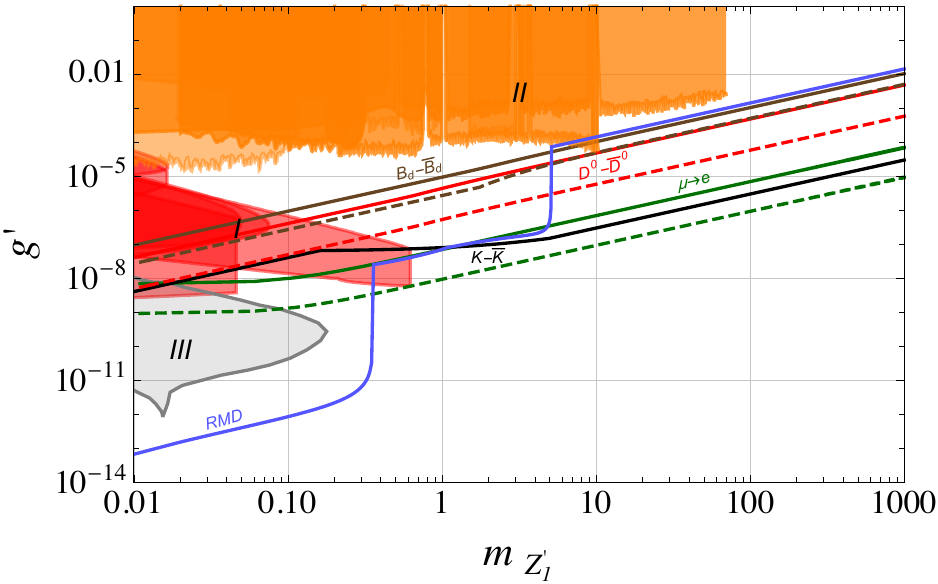}
		\includegraphics[width=0.49\linewidth]{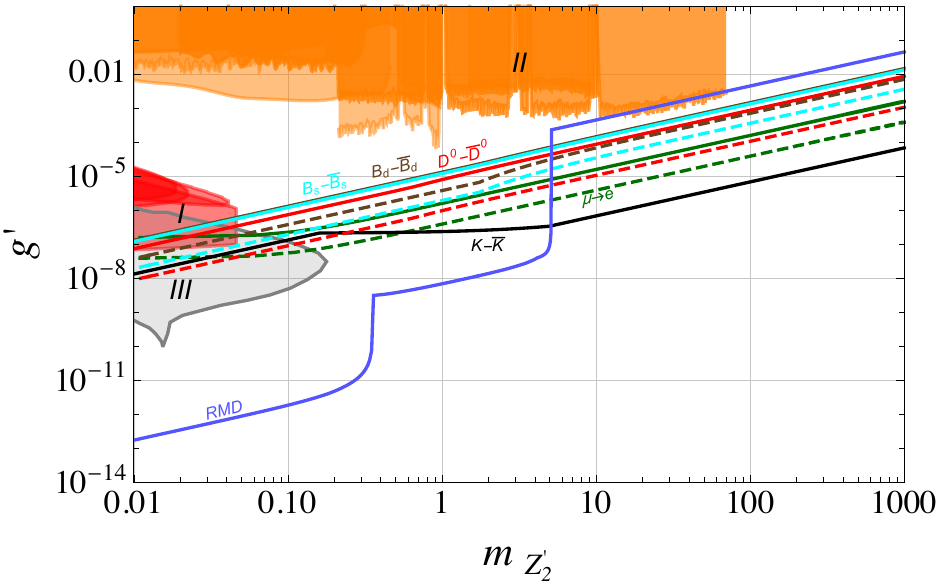}\\[2mm]
		\includegraphics[width=0.49\linewidth]{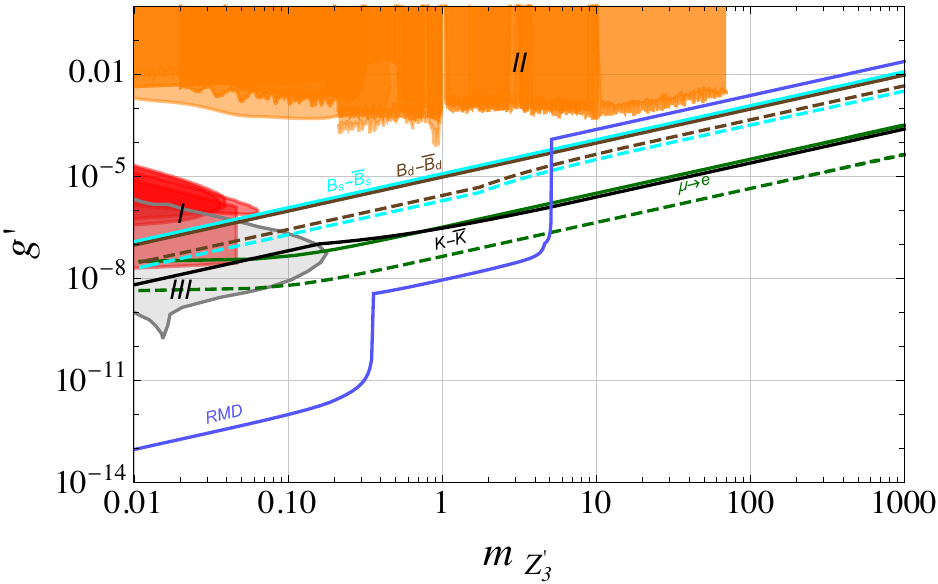}
		\caption{The strongest bounds on the parameter space for the $Z'_1$ (top left), $Z'_2$ (top right), and $Z'_3$ (bottom) gauge bosons in the $U(1)_{\rm FN}^3$ model. The color coding is the same as in Fig. \ref{fig:strongestbounds}.}. \label{fig:strongestboundsZ1}
	\end{center}
\end{figure}

Next, we extend the above analysis to the $G_{\rm FN}=U(1)_{\rm FN}^3$ benchmark, with the couplings of the three gauge bosons, $Z_i'$, $i=1,2,3$, listed in Appendix \ref{sec:app:decoupledFN}. The $Z_3'$ only couples to $d_R$ and $\ell_L, \nu_L$, see \eqref{eq:app:decoupl:ex}-\eqref{eq:app:decoupl:ex:cnuL3}. This follows immediately from the charge assignments in Eqs.~\eqref{eq:chain_lengths_decoupled} and \eqref{eq:NL:decoupled}, because $q_L^{(i)}$ and $u_R^{(i)}$ are not charged under $U(1)_3$. The large mixings in $d_R$ and lepton sectors result in an almost anarchic form of  $Z_3'$ couplings to $d_R$ and $\ell_L, \nu_L$, with large couplings to the 2nd and 3rd generation and only slightly suppressed couplings to the 1st generation SM fermions. The $Z_1'$ and $Z_2'$, on the other hand, predominantly couple to the first and second generation, respectively, with the exception of $Z_2'$ couplings to $d_R$ and $\ell_L$, $\nu_L$, where there are also large couplings to the third generation, again due to the large mixings, see Eqs.~\eqref{eq:cuL1}-\eqref{eq:cnuL2}. 

For all three $Z'_i$ the most stringent constraints arise from rare meson decays, $\mu \to e$ conversion, $K-\bar K$ mixing and SN constraints. This is the same as for the $U(1)_{\rm FN}$, Section \ref{sec:bounds:flavored:Z'}. However, the relative importance of these constrains has shifted because of the changes in the flavor patterns of the $Z'_i$ couplings, see Fig. \ref{fig:strongestboundsZ1}. 

In Fig. \ref{fig:strongestboundsZ1} we also show the constraints from beam-dump (red regions) and from $e^+e^-$ experiments (orange regions). Here some care is needed, since for $Z'_2$ and $Z'_3$ the 12 off-diagonal couplings can dominate over the diagonal 11 couplings, see, e.g., Eq.~\eqref{eq:app:decoupl:ex}. This means that we need to adjust our treatment of bounds from direct production of $Z'_i$ compared to what was done in Section \ref{sec:bounds:flavored:Z'}. For $m_e + m_\mu < m_{Z'} <2m_\mu$ the $Z'_i\to e\mu$ decay is now the dominant decay channel. In the $G_{\rm FN}=U(1)_{\rm FN}^3$ benchmark we have $\Gamma(Z'_2\to ee)/\Gamma(Z'_2\to e\mu)\simeq 0.16$ and $\Gamma(Z'_3\to ee)/\Gamma(Z'_3\to e\mu)\simeq 0.52$, which is then used to rescale appropriately the bounds from {\tt Darkcast}, cf. Section \ref{subsec:directZ'}. 
The beam dump constraints are stronger for  $Z_1'$ than they are for $Z_2'$ and $Z_3'$, because $Z_{2,3}'$ have suppressed couplings to the first generation quarks. This is also the reason that the regions excluded by SN constraints (gray regions in Fig.~\ref{fig:strongestboundsZ1}), are shifted to larger values of gauge coupling $g'$ for $Z_{2}'$ and $Z_3'$.

Fig.~\ref{fig:strongestboundsZ1} also gives the FCNC constraints from rare meson decays, $D-\bar D$, $B_d-\bar B_d$ and $B_s-\bar B_s$ mixing (blue, red, brown, cyan lines). 
The expressions for these can be taken directly  from Section \ref{sec:bounds:flavored:Z'}, with trivial relabeling corresponding to $Z'\to Z_i'$ replacements. For $Z_1'$ the most stringent constraints come from the FCNCs involving the first two generations, rare kaon decays, $\mu \to e$ conversion and $K-\bar K$ mixing, since the $Z_1'$ has appreciable 12 off-diagonal couplings. The $D-\bar D$ mixing, even though less stringent, can still be an important constraint, especially for future projections, while the FCNCs involving third generations lead to weaker bounds. For illustration we show the most stringent one, from $B_d-\bar B_d$ mixing. The $Z_2'$ and $Z_3'$ have much smaller 12 off-diagonal couplings. Even so, rare meson decays, $\mu \to e$ conversion and $K-\bar K$ mixing still lead to the most stringent constraints in our benchmark, but the FCNCs involving the third generation,  $B_d-\bar B_d$ and $B_s-\bar B_s$ mixing, are relatively more important.

In discussing the above constraints it is important to keep in mind that we show bounds for a single benchmark. We expect other $G_{\rm FN}=U(1)_{\rm FN}^3$ benchmarks to lead to a similar pattern of constraints. However, the relative strengths of different constraints may easily differ by relative ${\mathcal O}(1)$ factors (for instance for some benchmarks we would expect $\mu\to e$ conversion to be more stringent than $K-\bar K$ mixing even for heavy $Z_1'$, unlike what was found here). Fig.~\ref{fig:strongestboundsZ1} should thus be taken only as an illustration of how important different probes are for $G_{\rm FN}=U(1)_{\rm FN}^3$ inverted FN models.

\section{Conclusions}
\label{sec:conclusions}
We introduced a class of anomaly-free Froggatt-Nielsen (FN) models that can simultaneously explain the fermion mass hierarchy as well as the mixing patterns. These inverted  FN models differ from the traditional FN models in that the expansion parameter has the inverted form, $M/\langle \phi \rangle$, where $M$ is the vector-like mass parameter and $\langle \phi \rangle$ the vev of the flavon field. The observed pattern of masses and mixings is obtained for $\langle \phi \rangle\gg M$, while in the traditional FN models the opposite is required, $M\gg \langle \phi \rangle$. Different realizations of the inverted FN models differ in the choice of the FN horizontal group $G_{\rm FN}$ and the assignment of FN charges. The common feature, on the other hand, is that the fields that are chiral under the SM do not carry $G_{\rm FN}$ charges. These are the fields that couple directly to the Higgs, and also to the chains of vector-like fermions that are charged under $G_{\rm FN}$. 
This set-up gives zero modes that are localized toward the ends of chains, with exponentially suppressed overlaps with the zero node where the Higgs resides.  After electroweak symmetry breaking this results in the hierarchy of the SM fermion masses. The set-up also ensures that $G_{\rm FN}$ is anomaly free irrespective of the choice of the FN charges, making it easy to extend our work to other charge assignments or gauge groups. 

In the paper we explored in detail two choices for $G_{\rm FN}$: the "decoupled chains" model, where $G_{\rm FN} = U(1)_{\rm FN}^3$, and the "coupled chains" model, with $G_{\rm FN} = U(1)_{\rm FN}$. For both cases we showed using a numerical scan of input parameters that one can obtain the observed hierarchy of SM quark masses and CKM matrix elements given the appropriate choice of FN charges. The scan was over random ${\cal O}(1)$ complex flavon and Higgs Yukawa couplings and over random vector-like mass parameters, $M_a$, taken to be on average a factor $5$ smaller than $\langle \phi \rangle$. The obtained solutions for the pattern of quark masses and mixings do not contain any tunings -- the associated Barbieri-Giudice measure only reaches values of a few. Furthermore, the solutions simultaneously give the hierarchical quark masses and hierarchical mixing angles. Demanding that the quark masses are the  measured ones, the probability for the scan to give all three mixing angles below their SM values is about $8\%$ in both FN models (to be compared with the best case scenario of $0.5^3=12.5\%$, and much higher than $3 \cdot 10^{-6}$ for mixing angles that are distributed completely randomly). Similarly, we showed that the neutrino mass differences and PMNS matrix elements are well described by a completely anarchic form of a Weinberg dimension 5 operator. Since the Weinberg operator couples to chains of vector-like fermions the neutrino masses are still hierarchical, preferentially resulting in a normal ordering. 

The inverted FN models do share a  problem common to all FN models. Since one introduces a relatively large number of states charged under QCD, the theory is no longer asymptotically free in the UV, see, e.g.,~\cite{Alonso:2018bcg}. In the two models we presented there is a QCD Landau pole about four orders of magnitude above $\langle \phi\rangle$, i.e., for $\langle \phi\rangle\sim 10^7$ GeV this would be around $10^{11}$ GeV. One could potentially push this to even higher scales by adjusting the lengths of FN chains along with the size of $\langle \phi\rangle/M$ expansion parameter. Alternatively, for $\langle \phi\rangle\sim 10^{14}-10^{15}$ GeV, depending on the exact field content, the Landau pole is above the Planck scale. Another interesting possibility is that with an appropriate field content one may realize an asymptotically safe theory \cite{Sannino:2019sch}. The possible Landau pole in the $U(1)_{\rm FN}$ is less constraining;  for $g'<0.11$ at $\mu\simeq 10^7$ GeV the Landau pole in $g'$ is above the Planck scale.

In the second part of the paper we explored the phenomenological implications of the two inverted FN models. The structures we introduced leave imprints in FCNC transitions, which then leads to stringent bounds on the allowed values of $\langle \phi\rangle$ and $M_a$. We focused on the case where $G_{\rm FN}$ is gauged, resulting in three (one) gauge bosons $Z_i'$ ($Z'$) for $G_{\rm FN}=U(1)_{\rm FN}^3 (U(1)_{\rm FN})$. Since the FN charges have opposite signs for the weak $SU(2)_L$ doublets and the weak $SU(2)_L$ singlets, the $Z_i'$ have predominantly axial vector couplings to the SM fermions. The non-universal $G_{\rm FN}$ charges induce both flavor conserving and flavor violating couplings of $Z_i'$ to the SM fermions.   The most important FCNC contributions then come from tree level $Z_i'$ exchanges $K-\bar K$ mixing and $\mu\to e $ conversion bounds limiting $\langle \phi\rangle\gtrsim 10^7$ GeV, and consequently also the masses of vector-like fermions to roughly the same scale. The $Z_i'$, on the other hand, can be light, if $g'$ is small. For light $Z'$, with masses from several 10s MeV to several 100s MeV,  the rare meson decays and  beam dump searches are the most constraining, see Figs. \ref{fig:strongestbounds}, \ref{fig:plotonlyleptel} and \ref{fig:strongestboundsZ1} for the bounds on two representative benchmarks.

There are several ways in which our study could be extended in the future. Most immediately, one could explore the bounds on inverted FN models, which are not gauged so that there are no $Z'$ tree level contributions to the FCNCs. It would also be interesting to explore whether or not the inverted FN models can aid in exploring the experimental anomalies in $b\to s\ell^+\ell^-$ transitions, potentially along the lines of Ref. \cite{Grinstein:2018fgb}.

\acknowledgments We thank   Ben Allanach, Stefan Antusch, Fady Bishara,  Giancarlo d'Ambrosio, Gino Isidori, Jernej Kamenik, Nejc Ko\v snik, and Uli Nierste for insightful discussions, and Joachim Brod, Alex Kagan and Sourov Roy for useful comments on the manuscript.  MT and JZ acknowledge support in part by the DOE grant DE-SC0011784. JZ and AS thank American Slovenian Education Foundation for financial support. 
AS is supported by the Young Researchers Programme of the Slovenian Research Agency under the grant No. 50510. JZ thanks the theory group of the Jozef Stefan Institute for the hospitality and acknowledges the financial support from the Slovenian Research Agency (research core funding No. P1-0035).  JZ acknowledges financial support from LAPTh, where part of the research was performed, and especially for their hospitality.

\appendix
\section{Light $Z'$ contributions to $K-\bar K$ mixing}
\label{sec:app:KKbar}
In this appendix we give further details on the contributions from light $Z'$ exchanges to the $K-\bar K$ mixing. For $m_{Z'}\ll \Lambda_{\rm ChPT}\simeq 4\pi f_\pi\sim {\mathcal O}(1{\rm~GeV})$ we can use Chiral Perturbation Theory (ChPT) to describe these interactions.  We estimate the leading contributions from both axial and vector couplings of $Z'$. The former start at tree level, the latter at one loop. 

In order to construct the appropriate ChPT Lagrangian in the presence of flavor violating $Z'$ we use the spurion analysis \cite{Gasser:1984gg}. The QCD Lagrangian is now ${\cal L}_{{\rm QCD}+Z'}= \bar q (i \slashed \partial + g_s \slashed G^a T^a+ \chi_V \slashed Z' +\chi_A \slashed Z'\gamma_5) q-\bar q {\cal M}_q q$, where $q=(u,d,s)$, ${\cal M}_q=\diag(m_u,m_d,m_s)$,  and $\chi_{V,A}$ are $3\times3$ Hermitian matrices of the form
\beq
\chi_{V,A}=g' \begin{pmatrix}
0 & 0 & 0
\\
0 & c_{d_{V,A}}^{11} & c_{d_{V,A}}^{12}
\\
0 & c_{d_{V,A}}^{21} & c_{d_{V,A}}^{22}
\end{pmatrix},
\eeq
with $c_{d_{V,A}}^{ij}$ defined in \eqref{eq:VA:couplings}. The QCD$+Z'$ Lagrangian is formally invariant under $SU(3)_R\times SU(3)_L$ transformations, $q_{R,L}\to g_{R,L}(x) q_{R,L}$, if $\chi_{V,A} Z_\mu$, and ${\cal M}_q$ are promoted to spurions that transform as 
$v_\mu +a_\mu \to g_R(v_\mu +a_\mu) g_R^\dagger+i g_R\partial_\mu g_R^\dagger$, 
$v_\mu -a_\mu \to g_L(v_\mu -a_\mu) g_L^\dagger+i g_L\partial_\mu g_L^\dagger$, 
$s+ip \to g_R(s+ip) g_L^\dagger$,
and that take the values  $v^\mu=\chi_{V} Z_\mu'$, $a^\mu=\chi_{A} Z_\mu'$, $s={\cal M}_q$, $p=0$. 

The LO ChPT Lagrangian, including $Z'$ as the light degree of freedom, is therefore (we work in the unitary gauge for the $Z'$ gauge boson)
\beq\label{eq:coupl:LO}
\begin{split}
{\cal L}_{{\rm ChPT}+Z'}^{(2)} =& \frac{f^2}{4} {\rm Tr}\lp\nabla_\mu U \nabla^\mu U^\dagger \rp+2 B_0 \Tr\big[ (s-ip) U+ (s+ip) U^\dagger\big]
\\
&-\frac{1}{4}Z_{\mu\nu}'Z'{}^{\mu\nu}+\frac{m_{Z'}^2}{2} Z_\mu'Z'{}^\mu
\end{split}
\eeq
where $U=\exp(i \lambda^a \pi^a/f)$, with $\lambda^a$ the Gell-Mann matrices, $f$ is related to the meson decay constant,\footnote{We use the normalization $\langle 0| \bar q_1 \gamma_\mu q_2 |P(p)\rangle=i p_\mu f_P$.} $f\simeq f_\pi/\sqrt{2}=93$ MeV, the covariant derivative is
$\nabla_\mu U= \partial_\mu U - i (\chi_V+\chi_A) Z_\mu'  U + i U (\chi_V-\chi_A) Z_\mu'$, while $B_0$ is the low energy constant. Under chiral rotations the $U$ field transforms as $U\to g_R U g_L^\dagger$,  $\nabla_\mu U\to g_R \nabla_\mu U g_L^\dagger$. 

Since we consider radiative corrections due to the $Z'$ gauge bosons, the $Z'$ cannot be treated as merely an external field that enters into the spurions. In constructing the ChPT Lagrangian this means that there are two additional hermitian spurions, $Q_{R,L}=\chi_V\pm\chi_A$, that transform as $Q_{R,L}\to g_{R,L}(x) Q_{R,L}\,g_{R,L}^\dagger (x)$. This gives the following contributions to the LO ChPT Lagrangian,
\beq\label{eq:coupl:LO:QLR}
\begin{split}
{\cal L}_{Q_{R,L}}^{(2)} &= C_0^{(2)} \Tr\big[Q_RU Q_L U^\dagger\big]
\\
&+C_1^{(2)} \Tr\big[Q_RU Q_L U^\dagger\big] \Tr\big[(s+ip)U^\dagger\big]+{\rm h.c.}
\\
&+C_2^{(2)} \Tr\big[Q_RU Q_L U^\dagger(s+ip)U^\dagger\big]+{\rm h.c.}+\cdots.
\end{split}
\eeq
where we only show the terms that will be relevant for $K-\bar K$ mixing, where they will serve as counter-terms to one-loop corrections.  For this we also need part of the  ${\mathcal O}(p^4)$ Lagrangian, 
\beq\label{eq:coupl:NLO}
\begin{split}
{\cal L}_{Q_{R,L}}^{(4)} &\supset  C_0^{(4)} \Tr\big[Q_R(s+ip)U^\dagger \big] \Tr\big[Q_R(s+ip)U^\dagger \big]  +{\rm h.c.}
\\
&+C_1^{(4)} \Tr\big[Q_R(s+ip)U^\dagger \big] \Tr\big[Q_L U^\dagger (s+ip)\big]  +{\rm h.c.}
\\
&+C_2^{(4)} \Tr\big[Q_R(s+ip)U^\dagger Q_R  (s+ip)U^\dagger\big]  +{\rm h.c.}
\\
&+C_3^{(4)} \Tr\big[Q_L U^\dagger (s+ip) \big] \Tr\big[Q_L U^\dagger (s+ip)\big]  +{\rm h.c.}+\cdots.
\end{split}
\eeq

The ChPT Lagrangian is invariant under parity, $U\to U^\dagger$, $v^\mu\to -v^\mu$, $a^\mu\to a^\mu$, $s+ip\to s-ip$, $Q_L\leftrightarrow Q_R$. In addition, since the terms containing $Q_{R,L}$ can only arise from loops of $Z'$, the ChPT terms without external $Z'$ fields need to have even powers of $g'$. This is insured by requiring that the ChPT Lagrangian is invariant under the $Z_2$ transformation $Q_{L,R}\to - Q_{L,R}$. The chiral counting of the spurions is $Q_{R,L}\sim {\mathcal O}(p^0)$, $v_\mu, a_\mu,  \sim {\mathcal O}(p)$, $s+ip \sim {\mathcal O}(p^2)$. In principle one could thus have an arbitrary number of $Q_{R,L}$ insertions in ${\cal L}_{Q_{R,L}}^{(2),(4)}$. However, we work only to ${\mathcal O}(g'{}^2)$ and thus merely keep the terms that are $\sim Q_{R,L}^2$. 
Note that the first term in \eqref{eq:coupl:LO:QLR}, 
the analogue of the ChPT+QED Lagrangian \cite{Ecker:1988te,Urech:1994hd}, is in our counting ${\mathcal O}(p^0)$. However, its coefficient is $C_0^{(2)}\sim {\mathcal O}(m_{Z'}^2)$ and we thus include it as part of ${\cal L}_{Q_{R,L}}^{(2)}$.

In the usual ChPT the one-loop counterterms reside in the ${\mathcal O}(p^4)$ Lagrangian, while in the light $Z'$ case we need both the ${\cal L}_{Q_{R,L}}^{(2)}$ and ${\cal L}_{Q_{R,L}}^{(4)}$ Lagrangians.  This is easy to see from the  ChPT amplitude scaling, $M\sim p^\nu (p/m_{Z'})^{2 I_{Z'}}$, where $\nu=2+2L+\sum_i V_i (d_i-2)$, with $L$ the number of loops, $d_i$ the number of derivatives in $V_i$ vertices of type $i$, and $I_{Z'}$ the number of $Z'$ internal lines.  Each of the $p$ momenta in this scaling can be either of order ${\mathcal O}(m_{\pi, K,\eta})$ or ${\mathcal O}(m_{Z'})$, where the latter we can in general take to be parametrically smaller. The LO vertices with $Z'$ or $Z'^{2}$ have $d_i=1$ and $d_i=0$. The one loop $Z'$ contribution, $L=1, I_{Z'}=1$, thus scales as $M\sim p^4/m_{Z'}^2$, with $p\sim {\mathcal O}(m_{Z'})$ or $p\sim {\mathcal O}(m_K)$, showing that we need both ${\cal L}_{Q_{R,L}}^{(2)}$ and ${\cal L}_{Q_{R,L}}^{(4)}$ Lagrangians in order to capture all the counterterms.

\begin{figure}[t] 
	\begin{center}
	\begin{minipage}{0.35\linewidth}
	~~~~~~~~~~~~\includegraphics[width=0.79\linewidth]{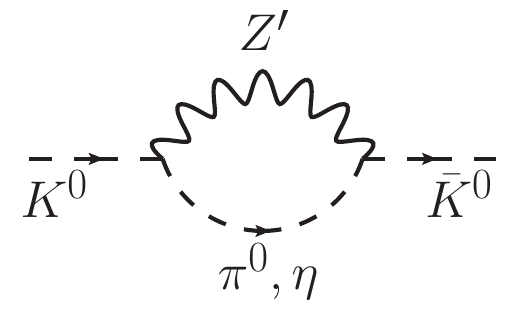}
	\end{minipage}~~~~
	\begin{minipage}{0.60\textwidth}
	\vspace{-0.8cm}
	\includegraphics[width=0.36\linewidth]{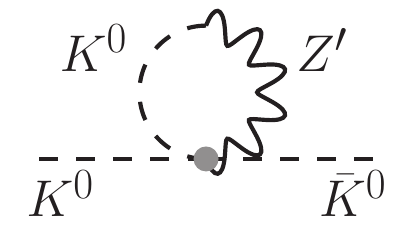}~~~~~
        \includegraphics[width=0.34\linewidth]{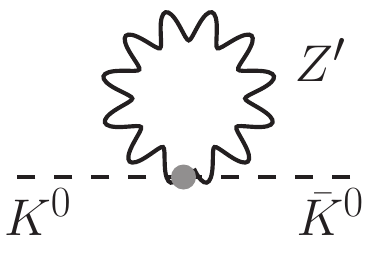}
        \end{minipage}
        	\vspace{-0.15cm}
		\caption{The one loop diagrams contributing to $K-\bar K$ mixing.}. \label{fig:Z':diagram:KKbar}
	\end{center}
\end{figure}

We are now ready to show the one-loop results for $K-\bar K$ mixing. Expanding \eqref{eq:coupl:LO} and \eqref{eq:coupl:LO:QLR} in meson fields 
 gives 
\beq\label{eq:coupl:LO:expand}
\begin{split}
{\cal L}_{{\rm ChPT}+Z'}^{(2)} + {\cal L}_{Q_{L,R}'}^{(2)}\supset &-f_K g' c_{d_A}^{12} Z'_\mu  \partial^\mu \bar K^0 +ig' c_{d_V}^{12}\Big(\sqrt{\tfrac{3}{2}}\eta \lrpartial_\mu \bar K^0-\tfrac{1}{\sqrt{2}} \pi^0 \lrpartial_\mu \bar K^0\Big)Z'{}^\mu 
\\
& +g'{}^2 \Big(Z'_\mu Z'^{\mu}-\frac{4}{f_K^2} C_0^{(2)}\Big) \big[ (c_{d_A}^{12})^2- (c_{d_V}^{12})^2\big] \big(\bar K^0\big)^2 
\\
& -g'{}^2 \frac{4}{f_K^2}  \Big(C_1^{(2)}m_{uds} + C_2^{(2)}m_{ds}\Big) \big[ (c_{d_A}^{12})^2- (c_{d_V}^{12})^2\big] \big(\bar K^0\big)^2 
\\
&+\frac{4}{3 f_K} g' c_{d_A}^{12}  Z'{}^\mu \big(K^0 \lrpartial_\mu \bar K^0{}\big) \bar K^0 +{\rm h.c.}
+\cdots,
\end{split}
\eeq
where we abbreviated $m_{uds}=m_u+m_d+m_s$, and $m_{ds}=m_d+m_s$. Expanding the NLO Lagrangian \eqref{eq:coupl:LO} in meson fields gives
\beq\label{eq:coupl:NLO:expand}
\begin{split}
{\cal L}_{Q_{L,R}'}^{(4)}\supset & -g'{}^2 \frac{4}{f_K^2} m_d^2 \Big(C_0^{(4)}+ C_2^{(4)}\Big) \Big( c_{d_A}^{12}+ c_{d_V}^{12}\Big)^2 \big(\bar K^0\big)^2 
\\
& +g'{}^2 \frac{4}{f_K^2}  m_d m_s C_1^{(4)} \big[ (c_{d_A}^{12})^2- (c_{d_V}^{12})^2\big] \big(\bar K^0\big)^2 
\\
& -g'{}^2 \frac{4}{f_K^2} m_s^2 C_3^{(4)} \Big( c_{d_A}^{12}- c_{d_V}^{12}\Big)^2 \big(\bar K^0\big)^2  +{\rm h.c.}
+\cdots.
\end{split}
\eeq
Above, we kept only the terms relevant for $K^0-\bar K^0$ mixing, defined $\phi_1 \lrpartial_\mu \phi_2= \phi_1 \partial_\mu \phi_2 -  (\partial_\mu \phi_1)  \phi_2$, and replaced $\sqrt2 f$ with the kaon decay constant, $f_K=155.6\pm0.4$ MeV, to account for the SU(3) breaking. The first term in \eqref{eq:coupl:LO:expand} is due to the axial vector coupling of the $Z'$ boson and leads to the tree level contribution to the $K^0-\bar K^0$ mixing amplitude, Eq. \eqref{eq:M12:KKbar:light}.  The second term is due to the vector coupling and contributes at one loop, see left diagram in Fig. \ref{fig:Z':diagram:KKbar}. The first and the last terms in \eqref{eq:coupl:LO:expand} contribute to the middle diagram in Fig. \ref{fig:Z':diagram:KKbar}, while the $Z'_\mu Z'{}^\mu$ term in the second line of \eqref{eq:coupl:LO:expand} gives the last diagram in Fig. \ref{fig:Z':diagram:KKbar}.

The one-loop contributions to $K-\bar K$ mixing are
\beq
\begin{split}
M_{12}^{1-{\rm loop}} =&\frac{g'{}^2 (c_{d_V}^{12})^2 }{32\pi^2 m_K m_{Z'}^2}\sum_{P=\pi^0,\eta} c_P^2 \Big\{\Big[
m_{Z'}^2A_0(m_P^2)-\big(m_K^2-m_P^2+m_{Z'}^2\big)A_0(m_{Z'}^2)
\\
&+\big(m_K^4-2m_K^2(m_P^2+m_{Z'}^2\big)+\big(m_P^2-m_{Z'}^2\big)^2\Big]B_0\big(m_K^2,m_P^2,m_{Z'}^2\big)\Big\}
\\
&+\frac{g'{}^2 (c_{d_A}^{12})^2 }{12\pi^2 m_K}\frac{m_K^2}{m_{Z'}^2}A_0(m_K^2)+\frac{g'{}^2}{16\pi^2 m_K} \Big[(c_{d_A}^{12})^2 -(c_{d_V}^{12})^2 \Big]\Big[m_{Z'}^2-\tfrac{3}{2} A_0(m_{Z'}^2)\Big],
\end{split}
\eeq
with $c_{\pi^0}=-1/\sqrt2$, $c_{\eta}=\sqrt{3/2}$, and $A_0$ and $B_0$ the Passarino-Veltman loop functions, see, e.g., \cite{Patel:2015tea}. 
The counter-terms give
\beq
\begin{split}
M_{12}^{\rm cntr.} =&\frac{2 g'{}^2  }{f_K^2 m_K}\Big\{ \Big(C_0^{(2)}+C_1^{(2)}m_{uds} + C_2^{(2)}m_{ds}-m_d m_s C_1^{(4)}\Big) \big[ (c_{d_A}^{12})^2- (c_{d_V}^{12})^2\big]
\\
&\quad + m_d^2 \Big(C_0^{(4)}+ C_2^{(4)}\Big) \Big( c_{d_A}^{12}+ c_{d_V}^{12}\Big)^2 
 +m_s^2 C_3^{(4)} \Big( c_{d_A}^{12}- c_{d_V}^{12}\Big)^2  +{\rm h.c.}
+\cdots\Big\}.
\end{split}
\eeq
Since there are no one-loop contributions of the form $\sim c_{d_A}^{12}+ c_{d_V}^{12}$, this implies a relation between the counter-tems in the last line, $m_d^2 \big(C_0^{(4)}+ C_2^{(4)}\big)=m_s^2 C_3^{(4)}$.

\section{Benchmarks}
\label{app:benchmarks}

In this appendix we give the details of the two benchmarks, one for the case of the decoupled chains with $U(1)_{\rm FN}^3$ horizontal symmetry, and one for the coupled chains with the $U(1)_{\rm FN}$ horizontal symmetry. 
\subsection{The benchmark for the decoupled FN chains}
\label{sec:app:decoupledFN}
The vector-like masses for our $U(1)_{\rm FN}^3$ benchmark are (in units of $10^7$ GeV)
\begin{align}
M^{q_1}&= \{-0.681-0.462 i,0.472\, -0.375 i,0.152\, -0.356 i\}, \\
M^{q_2}&= \{0.453\, -0.124 i,-0.284-0.218 i\},\\
M^{u_1}&= \{0.29,0.112\, -0.566 i,0.595,-0.431-0.29 i\}, \\
M^{u_2}&= \{0.439,-0.308+0.55 i\}, \\
M^{d_1}&= \{0.74,0.545\, -0.535 i,-0.211+0.288 i,0.197\, -0.744 i\}, \\
M^{d_2}&= \{0.656,-0.669-0.464 i,0.653\}, \\
M^{d_3}&= \{0.289\, -0.005 i,0.563\, -0.467 i,0.237\, +0.316 i\},\\
M^{L_1}&= \{-0.409-0.711 i,-0.466-0.244 i,0.274\, +0.416 i\}, \\
M^{L_2}&= \{-0.14+0.34 i,-0.725-0.204 i,-0.269+0.673 i\},\\
M^{L_3}&= \{0.494,0.434\, -0.278 i,0.073\, -0.505 i\},\\
M^{e_1}&= \{0.691\, +0.414 i,0.369,0.605,-0.056+0.357 i\},\\
M^{e_2}&= \{-0.324-0.233 i\},
\end{align}
where the entries are for different nodes, for instance for the first generation FN fermions that are quark doublets, $M^{q_1}=\{M^{q_1}_1, M^{q_1}_2,M^{q_1}_3\}$, etc. For Yukawa couplings between FN fermions and the flavon, we have (in units of $10^7$ GeV) 
\begin{align}
Y^{q_1}\langle \phi \rangle &= \{0.855\, -3.79 i,1.483\, -3.591 i,-1.058+3.737 i\}, \\
Y^{q_2}\langle \phi \rangle &= \{1.758\, -0.415 i,-2.965-1.881 i\},\\
Y^{u_1}\langle \phi \rangle &= \{4.021,-1.581+0.583 i,0.493\, -1.653 i,-3.996+0.27 i\},\\
Y^{u_2}\langle \phi \rangle &= \{1.855\, -0.007 i,-0.506+1.358 i\},\\
Y^{d_1}\langle \phi \rangle &= \{-2.099+2.57 i,-1.916-1.087 i,-3.804-1.739 i,-0.764-1.924 i\},\\
Y^{d_2}\langle \phi \rangle &= \{-1.72-3.53 i,-0.126-3.927 i,0.251\, -1.823 i\},\\
Y^{d_3}\langle \phi \rangle &= \{-1.21+2.662 i,-0.289+1.538 i,-0.851+1.307 i\},\\
Y^{L_1}\langle \phi \rangle &= \{-3.953+1.27 i,3.232\, -1.642 i,0.588\, -3.116 i\}, \\
Y^{L_2}\langle \phi \rangle &= \{-1.89-1.9 i,-2.044-1.056 i,-0.079-2.649 i\},\\
Y^{L_3}\langle \phi \rangle &= \{-4.053+0.833 i,-3.158+2.503 i,2.015\, +2.429 i\},\\
Y^{e_1}\langle \phi \rangle &= \{1.936,-2.288+1.128 i,-2.392+1.877 i,-3.371-0.651 i\},\\
Y^{e_2}\langle \phi \rangle &= \{4.59\},
\end{align}
where again the entries correspond to values on different nodes. Note that the ratios $|(Y^{f}\langle \phi \rangle)_{ij}/(M^f)_{ij}|$ are on average equal to $q=5$, but have a distribution that allows for ${\mathcal O}(1)$ deviations (in relative terms) from this value. 

The Yukawa couplings on the zero node, which couple chiral fermions to the Higgs, are
\beq
Y_0^u=
\left(
\begin{array}{ccc}
	-0.129+0.367 i & -0.58-0.087 i & 0.85\, -0.006 i \\
	-0.502+0.183 i & 0.425\, -0.006 i & -0.747-0.129 i \\
	0.272\, +0.249 i & 0.225\, -0.223 i & -0.025-0.543 i \\
\end{array}
\right),
\eeq

\beq
Y_0^d=
\left(
\begin{array}{ccc}
	0.714\, +0.006 i & 0.239\, -0.244 i & 0.475\, -0.008 i \\
	-0.328+0.351 i & 0.227\, +0.181 i & 0.727\, +0.008 i \\
	-0.084-0.309 i & 0.727\, -0.004 i & 0.343\, +0.002 i \\
\end{array}
\right),
\eeq

\beq
Y_0^\ell=
\left(
\begin{array}{ccc}
	-0.632-0.165 i & -0.748-0.194 i & 0.775\, -0.067 i \\
	0.661\, +0.221 i & 0.229\, -0.854 i & -0.508-0.632 i \\
	0.05\, +0.752 i & 0.503\, -0.054 i & 0.333\, +0.587 i \\
\end{array}
\right).
\eeq
The absolute values, $r_a$, for each of the above entries were taken to be in the interval $r_a\in [0.3,0.9]$, when constructing the benchmark.  Finally, the matrix of coefficients  in the neutrino mass term, Eq. \eqref{eq:Weinberg:op}, is
\beq
c_{ij}^\ell=
\left(
\begin{array}{ccc}
	0.61\, -0.063 i & -0.507-0.009 i & -0.381+0.285 i \\
	-0.507-0.009 i & -0.071+0.504 i & 0.016\, +0.532 i \\
	-0.381+0.285 i & 0.016\, +0.532 i & -0.47+0.308 i \\
\end{array}
\right).
\eeq

In the benchmark we set the kinetic mixing between different $Z_i'$ to zero, as we do the kinetic mixing of $Z_i'$ with hypercharge. The mass eigenstates, $Z_i'$, therefore correspond to the gauge bosons coupling to the $i-$th fermion generation in the flavor basis, that is, before the electroweak symmetry breaking. After fermion mass diagonalization that includes the electroweak symmetry breaking terms, the real parts of the Hermitian coupling matrices in Eq. \eqref{eq:Z':decoupled} are, for $Z_1'$, 
\begin{align}	
\Re(c_{u_{L,1}})=&
\begin{pmatrix}
-2.909 & -0.484 & 0.015 \\
-0.484 & -0.081 & 0.002 \\
0.015 & 0.002 & 0 \\
\end{pmatrix}, \label{eq:cuL1}
&\Re(c_{d_L,1})=&
\begin{pmatrix}
-2.977 & 0.193 & -0.006 \\
0.193 & -0.012 & 0 \\
-0.006 & 0 & 0 \\
\end{pmatrix},
\\
\Re(c_{u_{R,1}})=&
\begin{pmatrix}
3.978 & -0.062 & -0.003 \\
-0.062 & 0.001 & 0 \\
-0.003 & 0 & 0 \\
\end{pmatrix}, 
&\Re(c_{d_{R,1}})=&
\begin{pmatrix}
3.654 & 0.682 & -0.186 \\
0.682 & 0.198 & -0.002 \\
-0.186 & -0.002 & 0.024 \\
\end{pmatrix},
\\
\Re(c_{\ell_{L,1}})=&
\begin{pmatrix}
-2.165 & 1.138 & 0.676 \\
1.138 & -0.599 & -0.356 \\
0.676 & -0.356 & -0.211 \\
\end{pmatrix},
&\Re(c_{\ell_{R,1}})=&
\begin{pmatrix}
3.988 & 0.036 & -0.001 \\
0.036 & 0.001 & 0 \\
-0.001 & 0 & 0 \\
\end{pmatrix},
\\
\Re(c_{\nu_{L,1}})=&
\begin{pmatrix}
-2.341 & -0.907 & 0.023 \\
-0.907 & -0.518 & 0.148 \\
0.023 & 0.148 & -0.116 \\
\end{pmatrix},
\end{align}
while the nonzero imaginary entries for the couplings to quarks are $\Im[(c_{u_{R,1}})_{\{12,13\}}]=\{0.006, -0.001\}$,
$\Im[(c_{d_{R,1}})_{\{12,13,23\}}]=\{0.508,0.233,0.069\}$,  and for the couplings to leptons, $\Im[(c_{\ell_{R,1}})_{\{12,13\}}]=\{0.036,-0.002\}$, $\Im[(c_{\nu_L,1})_{\{12,13,23\}}]=\{-0.624, 0.521, 0.196\}$. The coefficients below the diagonal, $i>j$, are given by $\Im(c_a)_{ij}=-\Im (c_a)_{ji}$ since $(c_a)_{ij}=(c_a)_{ji}^*$.

For $Z_2'$ the real parts of the couplings in Eq. \eqref{eq:Z':decoupled} are
\begin{align}
\Re(c_{u_{L,2}})=&
\begin{pmatrix}
-0.054 & 0.322 & -0.012 \\
0.322 & -1.932 & 0.071 \\
-0.012 & 0.071 & -0.003 \\
\end{pmatrix},
&\Re(c_{u_{R,2}})=&
\begin{pmatrix}
0 & 0.028 & 0.001 \\
0.028 & 1.819 & 0.093 \\
0.001 & 0.093 & 0.005 \\
\end{pmatrix},
\\
\Re(c_{d_{L,2}})=&
\begin{pmatrix}
-0.008 & -0.128 & 0.001 \\
-0.128 & -1.979 & 0.015 \\
0.001 & 0.015 & -0.001 \\
\end{pmatrix},
&
\Re(c_{d_{R,2}})=&
\begin{pmatrix}
0.063 & -0.082 & 0.051 \\
-0.082 & 0.455 & 0.832 \\
0.051 & 0.832 & 2.361 \\
\end{pmatrix},
\\
\Re(c_{\ell_{L,2}})=&
\begin{pmatrix}
-0.178 & -0.319 & -0.034 \\
-0.319 & -1.134 & 0.889 \\
-0.034 & 0.889 & -1.604 \\
\end{pmatrix},
&\Re(c_{\ell_{R,2}})=&
\begin{pmatrix}
0 & -0.009 & 0 \\
-0.009 & 0.991 & -0.038 \\
0 & -0.038 & 0.001 \\
\end{pmatrix},
\\
\Re(c_{\nu_{L,2}})=&
\begin{pmatrix}
-0.537 & 0.582 & -0.139 \\
0.582 & -1.024 & 0.87 \\
-0.139 & 0.87 & -1.355 \\
\end{pmatrix},\label{eq:cnuL2} 
\end{align}
while the nonzero imaginary entries are $\Im[(c_{u_{L,2}})_{\{13,23\}}] =\{-0.002,0.009\}$, \\
$\Im[(c_{u_{R,2}})_{\{12,13,23\}}] =\{-0.003,-0.001,-0.036\}$, 
$\Im[(c_{d_{L,2}})_{\{13,23\}}] =\{-0.002,0.03\}$,
\\
$\Im[(c_{d_{R,2}})_{\{12,13,23\}}] =\{-0.148,-0.381,0.617\}$, 
$\Im[(c_{\ell_{L,2}})_{\{12,13,23\}}] =\{0.317,-0.533,-1.014\}$, 
$\Im[(c_{\ell_{R,2}})_{\{12,23\}}] =\{-0.009,0.003\}$, 
$\Im[(c_{\nu_{L,2}})_{\{12,13,23\}}] =\{0.459,-0.842, 0.793\}$, with 
$\Im(c_a)_{ij}=-\Im (c_a)_{ji}$.

For $Z_3'$ the only nonzero coupling matrices are 
\begin{align}
c_{d_R,3}=&\
\begin{pmatrix}
0.104 & -0.43-0.232 i & 0.089\, +0.211 i \\
-0.43+0.232 i & 2.305 & -0.84-0.677 i \\
0.089\, -0.211 i & -0.84+0.677 i & 0.505 \\
\end{pmatrix},\label{eq:app:decoupl:ex}
\\
c_{\ell_L,3}=&
\begin{pmatrix}
-0.616 & -0.797-0.317 i & -0.629+0.533 i \\
-0.797+0.317 i & -1.195 & -0.54+1.014 i \\
-0.629-0.533 i & -0.54-1.014 i & -1.105 \\
\end{pmatrix},
\\
c_{\nu_L,3}=&
\begin{pmatrix}
-0.085 & 0.307\, +0.153 i & 0.116\, +0.331 i \\
0.307\, -0.153 i & -1.384 & -1.016-0.985 i \\
0.116\, -0.331 i & -1.016+0.985 i & -1.447 \\
\end{pmatrix}.\label{eq:app:decoupl:ex:cnuL3}
\end{align}

\subsection{The benchmark for the coupled FN chains}
\label{sec:app:coupledFN}
\begin{figure}[t]
	\begin{center}
		\includegraphics[width=1\linewidth]{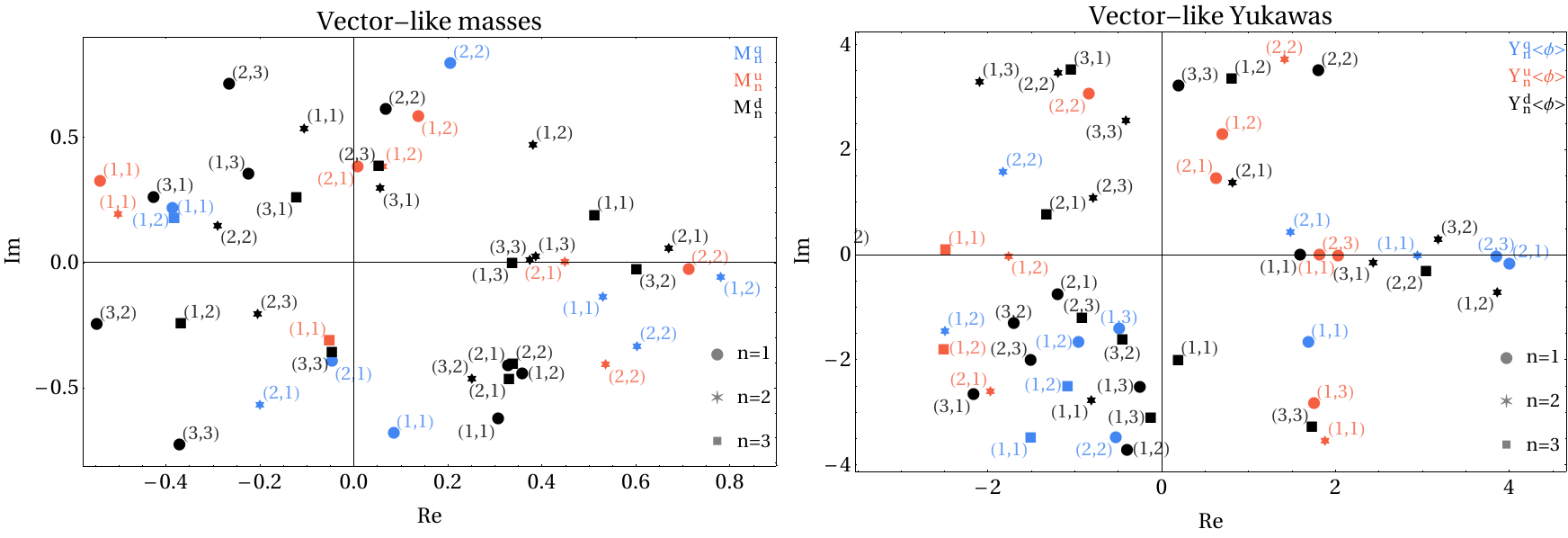}\\
		\includegraphics[width=1\linewidth]{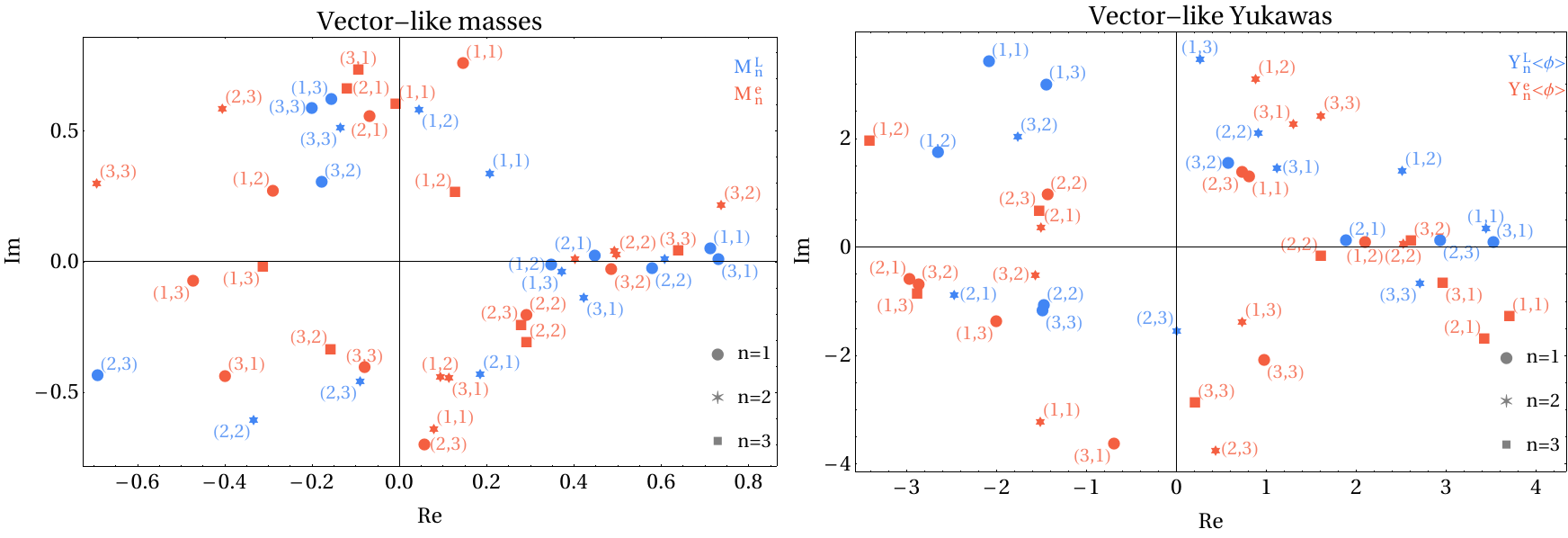}
		\caption{Numerical values for the entries in the vector-like fermion mass matrices for the  coupled FN chain benchmark in the case of quarks (top) and leptons (bottom).\label{fig:coupled_bench_mass_yukawa}
		}
	\end{center}
\end{figure}

The values of the vector-like mass matrices in complex plane for the $U(1)_{\rm FN}$ benchmark are shown in Fig.~\ref{fig:coupled_bench_mass_yukawa} (left) for $M^q_n$ in blue, $M_n^u$ in red, and $M_n^d$ in black, and in Fig.~\ref{fig:coupled_bench_mass_yukawa} (right) for $Y^q_n \langle \phi \rangle$ in blue, $Y_n^u \langle \phi \rangle$ in red, and $Y_n^d \langle \phi \rangle$ in black, in both cases in units of $10^7$ GeV. The values for the matrices on nodes $n=1(2,3)$ are denoted with a dot (star, square), with the labels denoting which element is being plotted, $(M_n^f)_{ij}\to (i,j)$. The Yukawa couplings on the zero node and the matrix of coefficients in the neutrino mass term are
\beq
Y_0^u=
\left(
\begin{array}{ccc}
	-0.003-0.542 i & 0.502\, +0.226 i & 0.14\, +0.262 i \\
	0.295\, +0.218 i & 0.706\, +0.004 i & 0.419\, +0.014 i \\
	0.28\, +0.279 i & -0.224+0.503 i & -0.607-0.484 i \\
\end{array}
\right),
\eeq
 
\beq
Y_0^d=
\left(
\begin{array}{ccc}
	0.016\, -0.437 i & -0.488+0.681 i & -0.224-0.135 i \\
	-0.59-0.591 i & 0.373\, -0.005 i & -0.407+0.068 i \\
	0.559\, -0.016 i & 0.014\, +0.421 i & 0.492\, +0.01 i \\
\end{array}
\right),
\eeq
\beq
Y_0^\ell=
\left(
\begin{array}{ccc}
	0.362\, -0.296 i & 0.326\, +0.572 i & -0.437+0.62 i \\
	0.298\, -0.005 i & 0.583\, +0.356 i & 0.114\, -0.341 i \\
	-0.092+0.312 i & -0.275+0.491 i & -0.368-0.28 i \\
\end{array}
\right),
\eeq
\beq
c_{ij}^\ell=
\left(
\begin{array}{ccc}
	-0.099+0.519 i & 0.56\, -0.041 i & -0.509+0.121 i \\
	0.56\, -0.041 i & -0.311-0.358 i & 0.403\, -0.719 i \\
	-0.509+0.121 i & 0.403\, -0.719 i & -0.699-0.324 i \\
\end{array}
\right).
\eeq

Fig.~\ref{fig:coupled_bench_mass_yukawa} shows that the entries are relatively uniformly distributed over the complex plane.  In constructing the benchmark we restricted the values of the amplitudes to lie within the same interval, $r\in [0.3,0.9]$, that was used in the numerical scan in Section \ref{sec:scan:coupled}.  The numerical values for the inputs are also available on request in the form of a {\tt Mathematica} notebook. 

After mass diagonalization, the real parts of the $Z'$ couplings in Eq. \eqref{eq:cfi:Z'} are 
\begin{align}
\label{eq:RecuL}
\Re(c_{u_L})&=
\begin{pmatrix}
-2.262 & 0.487 & -0.009 \\
0.487 & -2.665 & 0.012 \\
-0.009 & 0.012 & -0.001 \\
\end{pmatrix}, 
&\Re(c_{u_R})&=
\begin{pmatrix}
2.985 & 0.081 & 0 \\
0.081 & 1.918 & -0.093 \\
0 & -0.093 & 0.026 \\
\end{pmatrix},
\\
\Re(c_{d_L})&=
\begin{pmatrix}
-2.074 & -0.354 & -0.002 \\
-0.354 & -2.851 & -0.026 \\
-0.002 & -0.026 & -0.002 \\
\end{pmatrix},
&\Re(c_{d_R})&=
\begin{pmatrix}
2.987 & 0.007 & 0.017 \\
0.007 & 2.909 & -0.162 \\
0.017 & -0.162 & 2.494 \\
\end{pmatrix},
\\
\Re(c_{\ell_L})&=
\begin{pmatrix}
-1.982 & 0.024 & 0.007 \\
0.024 & -1.93 & -0.001 \\
0.007 & -0.001 & -1.936 \\
\end{pmatrix},
&\Re(c_{\ell_R})&=
\begin{pmatrix}
2.981 & 0.004 & 0.006 \\
0.004 & 2.956 & 0.017 \\
0.006 & 0.017 & 2.146 \\
\end{pmatrix},
\\
\label{eq:Recnu}
\Re(c_{\nu_L})&=
\begin{pmatrix}
-1.982 & 0.024 & 0.003 \\
0.024 & -1.935 & -0.001 \\
0.003 & -0.001 & -1.932 \\
\end{pmatrix},
\end{align}
while the imaginary parts of the couplings to quarks are $\Im[(c_{u_L})_{\{12,13,23\}}]=\{1.4,0.7,4.1\}\cdot 10^{-2}$,  $\Im[(c_{u_R})_{\{12,13,23\}}]=\{-3.6,1,20.3\}\cdot 10^{-2}$, $\Im[(c_{d_L})_{\{12,13,23\}}]=\{-1.1,-1.2,-7.1\}\cdot 10^{-2}$, $\Im[(c_{d_R})_{\{12,13,23\}}]=\{1.2 ,3.1,-6.3\}\cdot 10^{-2}$, while the imaginary parts of the couplings to leptons are $\Im[(c_{\ell_L})_{\{12,13,23\}}]=\{-0.2,-0.4, 1.1\}\cdot 10^{-2}$, $\Im[(c_{\ell_R})_{\{12,13,23\}}]=-\{0.3, -0.3,1.5\}\cdot 10^{-2}$, $\Im[(c_{\nu_L})_{\{12,13,23\}}]=\{-0.8, 0.7, 0.8\}\cdot 10^{-2}$. The elements below the diagonal are obtained from hermiticity of the coefficients,  $(c_a)_{ij}=(c_a)_{ji}^*$, so that $\Im(c_a)_{ij}=-\Im (c_a)_{ji}$.

\bibliographystyle{h-physrev}
\bibliography{FNbiblio}

\end{document}